\documentclass[aps,prd,onecolumn,nofootinbib]{revtex4-2} 
\usepackage{titlesec}
\titleformat*{\section}{\LARGE\bfseries}
\titleformat*{\subsection}{\Large\bfseries}
\titleformat*{\subsubsection}{\large\bfseries}

\usepackage{epsfig}
\usepackage{enumitem}

\usepackage{graphics}     
\usepackage{graphicx}     
\usepackage{subfigure}
\usepackage{longtable}     
\usepackage{url}          
\usepackage{bm}           
\usepackage{natbib}
\usepackage{textpos}
\usepackage{amsfonts,amsmath,amssymb,mathrsfs,bm,amsthm}
\usepackage{float}
\usepackage{booktabs}
\usepackage{array}
\usepackage{tabu}
\usepackage{dcolumn}
\usepackage{rotating}
\usepackage{ulem}

\newcommand{\RN}[1]{%
  \textup{\uppercase\expandafter{\romannumeral#1}}%
}

\usepackage{nicefrac}
\usepackage{amsthm}
\usepackage{color}
\usepackage{cancel}
\usepackage{appendix}
\usepackage{makecell}
\usepackage{upgreek}

\usepackage[colorlinks=true,linkcolor=black,citecolor=blue,urlcolor=blue,bookmarksopen]{hyperref}

\newcommand{\appropto}{\mathrel{\vcenter{
  \offinterlineskip\halign{\hfil$##$\cr
    \propto\cr\noalign{\kern2pt}\sim\cr\noalign{\kern-2pt}}}}}
\renewcommand{\v}[1]{\boldsymbol{#1}}		

\begin{document}

\title{\Large{
New bounds on macroscopic scalar-field topological defects \\from non-transient signatures due to environmental dependence \\and spatial variations of the fundamental constants
}}

\author{{\large\bf \rule[30pt]{0pt}{0pt}
Yevgeny~V.~Stadnik 
}}

\affiliation{{\large \rule[25pt]{0pt}{0pt}
Kavli Institute for the Physics and Mathematics of the Universe (WPI), The University of Tokyo Institutes for Advanced Study, The University of Tokyo, Kashiwa, Chiba 277-8583, Japan
}}

\raggedbottom


\maketitle

\vspace{3mm}
\begin{center}
\Large{\textbf{Abstract}}
\end{center}
\large

We point out that in models of macroscopic topological defects composed of one or more scalar fields that interact with standard-model fields via scalar-type couplings, the back-action of ambient matter on the scalar field(s) produces an environmental dependence of the fundamental constants of nature, as well as spatial variations of the fundamental constants in the vicinity of dense bodies such as Earth due to the formation of a ``bubble-like'' defect structure surrounding the dense body. 
In sufficiently dense environments, spontaneous symmetry breaking may be inhibited altogether for $\phi^2$ interactions, potentially delaying the cosmological production of topological defects to rather late times. 
We derive bounds on non-transient variations of the fundamental constants from torsion-pendulum experiments that search for equivalence-principle-violating forces, experiments comparing the frequencies of ground- and space-based atomic clocks, as well as ground-based clocks at different heights in the recent Tokyo Skytree experiment, and measurements comparing atomic and molecular transition frequencies in terrestrial and low-density astrophysical environments. 
Our results constrain the present-day mass-energy fraction of the Universe due to a network of infinite domain walls produced shortly after the BBN or CMB epochs to be $\Omega_{\textrm{walls},0} \ll 10^{-10}$ for the symmetron model with a $\phi^4$ potential and $\phi^2$ interactions, improving over CMB quadrupolar temperature anisotropy bounds by at least 5 orders of magnitude. 
Our newly derived bounds on domain walls with $\phi^2$ interactions via their effects of non-transient variations of the fundamental constants are significantly more stringent than previously reported clock- and cavity-based limits on passing domain walls via transient signatures (regardless of their possible degradation due to previously neglected effects of the strongly repulsive potential generated by Earth on the passing domain walls) and previous bounds from different types of non-transient signatures (by about 10 orders of magnitude for wall thicknesses comparable to the size of Earth), under the same set of assumptions.

\vspace{200mm}

\large

\tableofcontents
\vspace{200mm}

\section{Introduction}
\label{Sec:Intro}
Astrophysical and cosmological observations indicate that about $95\%$ of the matter-energy content of the Universe is in the form of dark components, the identities and properties of which remain a mystery \cite{PDG_2018_review}. 
Topological defects are solitonic configurations of fields that acquire stability via a non-trivial field topology and may be produced during a cosmological phase transition associated with the spontaneous breaking of an underlying symmetry \cite{Vilenkin_1985_TDs}. 
Zero-dimensional monopoles are a good candidate to explain the observed dark matter, while one-dimensional strings and two-dimensional walls may comprise a sub-dominant fraction of the dark components \cite{Spergel_1989_walls,Hindmarsh_2008_strings}. 
The transverse extent of such topological defects may be macroscopic. 
This possibility motivated a number of recent proposals to search for such macroscopic objects using networks of terrestrial detectors, such as magnetometers \cite{Pospelov_2013_TDM-magnetometers}, clocks \cite{Derevianko_2014_TDM-clocks} (see also the earlier work \cite{Olive-Wall_2011_TD-clock}), cavities and laser interferometers \cite{Stadnik_2015_DM-LIFO,Stadnik_2016_DM-cavities,Stadnik_2019_DM-LIFO}, as well as astrophysical networks of pulsars \cite{Stadnik_2014_TDM-pulsars}. 

Several clock- and cavity-based searches for transient variations of the fundamental constants induced by the passage of domain walls have been performed recently \cite{Wcislo_2016_TDM-cavity,Roberts_2017_TDM-GPS,Wcislo_2018_TDM-cavity,Roberts_2019_TDM-clocks}. 
In these papers, the authors assumed that the domain walls are composed of a scalar field $\phi$ that interacts with standard-model fields via quadratic-in-$\phi$ couplings and reported new limits on such objects that improved over other known bounds by as much as ten orders of magnitude. 
In our present work, we point out that a number of important non-transient signatures in these topological defect models with $\phi^2$ interactions had been overlooked in the previous literature and that, when these non-transient signatures are taken into account, existing data from different types of experiments (most notably torsion-pendulum experiments) give significantly more stringent bounds than those reported in Refs.~\cite{Wcislo_2016_TDM-cavity,Roberts_2017_TDM-GPS,Wcislo_2018_TDM-cavity,Roberts_2019_TDM-clocks}. 

The crucial point that has been overlooked in the earlier literature concerning models of scalar-field topological defects with $\phi^2$ interactions (or even linear-in-$\phi$ interactions) is that one generally cannot neglect the back-action of ambient matter on the scalar field(s). 
Indeed, the terrestrial energy density is $\sim 10^{30}$ times greater than the critical energy density of the Universe 
--- an energy density scale which topological defects may \textit{a priori} be expected to be associated with --- 
and so one cannot necessarily neglect the back-action of ambient matter on the scalar field \textit{even when the interaction constant is small}. 
Similar back-action effects of ambient matter on scalar fields are known to arise in various scalar-field models with $\phi^2$ interactions \cite{Damour-Farese_1993-phi2,Damour-Farese_1996-phi2,Pietroni_2005-symmetron,Olive-Pospelov_2008-FCs+symmetron,Khoury_2010-symmetron,Updahye_2013-symmetron,Khoury_2011-symmetron,Jaffe_2017-symmetron,Abele_2018-symmetron,Burrage_2018-symmetron,Brax-Fichet_2019-symmetron,Vardanyan_2019-symmetron,Suyama_2015-asymmetron,Pretorius_2016-asymmetron,Popchev_2016-asymmetron,Doneva_2016-asymmetron,Suyama_2017-asymmetron}, including models of oscillating scalar dark matter fields with $\phi^2$ interactions \cite{Stadnik_2015_DM-LIFO,Stadnik_2015_DM-VFCs,Stadnik_2016_DM-clocks,Blas-Sibiryakov_2017_DM-binaries,Hees_2018_DM-EP,Soerensen_2018_DM-BBN}. 
In models of scalar-field topological defects with $\phi^2$ interactions, the back-action of ambient matter on the scalar field(s) produces an environmental dependence of the fundamental constants, as well as spatial variations of the fundamental constants in the vicinity of a dense body such as Earth due to the formation of a ``bubble-like'' defect structure surrounding the dense body. 
These effects are most striking when spontaneous symmetry breaking becomes switched off altogether in sufficiently dense environments, which may potentially delay the cosmological production of topological defects to rather late times. 
Additionally, the potential generated by a dense body such as Earth can affect the propagation of a passing topological defect. 

In this paper, we derive bounds on non-transient variations of the fundamental constants arising in models of scalar-field topological defects with $\phi^2$ interactions by using existing data from torsion-pendulum experiments that search for equivalence-principle-violating forces, experiments comparing the frequencies of ground- and space-based atomic clocks as well as ground-based clocks at different heights, and measurements comparing atomic and molecular transition frequencies in terrestrial and low-density astrophysical environments. 
We find that existing torsion-pendulum data and recent clock comparison measurements at different heights within Tokyo Skytree, as well as existing laboratory and astrophysical spectroscopy data, place strong constraints on the maximum allowable mass-energy content stored in a network of ``infinite'' domain walls produced shortly after the big bang nucleosynthesis (BBN) or cosmic microwave background (CMB) epochs. 
Additionally, we point out that these same torsion-pendulum data and clock comparison measurements at different heights within Tokyo Skytree give significantly more stringent bounds on domain walls via the effects of non-transient variations of the fundamental constants, compared with previous bounds from astrophysical and laboratory measurements via different types of non-transient signatures, as well as recent clock- and cavity-based searches for transient signatures of passing domain walls. 
We also discuss phenomenological aspects of topological defect models with different types of interactions and models of non-topological solitons.

The structure of our paper is as follows. 
In Sec.~\ref{Sec:Main}, we consider the theory, cosmology and phenomenology of macroscopic topological defects composed of scalar fields that interact with standard-model fields via $\phi^2$ couplings, paying particular attention to the role of back-action of ambient matter on the scalar field. 
We point out the emergence of non-transient signatures in this case which are associated with environmental dependence and spatial variations of the fundamental constants, and derive bounds on these non-transient variations of the fundamental constants using existing data from various experiments and measurements. 
We also derive bounds on the associated bubble-like defect structures surrounding Earth and the Sun from the consideration of their gravitational effects. 
In Sec.~\ref{Sec:TDM}, we discuss phenomenological aspects of terrestrial searches for ``dark matter'' networks of topological defects. 
We demonstrate that existing torsion-pendulum data give significantly more stringent bounds on models of domain walls with $\phi^2$ interactions via non-transient signatures compared with previously reported limits from clock and cavity experiments that sought for transient signatures, with the recent clock comparison measurements at different heights within Tokyo Skytree also probing (via non-transient signatures) regions of parameter space that were inaccessible to previous searches for transient signatures. 
In Sec.~\ref{Sec:Discussion}, we summarise our findings and discuss our results in a broader context, including discussion of phenomenological aspects of topological defect models with different types of interactions and models of non-topological solitons. 

Throughout this work, unless explicitly stated otherwise, we adopt the natural system of units $\hbar = c =1$, where $\hbar$ is the reduced Planck constant and $c$ is the speed of light in vacuum.

\vspace{200mm}

\section{Macroscopic topological defects}
\label{Sec:Main}

\subsection{Theory}
\label{Sec:Main_Theory}
\begin{figure*}[t!]
\centering
\includegraphics[width=8.5cm]{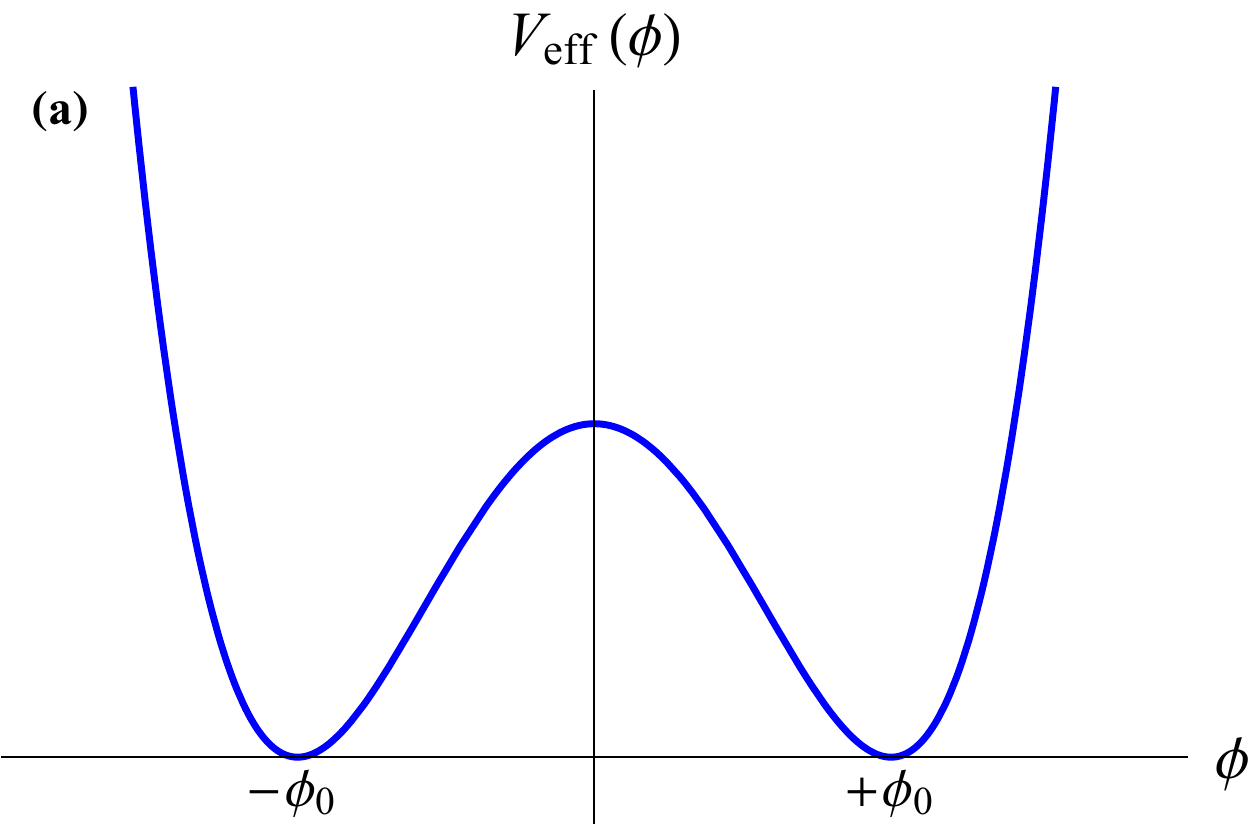}
\hspace{5mm}
\includegraphics[width=8.5cm]{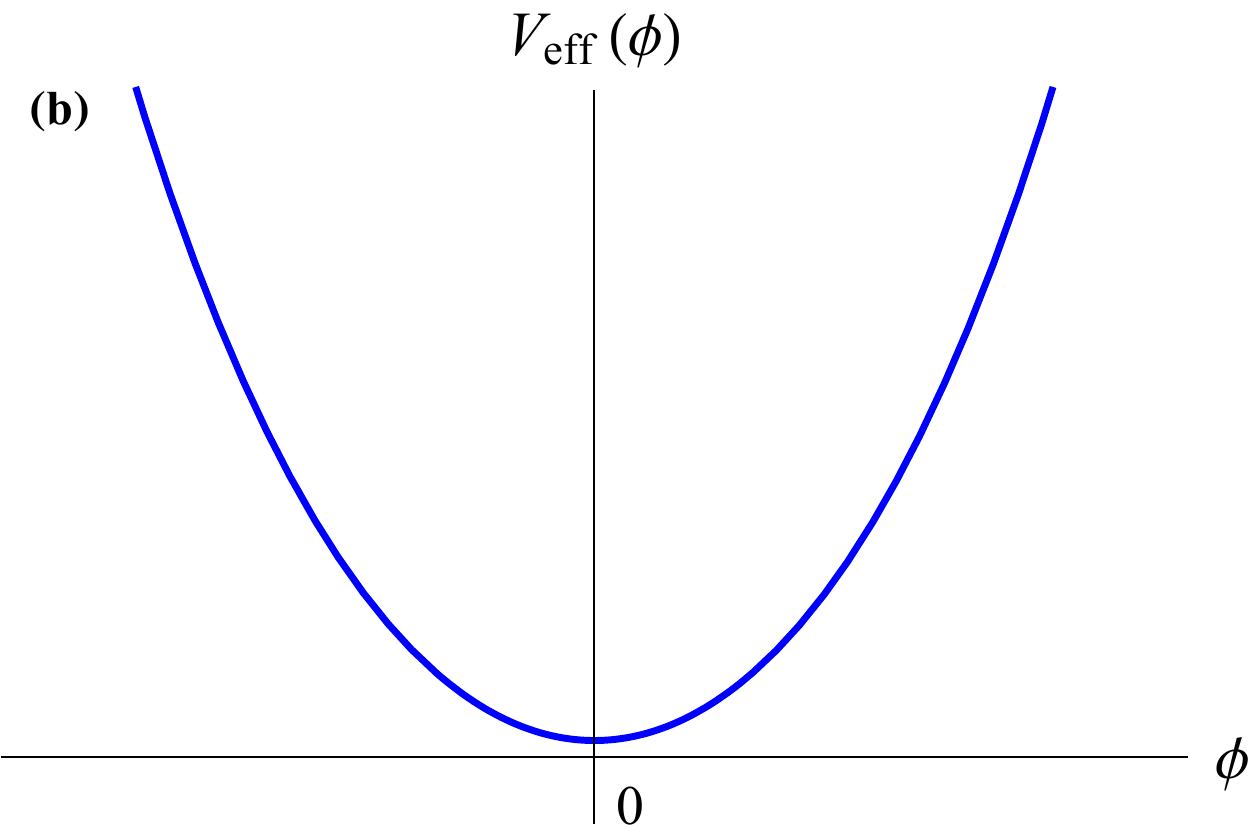}
\caption{ \normalsize (Colour online) 
Form of the effective potential (\ref{phi4_effective-potential}) in a \textbf{(a)} low-density environment when $2 \rho_X / (\Lambda'_X)^2 \ll \lambda \phi_0^2$ and \textbf{(b)} high-density environment when $2 \rho_X / (\Lambda'_X)^2 \gg \lambda \phi_0^2$. 
}
\label{Fig:Symmetron_potentials}
\end{figure*}

Topological defects are solitonic configurations of fields that acquire stability via a non-trivial field topology and may be produced during a cosmological phase transition associated with the spontaneous breaking of an underlying symmetry \cite{Vilenkin_1985_TDs}. 
As the simplest example, consider a single real scalar field $\phi$ with the following $\phi^4$ potential:
\begin{equation}
\label{phi4_potential}
V ( \phi ) = \frac{\lambda}{4} \left( \phi^2 - \phi_0^2 \right)^2  \, , 
\end{equation}
where $\lambda$ is a dimensionless parameter. 
The potential in (\ref{phi4_potential}) is invariant under the discrete $Z_2$ symmetry operation $\phi \to -\phi$. 
The field $\phi$ may settle in either of the two energetically equivalent minima at $\phi = \pm \phi_0$, which are separated by a potential barrier of height $\lambda \phi_0^4 / 4$, see Fig.~\ref{Fig:Symmetron_potentials}(a). 
In this case, the $Z_2$ symmetry is spontaneously broken, since the vacuum states associated with these minima do not exhibit the original symmetry present in the potential. 
If there exist two spatially separated regions of space with topologically distinct vacua, then a domain wall forms between the two vacua, with the following transverse ``kink'' profile \cite{Zeldovich-Okun_1974-DWs}: 
\begin{equation}
\label{wall_kink_solution}
\phi \left( x \right) = \phi_0 \tanh \left( x / d \right)  \, , 
\end{equation}
where the transverse size of the wall is set by: 
\begin{equation}
\label{wall_kink_thickness}
d = \sqrt{\frac{2}{\lambda}} \frac{1}{\phi_0}  \, . 
\end{equation}
The regions on either side of the wall are known as domains, by analogy with magnetic domains in ferromagnetic materials. 

Domain-wall transverse profiles akin to (\ref{wall_kink_solution}) also arise for other potentials, such as the Sine-Gordon potential \cite{Spergel_1989_walls}. 
With more than one scalar field and more complicated potentials, more complex topological wall structures become possible \cite{Bazeia2002_DWs-review}. 
Topological strings and monopoles require more complicated setups than topological walls. 
In particular, the minimal model admitting topological strings involves the Abelian U(1) gauge group with an associated complex scalar field $\phi$ and the following $\phi^4$-type potential \cite{Abrikosov_1957-vortices,Nielsen-Olesen_1973-strings}: 
\begin{equation}
\label{phi4_potential-strings}
V ( \phi ) = \lambda \left( \left| \phi \right|^2  - \frac{\phi_0^2}{2} \right)^2  \, , 
\end{equation}
while the minimal model admitting topological monopoles involves the non-Abelian SU(2) gauge group with an associated triplet of real scalar fields $\phi^a$, where the index $a=1,2,3$, and the following $\phi^4$-type potential \cite{Hooft_1974-monopoles,Polyakov_1974-monopoles}: 
\begin{equation}
\label{phi4_potential-monopoles}
V ( \phi^a ) = \frac{\lambda}{4} \left( \phi^a \phi^a - \phi_0^2 \right)^2  \, . 
\end{equation}

In our present paper, we focus on the simplest model admitting topological defects, namely ``infinite'' domain walls that stretch across the horizon size and arise from the $\phi^4$ potential in (\ref{phi4_potential}). 
There is no loss of generality in focusing on this minimal model, which was also the only model of topological domain walls explicitly presented in the theory paper \cite{Derevianko_2014_TDM-clocks} (see the Supplementary Information therein\footnote{\normalsize Let us remark that the single-scalar-field configuration described in the main text of \cite{Derevianko_2014_TDM-clocks}, based on a prescription of a single real scalar field $\phi$, which takes a non-zero value inside the ``defect'' object and tends to zero outside of the object, does \textit{not} describe a \textit{topological} defect. Indeed, models of topological strings and monopoles require at least one complex scalar field or several real scalar fields, respectively, as well as associated gauge fields. In the case of wall-like objects, the prescription in \cite{Derevianko_2014_TDM-clocks} implies that the vacua on either side of the wall ``defect'' are topologically equivalent (i.e., the vacua are identical and are not separated by a potential barrier) and hence this type of wall ``defect'' would lack topological stability.}), 
on which recent clock-based searches for transient signatures of passing domain walls were predicated. 
Indeed, the effects of back-action of ambient matter on the scalar field(s) comprising a topological soliton (including walls, strings and monopoles) which we study in our present paper are quite generic. 
Let us consider the following quadratic-in-$\phi$ interactions of the scalar field with the standard-model fields: 
\begin{equation}
\label{quad_portal}
\mathcal{L}_\textrm{int}^\textrm{quad} = \left(\frac{\phi}{\Lambda'_\gamma}\right)^2 \frac{F_{\mu\nu}F^{\mu\nu}}{4} - \sum_f \left(\frac{\phi}{\Lambda'_f}\right)^2 m_f \bar{f}f  \, , 
\end{equation}
where the first term represents the interaction of the scalar field with the electromagnetic field tensor $F$, while the second term represents the interaction of the scalar field with the standard-model fermion fields $f$, with $m_f$ being the ``standard'' mass of the fermion and $\bar{f} = f^\dagger \gamma^0$ the Dirac adjoint. 
Note that the choice of signs in (\ref{quad_portal}) is identical to those used in Ref.~\cite{Derevianko_2014_TDM-clocks}. 

When the ambient-matter fields are associated with non-relativistic atoms, we can present Eq.~(\ref{quad_portal}) in the following convenient form: 
\begin{equation}
\label{quad_portal-atoms}
\mathcal{L}_\textrm{int}^\textrm{quad} = - \sum_{X=\gamma,e,N} \frac{\rho_X \phi^2}{\left( \Lambda'_X \right)^2}  \, , 
\end{equation}
where $\rho_\gamma = |\v{E}|^2/2 \approx -F_{\mu\nu}F^{\mu\nu}/4$ is the Coulomb energy density of a non-relativistic nucleus, and $\rho_e$ and $\rho_N$ are the electron and nucleon energy densities, respectively. 
The effective potential experienced by the scalar field in the presence of ambient matter, therefore, reads: 
\begin{align}
V_\textrm{eff} \left( \phi \right) &= V ( \phi ) + \sum_{X=\gamma,e,N} \frac{\rho_X \phi^2}{\left( \Lambda'_X \right)^2}  \label{effective-potential_generic} \\       
&= \frac{\lambda}{4} \left( \phi^2 - \phi_0^2 \right)^2 + \sum_{X=\gamma,e,N} \frac{\rho_X \phi^2}{\left( \Lambda'_X \right)^2}  \, .   \label{phi4_effective-potential}
\end{align}
In a low-density environment when $2 \rho_X / (\Lambda'_X)^2 \ll \lambda \phi_0^2$, the effective potential in (\ref{phi4_effective-potential}) has two minima at $\phi \approx \pm \phi_0$ as shown in Fig.~\ref{Fig:Symmetron_potentials}(a), while in a high-density environment when $2 \rho_X / (\Lambda'_X)^2 \gg \lambda \phi_0^2$, the effective potential has only a single minimum at $\phi = 0$ as shown in Fig.~\ref{Fig:Symmetron_potentials}(b). 
The inhibition of spontaneous symmetry breaking in a high-density environment is a characteristic feature of the symmetron model, in which the effective potential takes the form (\ref{phi4_effective-potential}) and which has been the subject of numerous earlier works \cite{Pietroni_2005-symmetron,Olive-Pospelov_2008-FCs+symmetron,Khoury_2010-symmetron,Updahye_2013-symmetron,Khoury_2011-symmetron,Jaffe_2017-symmetron,Abele_2018-symmetron,Burrage_2018-symmetron,Brax-Fichet_2019-symmetron,Vardanyan_2019-symmetron}, but predominantly in a different context where cosmological domain walls are presumed absent. 

When the dense region of ordinary matter has a finite size and is surrounded by a large region of low-density ambient matter, a high density in the dense region is insufficient by itself to ensure that the scalar field will attain its minimum near $\phi = 0$ inside the dense region. 
The characteristic length scale, $d'$, over which the scalar field appreciably changes inside the dense region of ordinary matter must also be small compared to the size of the dense region. 
In the case of a spherical dense region of radius $R$, we thus require $d' \ll R$. 
We refer to this case as the ``strong screening'' regime, since the scalar field amplitude inside the dense region is suppressed compared to its value far away from the dense region. 
In the limit when $\rho_X / (\Lambda'_X)^2 \gg \lambda \phi_0^2$, we have $d' \ll d$, where: 
\begin{equation}
\label{wall_thickness-dense_region}
d' \approx \frac{\Lambda'_X}{\sqrt{2 \rho_X}}  \, . 
\end{equation}
Otherwise, the scalar field amplitude inside the dense region will only deviate slightly from its value far away from the dense region and the scalar field will not attain a minimum near $\phi = 0$ inside the dense region. 
We refer to this case as the ``weak screening'' regime, since the scalar field amplitude is practically unaffected by the presence of the dense region. 
In both the strong screening and weak screening regimes, the scalar field profile is deformed in the vicinity of a dense body, such as Earth, due to the back-action of ambient matter on the scalar field. 

\begin{figure*}[t!]
\centering
\includegraphics[width=8.5cm]{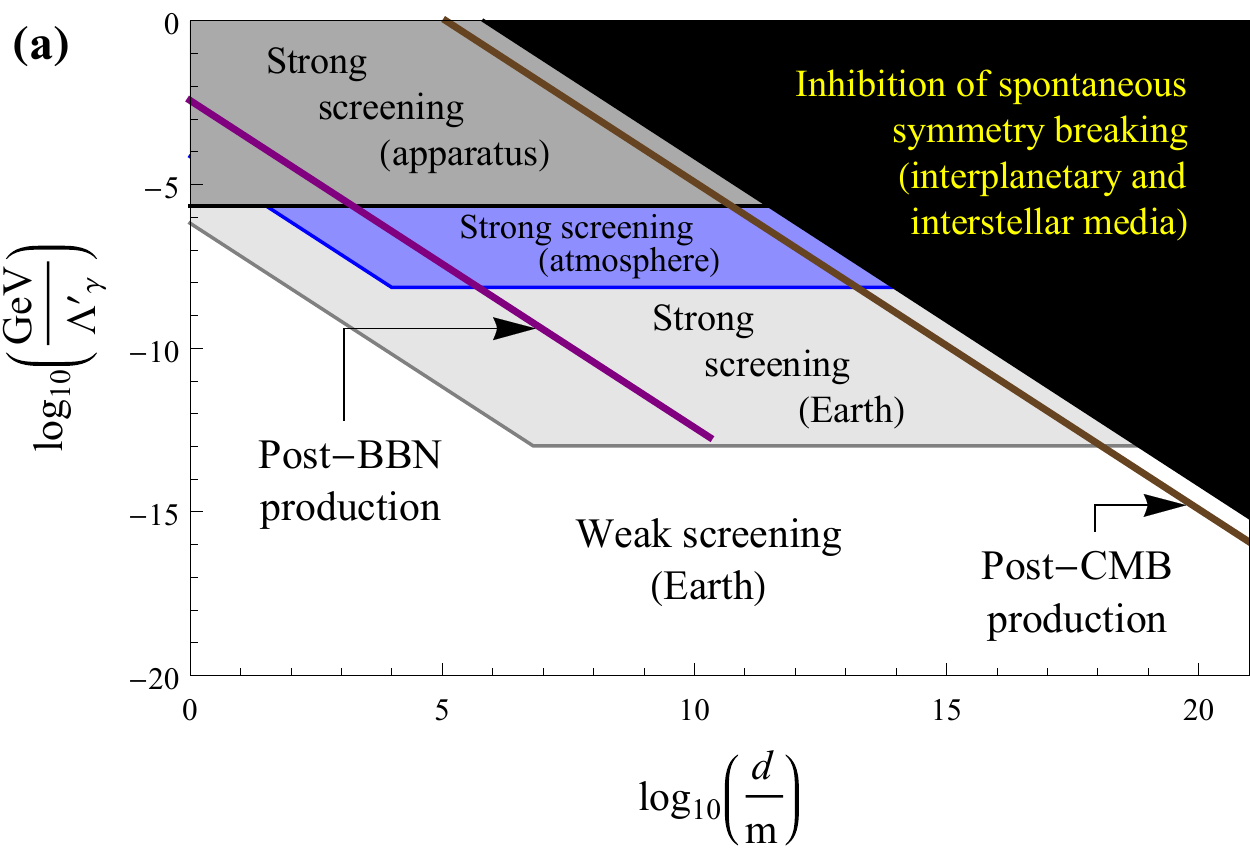}
\hspace{5mm}
\vspace{3mm}
\includegraphics[width=8.5cm]{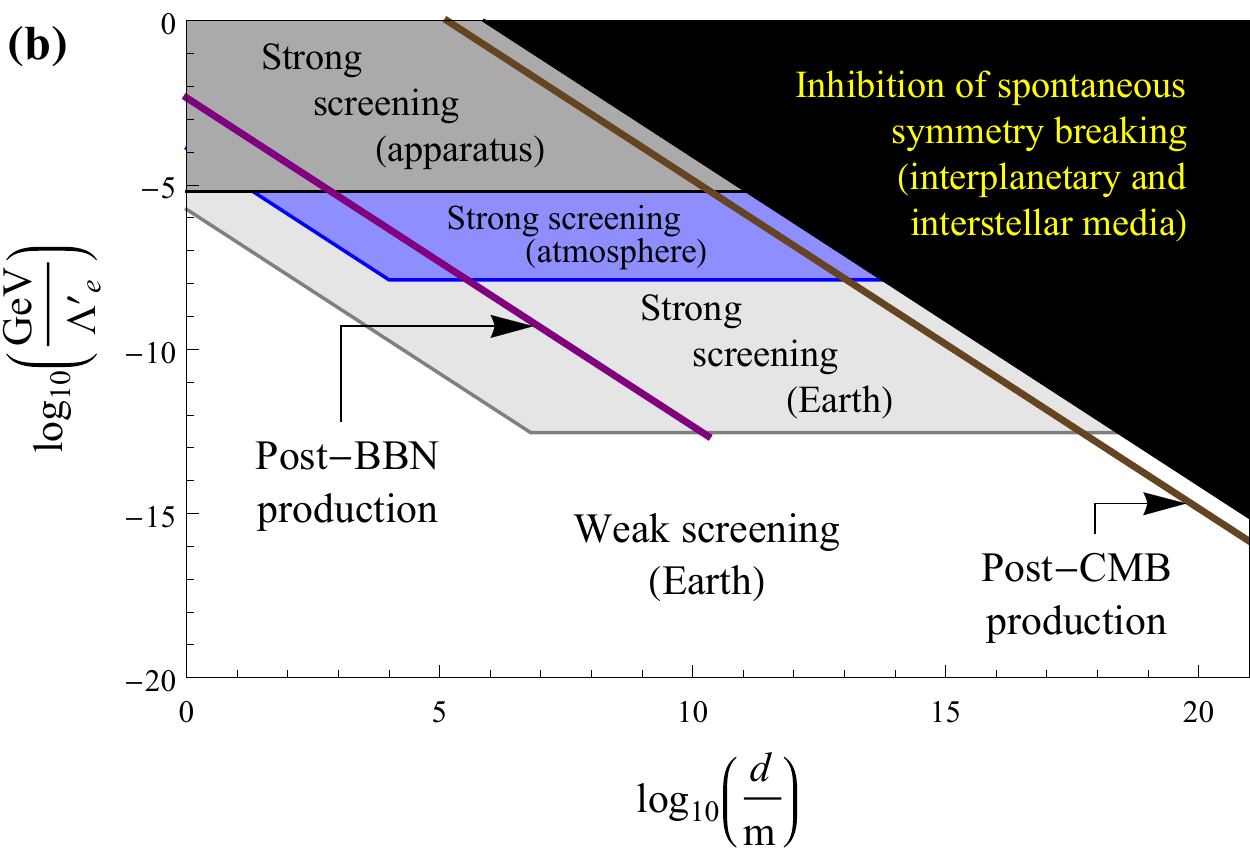}
\includegraphics[width=8.5cm]{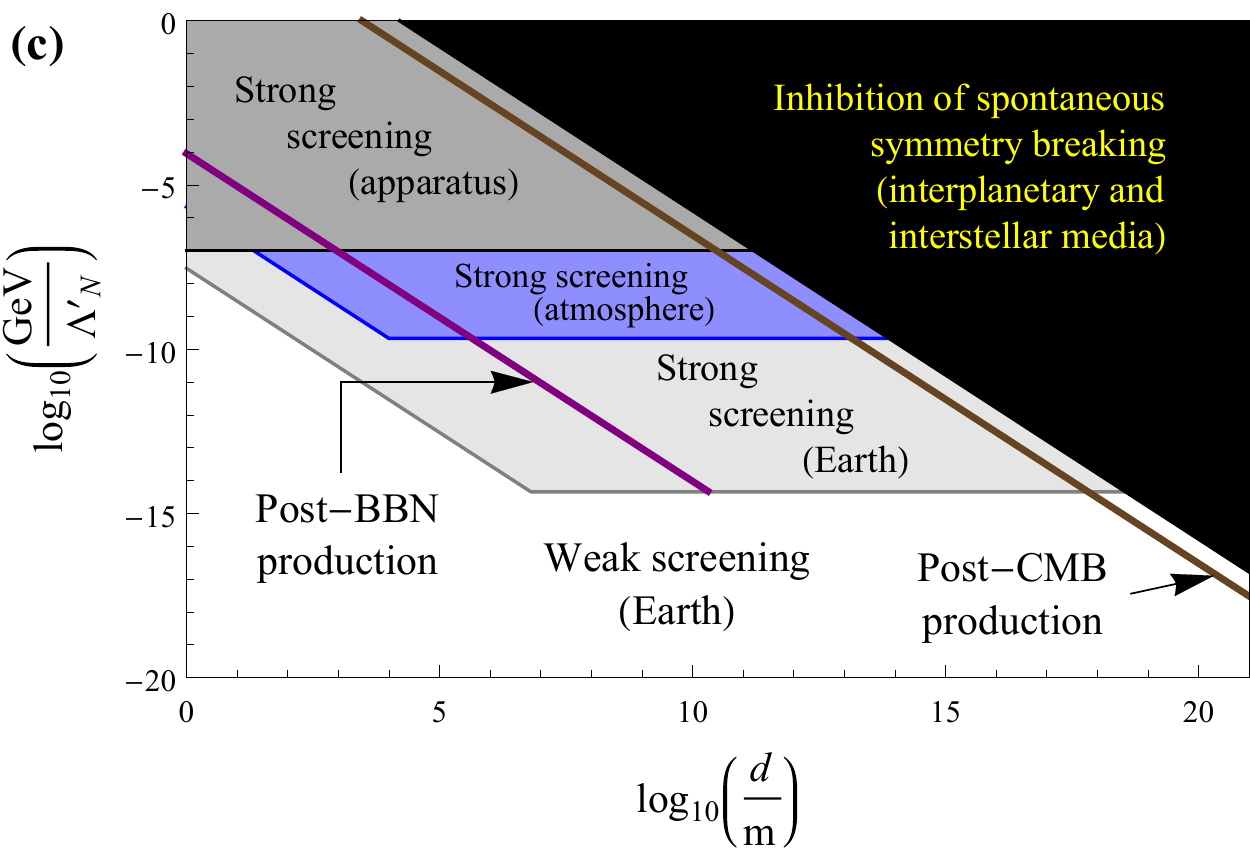}
\caption{ \normalsize (Colour online) 
Regions of domain-wall-model parameter spaces for the quadratic interactions of a scalar field $\phi$ with the \textbf{(a)} electromagnetic field (photon), \textbf{(b)} electron and \textbf{(c)} nucleons, as defined in (\ref{quad_portal}), 
in which the scalar field is weakly screened near the surface of and inside Earth (white region), 
strongly screened near the surface of and inside Earth (light grey region), 
strongly screened by Earth's atmosphere (blue region), 
strongly screened inside an apparatus or satellite of size $\sim 60~\textrm{cm}$ and with a density comparable to Earth's average density (dark grey region), 
and where spontaneous symmetry breaking is inhibited altogether by the interplanetary and interstellar media (black region). 
The purple and brown lines denote the parameters for which scalar-field domain walls can be cosmologically produced shortly after the BBN or CMB epochs, respectively. 
}
\label{Fig:Screening_regions}
\end{figure*}

To gain an understanding of the domain-wall parameter values for which the scalar field is strongly or weakly screened near the surface of and inside Earth, we recall that the main mass-energy contributions in an electrically neutral atom containing $A$ nucleons and $Z$ electrons are as follows: 
\begin{equation}
\label{atom_mass-energy_contributions}
M_\textrm{atom} \approx A m_N + Z m_e + \frac{a_C Z (Z-1)}{A^{1/3}} + Z a_p + (A-Z) a_n  \, . 
\end{equation}
The first two terms in (\ref{atom_mass-energy_contributions}) correspond to the nucleon and electron mass-energies, respectively. 
The third term corresponds to the energy associated with the electrostatic repulsion between protons in a spherical nucleus of uniform electric-charge density, with the coefficient $a_C \approx 0.7~\textrm{MeV}$ scaling proportionally to the electromagnetic fine-structure constant $\alpha$. 
The final two terms correspond to the electromagnetic energies of the proton and neutron, respectively, with the coefficients $a_p \approx +0.63~\textrm{MeV}$ and $a_n \approx -0.13~\textrm{MeV}$ derived from the application of the Cottingham formula \cite{Cottingham_1963} to electron-proton scattering \cite{Gasser-Leutwyler_1982}. 
Assuming that the elemental composition of Earth's interior is a 1:1:1 ratio of $^{24}\textrm{Mg}^{16}\textrm{O}$, $^{28}\textrm{Si}^{16}\textrm{O}_2$ and $^{56}\textrm{Fe}$ by number, we find the fractional mass-energy contributions to Earth's mass-energy content from the electromagnetic, electron-mass and nucleon-mass components as summarised in Table \ref{tab:frac_mass-energies}. 
Furthermore, noting that Earth's radius is $R_\oplus \approx 6400~\textrm{km}$ and Earth's average density is $\rho_\oplus \approx 5.5~\textrm{g/cm}^3$, we find the regions of domain-wall parameters for which the scalar field is strongly or weakly screened near the surface of and inside Earth, as shown in Fig.~\ref{Fig:Screening_regions} by the light grey and white regions, respectively. 
For simplicity, in the present work we neglect Earth's internal structure and density variations therein.

\begin{table}[t!]
\centering
\caption{\normalsize  
Summary of the fractional mass-energy contributions $K'_X$ due to electromagnetic ($X = \gamma$), electron-mass ($X = e$) and nucleon-mass ($X = N$) components in various systems, and analogous fractional mass-energy differences $\Delta K'_X$ in various test-mass pairs. 
We have assumed that the elemental composition of Earth's interior is a 1:1:1 ratio of $^{24}\textrm{Mg}^{16}\textrm{O}$, $^{28}\textrm{Si}^{16}\textrm{O}_2$ and $^{56}\textrm{Fe}$ by number, and that the elemental composition of Earth's atmosphere is a 4:1 ratio of $^{14}\textrm{N}_2$ and $^{16}\textrm{O}_2$ by number. 
We have assumed that the elemental composition of the interplanetary and interstellar media is $75\%$ $^{1}$H and $25\%$ $^{4}$He by mass, and have neglected the effects of stellar nucleosynthesis. 
}
\label{tab:frac_mass-energies}
\vspace{5mm}
\large
\begin{tabular}{ |c|c|c|c| }%
\hline
 System & $K'_\gamma$ & $K'_e$ & $K'_N$  \\ \hline 
 Earth's interior & $1.9 \times 10^{-3}$ & $2.4 \times 10^{-4}$ & $1.0$  \\ \hline 
 Earth's atmosphere & $9.5 \times 10^{-4}$ & $2.7 \times 10^{-4}$ & $1.0$  \\ \hline 
 Interplanetary and interstellar media & $6.3 \times 10^{-4}$ & $4.4 \times 10^{-4}$ & $1.0$  \\ \hline 
\multicolumn{1}{c}{} & \multicolumn{1}{c}{} & \multicolumn{1}{c}{} & \multicolumn{1}{c}{}  \\ 
\end{tabular}
\begin{tabular}{ |c|c|c|c| }%
\hline
 Test-mass pair & $\Delta K'_\gamma$ & $\Delta K'_e$ & $\Delta K'_N$  \\ \hline 
 Be-Ti & $-1.5 \times 10^{-3}$ & $-7.2 \times 10^{-6}$ & $+1.5 \times 10^{-3}$  \\ \hline 
 Be-Al & $-9.5 \times 10^{-4}$ & $-2.0 \times 10^{-5}$ & $+9.7 \times 10^{-4}$  \\ \hline 
 Pt-Ti & $+2.0 \times 10^{-3}$ & $-3.2 \times 10^{-5}$ & $-1.9 \times 10^{-3}$  \\ \hline 
\multicolumn{1}{c}{} & \multicolumn{1}{c}{} & \multicolumn{1}{c}{} & \multicolumn{1}{c}{}  \\ 
\end{tabular}
\end{table}

For sufficiently strong interactions of the form (\ref{quad_portal},\ref{quad_portal-atoms}), the scalar field may even become strongly screened inside the main components of an experimental apparatus and thus evade detection altogether. 
The transition from the weak screening regime to the strong screening regime in this case depends on the details of the apparatus, including the sizes, materials and geometries of the apparatus components, as well as the details of the surrounding shielding and laboratory environment. 
For the experiments of interest in our present paper, we shall simply model the whole apparatus or satellite as a uniform sphere of radius $R \sim 30~\textrm{cm}$ and with a density comparable to Earth's average density. 
In this case, the regions of domain-wall parameters for which the scalar field is strongly screened inside the apparatus or satellite are shown in Fig.~\ref{Fig:Screening_regions} by the dark grey regions. 
In order to probe these dark grey regions, measurements should generally be performed using apparata with sufficiently small components and/or systems with sufficiently low densities.

The scalar field may also be strongly screened by Earth's atmosphere, in which case the scalar field would evade detection by ground-based experiments on Earth. 
Treating Earth's atmosphere as a 4:1 ratio of $^{14}\textrm{N}_2$ and $^{16}\textrm{O}_2$ by number, with a constant density of $\approx 10^{-3}~\textrm{g/cm}^3$ and extending out from Earth's surface to an altitude of $\approx 10~\textrm{km}$, 
we find the regions of domain-wall parameters for which the scalar field is strongly screened by Earth's atmosphere, as shown in Fig.~\ref{Fig:Screening_regions} by the blue region. 
The density of Earth's atmosphere decreases rapidly with increasing height. 
At an altitude of $\approx 10^3~\textrm{km}$, above which hydrogen and helium become the main components of Earth's outer atmosphere, the particle density is $\sim 10^4~\textrm{cm}^{-3}$ and the transition to the strong screening regime by Earth's outer atmosphere requires even stronger interactions than for the transition to the regime of strong screening of the scalar field inside a satellite. 
Additionally, for sufficiently strong interactions, spontaneous symmetry breaking may be inhibited altogether within the Solar System and Galaxy by the interplanetary and interstellar media, respectively, in which case $\phi = 0$ in all local experiments. 
The transition to the interplanetary medium occurs at an altitude of $\sim 10^4~\textrm{km}$, where the particle density (which varies with time) is typically $\sim 10~\textrm{cm}^{-3}$. 
The transition from the interplanetary medium to the interstellar medium occurs at a distance of $\sim 100$ astronomical units from the Sun. 
Since the average energy density of the interstellar medium in our local galactic neighbourhood is comparable to the typical interplanetary medium energy density near Earth, for simplicity we neglect variations in the densities of the interplanetary and interstellar media within the Solar System and Galaxy (unless explicitly stated otherwise). 
Furthermore, since we will be restricting our attention to domain-wall transverse size parameters $d$ less than the size of our Galaxy ($\sim 30~\textrm{kpc}$ across), we shall also neglect the finite size of our Galaxy and the surrounding intergalactic medium (which has a much lower average energy density than those of the interplanetary and interstellar media). 
Assuming that the elemental composition of the interplanetary and interstellar media is $75\%$ $^{1}$H and $25\%$ $^{4}$He by mass, we find the regions of domain-wall parameters for which spontaneous symmetry breaking is inhibited altogether by the interplanetary and interstellar media, as shown in Fig.~\ref{Fig:Screening_regions} by the black region.

In addition to the back-action effects of ambient matter on the scalar field discussed above, the interactions in Eq.~(\ref{quad_portal}) also cause the scalar field to affect the properties of ordinary matter via apparent variations of the physical ``constants'', including changes in the strengths of fundamental interactions and variations in particle masses. 
Comparing the terms in Eq.~(\ref{quad_portal}) with the relevant terms in the standard-model Lagrangian: 
\begin{equation}
\label{SM_Lagrangian}
\mathcal{L}_\textrm{SM} \supset - \frac{F_{\mu\nu}F^{\mu\nu}}{4} - \sum_f  q_f J_\mu A^\mu - \sum_f  m_f \bar{f}f  \, ,
\end{equation}
where $q_f$ is the electric charge carried by the fermion $f$, $J^\mu = \bar{f} \gamma^\mu f$ is the electromagnetic 4-current and $A^\mu$ is the electromagnetic 4-potential, 
we see that the interactions in (\ref{quad_portal}) effectively alter the electromagnetic fine-structure constant $\alpha$ and fermion masses according to: 
\begin{gather}
\alpha \to \frac{\alpha}{1 - \left(\phi/\Lambda'_\gamma\right)^2} \approx \alpha \left[ 1 + \left(\frac{\phi}{\Lambda'_\gamma}\right)^2 \right]  \, ,  \label{quad_VFCs-alpha} \\ 
m_f \to m_f \left[1 + \left(\frac{\phi}{\Lambda'_f}\right)^2 \right] \, . \label{quad_VFCs-fermions} 
\end{gather}

Since the scalar field profile is deformed in the vicinity of a dense body, such as Earth, due to the back-action of ordinary matter on the scalar field, and because more generally the scalar field amplitude depends on the local ambient matter density, the apparent values of the fundamental constants will therefore depend on the environment in which measurements are performed and there will also be spatial variations of the fundamental constants in the vicinity of dense bodies. 
The phenomenology of these environmental dependencies and spatial variations of the fundamental constants, caused by the reciprocal effects of ambient matter and the scalar field on one another, is the central focus of our present work. 
Although we focus on the interactions (\ref{quad_portal}) in our present work, we note that similar types of reciprocal effects of ambient matter and the scalar field on one another also arise for other types of interactions (see Sec.~\ref{Sec:Discussion}).

\subsection{Cosmology}
\label{Sec:Main_Cosmology}
A generic mechanism for the production of topological defects is the so-called Kibble mechanism \cite{Kibble_1976}. 
The Kibble mechanism operates during the cosmological phase transition epoch during which the underlying symmetry becomes spontaneously broken and the scalar field develops a non-zero expectation value. 
The phase transition proceeds independently in causally disconnected regions of Hubble size $l_H \sim 1/H$, where $H$ is the Hubble parameter as a function of time ($H \sim 1/t$ in a matter- or radiation-dominated Universe). 
Inside any given Hubble region, the scalar field configuration is correlated up to some correlation length $l_\textrm{cor}$, which cannot exceed the Hubble size, $l_\textrm{cor} \lesssim l_H$. 
Within a particular correlation volume, the system settles in one and the same vacuum. 
The vacuum states in different correlation volumes are uncorrelated and so a topological defect may form at the boundary between adjacent correlation volumes. 
On average, $\mathcal{O}(1)$ topological defect is produced per correlation volume $\sim l_\textrm{cor}^3$, meaning that there should be at least $\mathcal{O}(1)$ topological defects per Hubble volume $\sim l_H^3$. 

The interactions (\ref{quad_portal},\ref{quad_portal-atoms}) provide a natural mechanism to cosmologically produce topological defects at rather late times via the Kibble mechanism, distinct from inflationary production scenarios (see, e.g., \cite{Dolgov-Rudenko_2018-DW_cosmo}). 
At early times when $2 \rho_X / (\Lambda'_X)^2 \gg \lambda \phi_0^2$, there is no spontaneous symmetry breaking and the scalar field remains at $\phi = 0$. 
Then at the later critical time when $2 \rho_X / (\Lambda'_X)^2 = \lambda \phi_0^2$, the cosmological phase transition can commence and the scalar field can start to roll towards one of the two new minima of the effective potential (\ref{phi4_effective-potential}). 
The energy density of either a non-relativistic or relativistic component decreases with time at least as rapidly as $\rho_X \propto t^{-3/2}$, regardless of whether the Universe is matter or radiation dominated, and so the transition from the regime $2 \rho_X / (\Lambda'_X)^2 \gg \lambda \phi_0^2$ to $2 \rho_X / (\Lambda'_X)^2 \ll \lambda \phi_0^2$ occurs within $\mathcal{O}(1)$ Hubble time $t_H \sim 1/H$. 
In other words, the phase transition occurs rather quickly. 
This phase transition is second order, since the scalar field smoothly develops a non-zero expectation value during the transition, rather than abruptly (first-order phase transition). 

The evolution of a scalar field in an expanding Universe is governed by the following time-dependent differential equation: 
\begin{equation}
\label{scalar_field_evolution_DE}
\ddot{\phi} + 3H(t) \dot{\phi} - \v{\nabla}^2 \phi = -\frac{\partial V_\textrm{eff}}{\partial \phi}  \, , 
\end{equation}
where $\v{\nabla}$ is the spatial gradient operator with respect to proper (non-expanding) spatial coordinates. 
In order for the scalar field to roll down from $\phi = 0$ to one of the two new minima at $\phi = \pm \phi_0$ within $\mathcal{O}(1)$ Hubble time, the scalar-field amplitude must change by the amount $|\delta \phi| \sim |\dot{\phi}/H| \sim \phi_0$ in the same time interval. 
This ``fast roll'' regime requires $|H \dot{\phi}| \ll |\partial V_\textrm{eff} / \partial \phi| \sim |\v{\nabla}^2 \phi|$ during the formation of the topological defect. 
For the effective potential (\ref{phi4_effective-potential}), this is equivalent to the requirement $d \ll l_H$, which corresponds to the case when the topological defects are well separated and evolve adiabatically. 
Otherwise, the scalar field remains stuck near $\phi = 0$ due to ``Hubble friction'' and topological defects do not form. 
In other words, the maximum transverse size of a topological defect produced during a phase transition at time $t_*$ is $d_\textrm{max} \sim l_H(t_*)$. 
In Fig.~\ref{Fig:Screening_regions}, we show the regions of domain-wall parameters in the case of interactions (\ref{quad_portal}) for which domain walls are produced shortly after the BBN (redshift $z \sim 10^8$) or CMB ($z \approx 10^3$) epochs by the purple and brown lines, respectively. 
We focus on these two cosmologically interesting epochs, but note that domain walls may also be produced via the same mechanism at different times. 

The subsequent evolution of the network of topological defects depends on the dynamics of the defect network. 
We focus on a network of infinite domain walls arising from the $\phi^4$ potential (\ref{phi4_potential}). 
In the limiting case that the initial network of domain walls undergoes little or no dynamical evolution after formation, the total proper area of the walls simply grows as $\propto a^2$ with the scale factor $a(t)$ and so the average energy density associated with the walls scales as $\rho_\textrm{walls} \propto a^2/a^3 \propto a^{-1}$. 
Numerical simulations \cite{Spergel_1989_walls,Hindmarsh2003walls,Avelino2004walls,Avelino2005walls}, however, show no evidence of such ``frustrated'' domain-wall networks, but instead reveal that only $\mathcal{O}(1)$ domain walls survive per Hubble volume, with the average energy density associated with the walls scaling approximately as $\rho_\textrm{walls} \propto t^{-1}$ (which decreases faster with time than $\rho_\textrm{walls} \propto a^{-1}$ in a matter- or radiation-dominated Universe). 
Numerical simulations with more complicated potentials and more complicated types of hybrid wall networks \cite{Avelino2006walls,Avelino2010walls} also find no evidence of frustrated wall networks. 
The physical explanation for the apparent lack of frustrated domain-wall networks is that the collisions of domain walls lead to wall reconnection and annihilation, thereby ``clearing'' the volume around the few surviving domain walls and hence decreasing the energy stored in the wall network. 
We note that, in the presence of interactions (\ref{quad_portal},\ref{quad_portal-atoms}), the average energy density of the wall network may decrease even faster than $\rho_\textrm{walls} \propto t^{-1}$, due to the collisions of walls with dense bodies of ordinary matter potentially leading to the loss of wall energy via ``pinching off'' of parts of the walls and their subsequent radiation (see Sec.~\ref{Sec:TDM_Transients}). 

To determine the relationship between the average energy density associated with a network of domain walls and the wall parameters, we note that the energy density inside a single wall is given by $\rho_\textrm{inside} \sim \phi_0^2/d^2$. 
The energy associated with a network of $N_\textrm{walls}$ walls is hence given by $E_\textrm{walls} \sim \rho_\textrm{walls} l_H^3 \sim N_\textrm{walls} \rho_\textrm{inside} d l_H^2$. 
We thus arrive at the following relation: 
\begin{equation}
\label{DW_amplitude-infinite_network}
\phi_0^2 \sim \frac{\rho_\textrm{walls} d}{N_\textrm{walls} H}  \, . 
\end{equation}
The existence of a present-day network of infinite domain walls would give rise to a quadrupolar temperature anisotropy in the CMB via the Sachs-Wolfe effect \cite{Sachs-Wolfe1967}, with an amplitude of $(\delta T / T)_\textrm{quad} \sim \Omega_{\textrm{walls},0} = \rho_{\textrm{walls},0}/\rho_{c,0}$ \cite{Spergel_1989_walls}, where $\Omega_{\textrm{walls},0}$ is the present-day mass-energy fraction of the Universe due to the wall network and $\rho_{c,0} \sim 10^{-29}~\textrm{g/cm}^{3}$ is the present-day critical density of the Universe. 
From the Planck 2018 temperature power spectrum \cite{Planck2018}, the size of an extra source of quadrupolar temperature anisotropy is bounded to be $(\delta T / T)_\textrm{quad} \lesssim 10^{-5}$, and so we have the following bound: 
\begin{equation}
\label{Omega_walls_bound}
\Omega_{\textrm{walls},0} \lesssim 10^{-5}  \, . 
\end{equation}
We stress that the bound (\ref{Omega_walls_bound}) is based purely on the gravitational effects of domain walls of cosmological origin. 
Regarding the number of domain walls at the present epoch, the earliest numerical simulations in Ref.~\cite{Spergel_1989_walls} indicated that $N_\textrm{walls}$ grows logarithmically in time according to the empirical relation $N_\textrm{walls} \approx 1 + 0.2 \ln(t/d)$, which gives $N_{\textrm{walls},0} \sim \mathcal{O}(10)$ for domain walls with the macroscopic transverse size parameters of interest in our present work. 
The authors of \cite{Spergel_1989_walls} attribute this apparent logarithmic growth in time to the self-avoiding random-walk nature of the wall collisions and subsequent reconnections and annihilations, and suggest that the walls adopt a ``stacked'' formation over time to avoid collisions with one another. 
The more recent numerical simulations in Refs.~\cite{Hindmarsh2003walls,Avelino2004walls,Avelino2005walls}, however, do not provide strong evidence for this logarithmic factor and indicate that only $\mathcal{O}(1)$ walls survive to the present epoch. 
We shall make the minimalistic assumption that $N_{\textrm{walls}} \sim \mathcal{O}(1)$ at all times, which leads to slightly stronger limits in Sec.~\ref{Sec:Main_Phenomenology} than for $N_{\textrm{walls},0} \sim \mathcal{O}(10)$. 

\subsection{Phenomenology}
\label{Sec:Main_Phenomenology}
We begin by discussing non-transient signatures that are independent of the parameter $\phi_0$ appearing in (\ref{phi4_potential},\ref{phi4_effective-potential}). 
First of all, the interactions in (\ref{quad_portal}) allow the emission of $\phi$ particles from hot environments, such as the interiors of stars. 
In particular, pairs of $\phi$ particles can be produced via the pair annihilation of photons $\gamma \gamma \to \phi \phi$, electron-positron annihilation $e^- e^+ \to \phi \phi$, the Compton-type process $e \gamma \to e \phi \phi$, and the nucleon-bremsstrahlung-type process $N N \to N N \phi \phi$. 
Constraints on the emission of new particles from the core of supernova 1987A lead to the following bounds on the interactions (\ref{quad_portal}) \cite{Olive-Pospelov_2008-FCs+symmetron}: 
\begin{gather}
\Lambda'_\gamma \gtrsim 3~\textrm{TeV}  \, ,  \label{SNe_bound-photon} \\ 
\Lambda'_e \gtrsim 400~\textrm{GeV}  \, ,  \label{SNe_bound-electron} \\ 
\Lambda'_N \gtrsim 15~\textrm{TeV}  \, .  \label{SNe_bound-nucleon} 
\end{gather}
Let us remark that such emission processes require the effective mass of the $\phi$ particles to be less than the core temperature of the supernova $T_\textrm{core} \sim 30~\textrm{MeV}$, otherwise the $\phi$ particles will not be emitted from the supernova due to the law of conservation of energy. 
In a very dense environment, $m_\textrm{eff}^2 \approx 2 \rho_X / (\Lambda'_X)^2$, which for a supernova core density of $\rho_\textrm{core} \sim 3 \times 10^{14}~\textrm{g/cm}^3$ requires $\Lambda'_N \gtrsim \textrm{GeV}$. 
Additionally, the interactions in (\ref{quad_portal}) induce quantum forces mediated by the exchange of a pair of $\phi$ particles between two bodies. 
These quantum forces are constrained by short-range tests of gravity \cite{Adelberger2007Short-range_gravity-A,Adelberger2007Short-range_gravity-B}, leading to the following bound on the nucleon interaction in (\ref{quad_portal}) \cite{Olive-Pospelov_2008-FCs+symmetron,Brax-Fichet_2019-symmetron}: 
\begin{equation}
\label{quantum_forces_bound-nucleon}
\Lambda'_N > 2~\textrm{TeV}  \, , 
\end{equation}
with weaker bounds on the electromagnetic and electron interactions. 

We shall now demonstrate, however, that there are generally much stronger bounds on the interactions (\ref{quad_portal}) for macroscopic domain walls, produced under the minimal cosmological assumptions outlined in Sec.~\ref{Sec:Main_Cosmology} above, stemming from different types of non-transient signatures that depend on the parameter $\phi_0$ and arise from an environmental dependence and spatial variations of the fundamental constants due to the reciprocal effects of ambient matter and the domain-wall scalar field on one another. 
The scalar field profile is deformed in the vicinity of a dense body due to the back-action of ambient matter on the scalar field via the interactions (\ref{quad_portal},\ref{quad_portal-atoms}). 
The geometrical features of the scalar field deformation are determined by the geometry of the dense body. 
Around a spherical dense body, such as Earth, the scalar field deformation resembles a spherical ``bubble-like'' defect structure. 
We derive the scalar field profile around a homogeneous spherical body of radius $R$ in Appendix~\ref{Sec:Appendix}. 
In the weak screening regime when $2 \rho_X / (\Lambda'_X)^2 \ll \lambda \phi_0^2$ or $d' \gg R$, where $d'$ is given by Eq.~(\ref{wall_thickness-dense_region}) in the limiting case when $\rho_X / (\Lambda'_X)^2 \gg \lambda \phi_0^2$, the scalar field profile outside of the spherical body reads: 
\begin{equation}
\label{scalar_profile_weak_screening-A}
\phi_\textrm{out}(r) \approx \phi_0 - \frac{\phi_0 \rho_X d^2 R}{4 \left( \Lambda'_X \right)^2 r} \exp \left[ - \frac{2 \left( r - R \right)}{d} \right]  ~~~\textrm{for} ~ d \ll R  \, , 
\end{equation}
\begin{equation}
\label{scalar_profile_weak_screening-B}
\phi_\textrm{out}(r) \approx \phi_0 - \frac{2 \phi_0 \rho_X R^3}{3 \left( \Lambda'_X \right)^2 r} \exp \left[ - \frac{2 \left( r - R \right)}{d} \right]  ~~~\textrm{for} ~ d \gg R  \, . 
\end{equation}
On the other hand, in the strong screening regime when $2 \rho_X / (\Lambda'_X)^2 \gg \lambda \phi_0^2$ and $d' \ll R$, the scalar field profile outside of the spherical body reads as follows in the limiting cases when $d \ll R$ or $d \gg R$: 
\begin{equation}
\label{scalar_profile_strong_screening-A}
\phi_\textrm{out}(r) \sim \frac{\phi_0 d' + \phi_0 R (1 - R/r)}{\textrm{min} \left( d , R \right)}  ~~~\textrm{for} ~ r - R \ll \textrm{min} \left( d , R \right)  \, , 
\end{equation}
\begin{equation}
\label{scalar_profile_strong_screening-Bx}
\phi_\textrm{out}(r) - \phi_0 \sim - \frac{\phi_0 R}{r} \exp \left[ - \frac{2 \left( r - r_* \right)}{d} \right] ~~~\textrm{for} ~ r - R \gg \textrm{min} \left( d , R \right)  \, , 
\end{equation}
where $r_* - R \sim \textrm{min}(d,R)$. 

\begin{figure*}[t!]
\centering
\includegraphics[width=8.5cm]{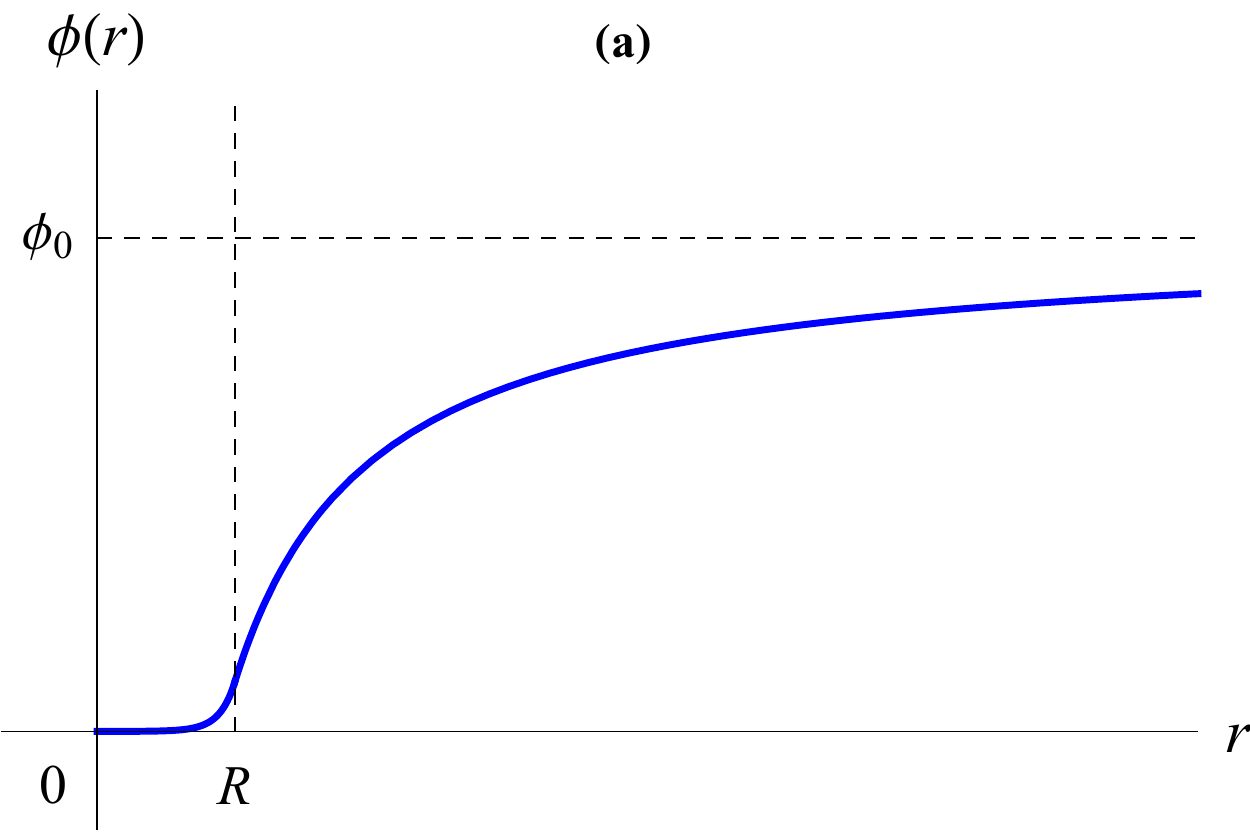}
\hspace{5mm}
\includegraphics[width=8.5cm]{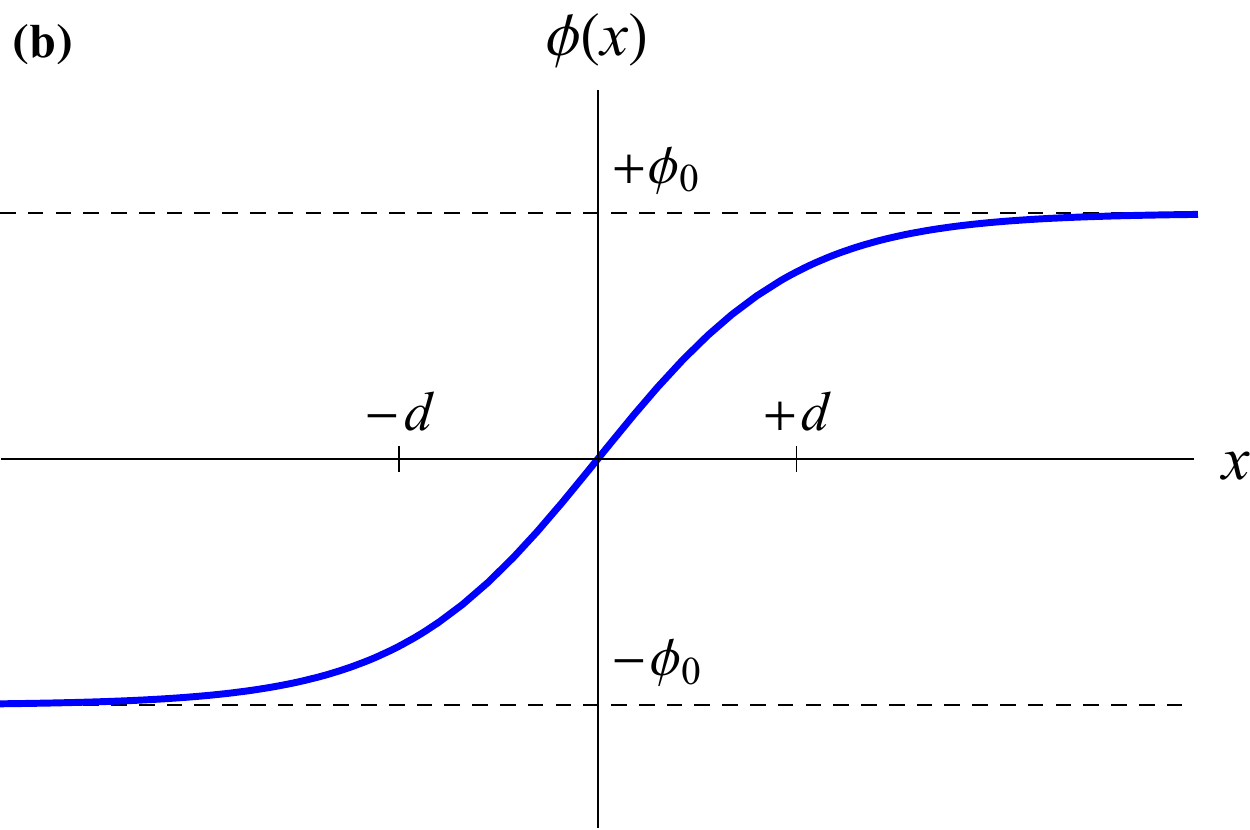}
\caption{ \normalsize (Colour online) 
Comparison of the forms of the scalar field profile for 
\textbf{(a)} a spherical bubble-like defect structure surrounding a dense spherical body in the case when the scalar field is strongly screened near the surface of and inside the dense body 
and \textbf{(b)} a domain wall in vacuum, as described by Eq.~(\ref{wall_kink_solution}) and arising in the same scalar-field model. 
}
\label{Fig:Scalar-field_profiles_comparison}
\end{figure*}

\begin{figure*}
\centering
\includegraphics[width=8.5cm]{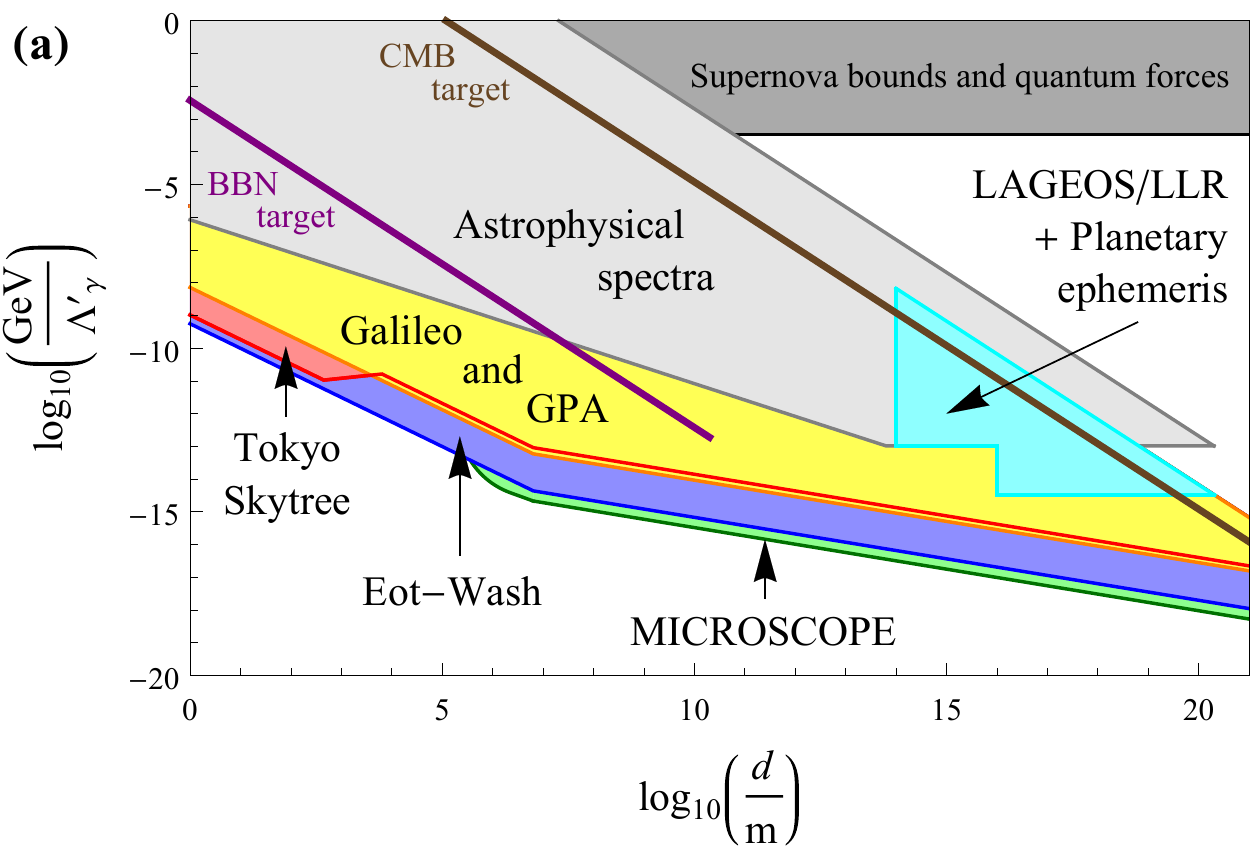}
\hspace{5mm}
\vspace{3mm}
\includegraphics[width=8.5cm]{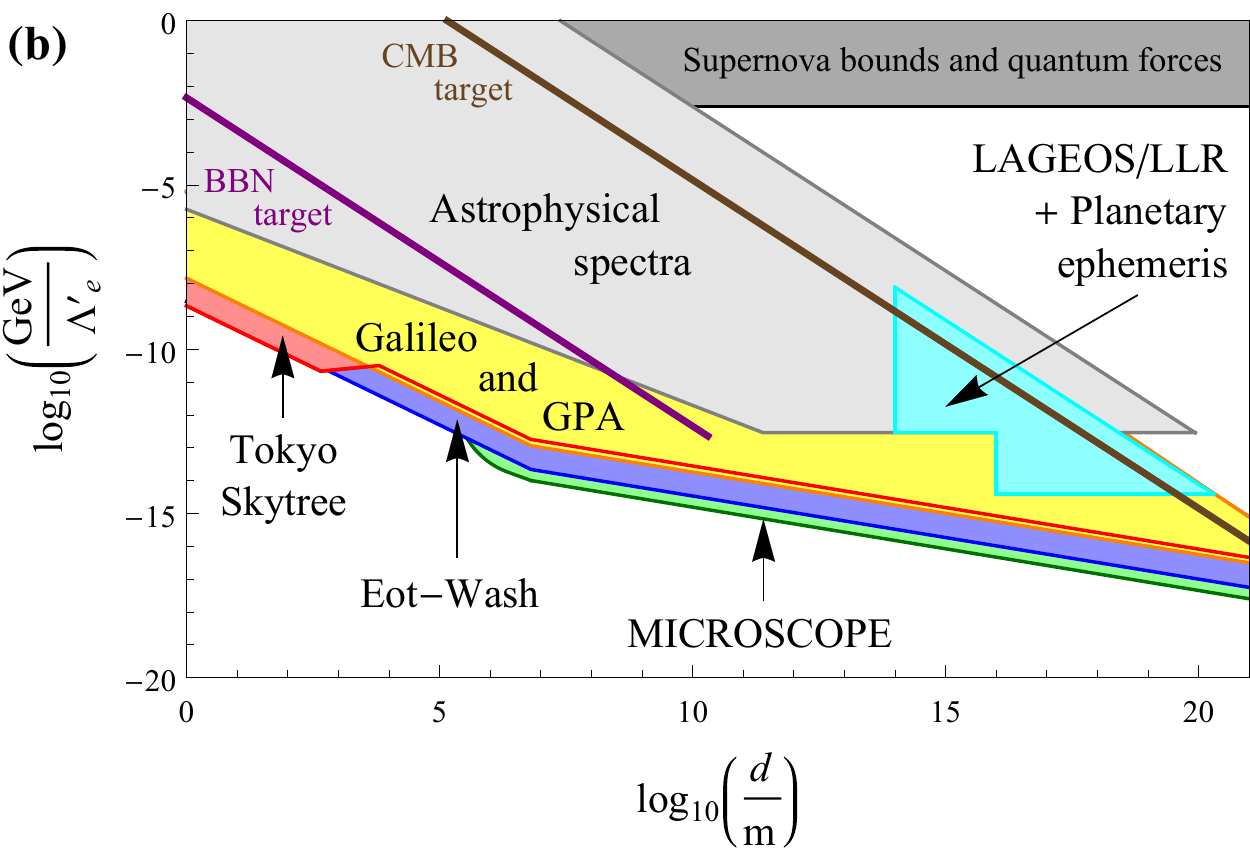}
\includegraphics[width=8.5cm]{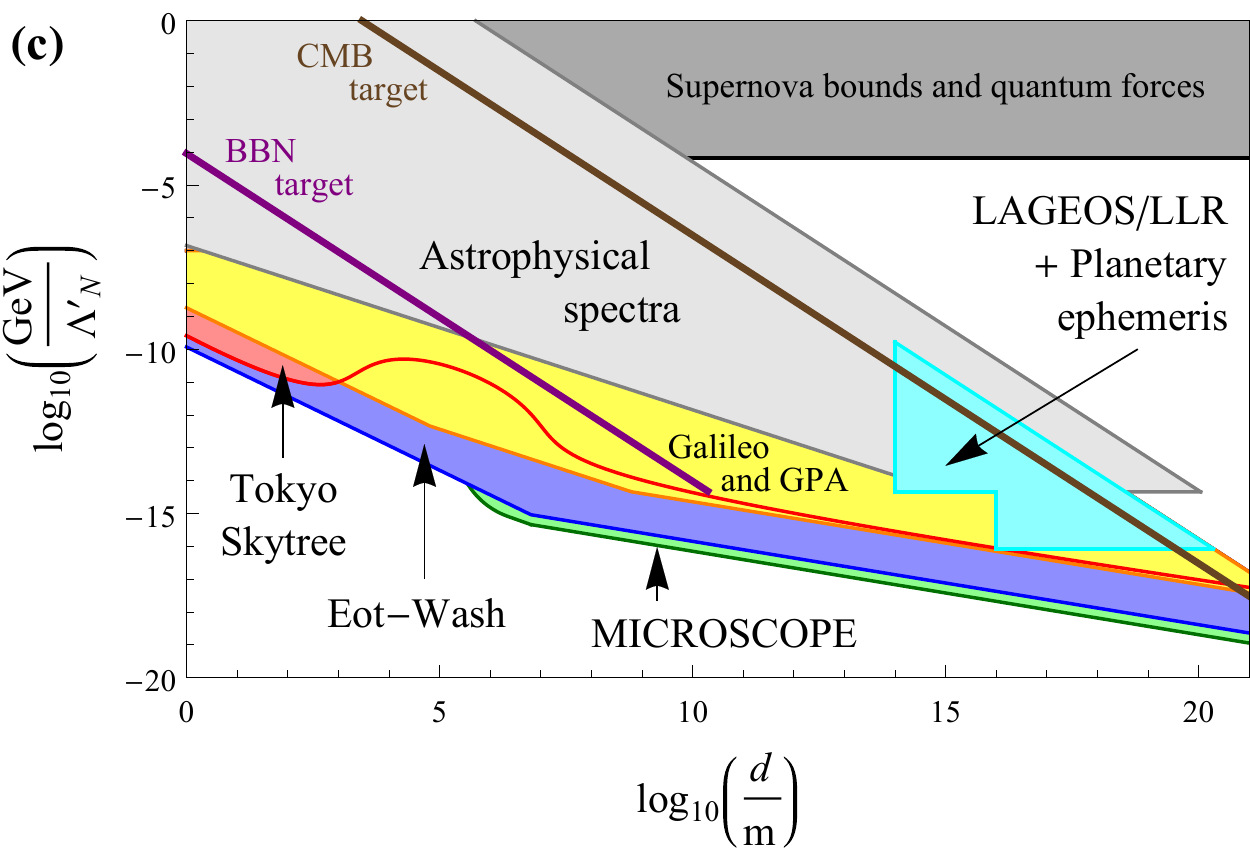}
\caption{ \normalsize (Colour online) 
Constraints on the quadratic interactions of a domain-wall-model scalar field $\phi$ with the \textbf{(a)} electromagnetic field (photon), \textbf{(b)} electron and \textbf{(c)} nucleons, as defined in (\ref{quad_portal}), as a function of the domain-wall transverse size parameter $d = \sqrt{2 / ( \lambda \phi_0^2 )}$, assuming that the domain-wall network has a present-day average energy density of $\rho_{\textrm{walls},0} \sim 10^{-5} \rho_{c,0} \sim 10^{-34}~\textrm{g/cm}^3$ and contains $\mathcal{O}(1)$ walls at all times. 
The following regions of parameters are excluded by our present work from the consideration of:~comparison of Sr optical clocks at different heights in the Tokyo Skytree experiment (red region), 
comparison of ground- and space-based H microwave clocks in the Gravity Probe A (GPA) mission and on board the Galileo satellites (yellow region), 
Be-Ti and Be-Al torsion pendula in the ground-based E$\ddot{\textrm{o}}$t-Wash experiments (blue region), 
Pt-Ti torsion pendulum in the space-based MICROSCOPE mission (green region), comparisons of atomic and molecular spectra in ground-based laboratory and low-density astrophysical environments (light grey region), 
and comparison of orbital position data of the Earth-bound LAGEOS satellite and those of lunar orbiters with lunar laser ranging data, as well as separate analyses of planetary ephemeris data (cyan region). 
The region in dark grey corresponds to existing constraints from supernova energy-loss bounds \cite{Olive-Pospelov_2008-FCs+symmetron} and short-range tests of gravity \cite{Olive-Pospelov_2008-FCs+symmetron,Brax-Fichet_2019-symmetron}. 
The purple and brown lines denote the parameters for which scalar-field domain walls can be cosmologically produced shortly after the BBN or CMB epochs, respectively. 
See the main text for more details. 
}
\label{Fig:Infinite_DW_Network-quadratic}
\end{figure*}

The form of the scalar field profile for a spherical bubble-like defect structure surrounding a dense spherical body in the case when the scalar field is strongly screened near the surface of and inside the dense body is shown in Fig.~\ref{Fig:Scalar-field_profiles_comparison}(a) for $d \gg R$; 
for comparison, in Fig.~\ref{Fig:Scalar-field_profiles_comparison}(b), we show the form of the scalar field profile of a domain wall in vacuum, as described by Eq.~(\ref{wall_kink_solution}) and arising in the same scalar-field model. 
We note that, unlike the domain walls in (\ref{wall_kink_solution}) that propagate freely in vacuum, the bubble-like defect structures in Eqs. (\ref{scalar_profile_weak_screening-A}) -- (\ref{scalar_profile_strong_screening-Bx}) are permanently affixed to and follow the motion of the dense body. 
In the case when the scalar field is strongly screened near the surface of and inside the dense body, the scalar field bubble described by Eqs.~(\ref{scalar_profile_strong_screening-A},\ref{scalar_profile_strong_screening-Bx}) starts to form when the size of the dense proto-body exceeds the critical size $\sim d'$ during the formation of the dense body. 
In arriving at Eqs.~(\ref{scalar_profile_weak_screening-A}), (\ref{scalar_profile_weak_screening-B}), (\ref{scalar_profile_strong_screening-A}) and (\ref{scalar_profile_strong_screening-Bx}), we have assumed that the dense spherical body is located in a domain with vacuum value $+ \phi_0$ [the overall signs of $\phi_\textrm{out}$ in Eqs.~(\ref{scalar_profile_weak_screening-A}), (\ref{scalar_profile_weak_screening-B}), (\ref{scalar_profile_strong_screening-A}) and (\ref{scalar_profile_strong_screening-Bx}) are reversed if the dense body is located in a domain with vacuum value $-\phi_0$] and far away from a wall boundary joining two domains. 
Since we expect the present-day network of infinite domain walls to contain only $\mathcal{O}(1)$ walls (see Sec.~\ref{Sec:Main_Cosmology}), the probability for a dense body such as Earth to be located near a wall boundary at any given time is extremely small. 
The deformed scalar field profiles around Earth in Eqs.~(\ref{scalar_profile_weak_screening-A}), (\ref{scalar_profile_weak_screening-B}), (\ref{scalar_profile_strong_screening-A}) and (\ref{scalar_profile_strong_screening-Bx}) lead to a number of interesting non-transient signatures via the apparent variations of the fundamental constants in (\ref{quad_VFCs-alpha},\ref{quad_VFCs-fermions}). 
We elucidate these signatures in subsections \ref{Sec:Main_TP}, \ref{Sec:Main_Clocks} and \ref{Sec:Main_Lab-astro-spectra} below. 
Additionally, the gravitational effects of these bubble-like defect structures surrounding Earth and the Sun lead to another set of bounds, as we discuss in subsection \ref{Sec:Main_Grav-effects} further below. 
We summarise our results in Fig.~\ref{Fig:Infinite_DW_Network-quadratic}, in which we assume the relation (\ref{DW_amplitude-infinite_network}) with $N_\textrm{walls} \sim \mathcal{O}(1)$ and the maximum allowable present-day average energy density $\rho_{\textrm{walls},0} \sim 10^{-5} \rho_{c,0} \sim 10^{-34}~\textrm{g/cm}^3$, see Eq.~(\ref{Omega_walls_bound}). 
Unless explicitly stated otherwise, we assume that all the relevant ground-based laboratory apparata are located within $\sim 1~\textrm{m}$ of Earth's surface.

\subsubsection{Torsion pendula}
\label{Sec:Main_TP}
The spherically-symmetric scalar field profiles in Eqs.~(\ref{scalar_profile_weak_screening-A}), (\ref{scalar_profile_weak_screening-B}), (\ref{scalar_profile_strong_screening-A}) and (\ref{scalar_profile_strong_screening-Bx}), induced by the interactions (\ref{quad_portal},\ref{quad_portal-atoms}), give rise to radially-directed spatial gradients in $\alpha$ and the fermion masses, see Eqs.~(\ref{quad_VFCs-alpha},\ref{quad_VFCs-fermions}). 
By analogy with the acceleration that a test particle or test mass experiences in the presence of a spatial gradient in a potential, spatial gradients in $\alpha$ and the particle masses give rise to accelerations on test particles and test masses of mass $M_\textrm{test}$: 
\begin{equation}
\label{test-mass_acceleration-generic}
\delta \v{a}_\textrm{test} = - \frac{\v{\nabla} M_\textrm{test}}{M_\textrm{test}}  \, . 
\end{equation}
Since the mass-energy contributions due to electromagnetic, electron-mass and nucleon-mass components generally differ for test masses of different material compositions, see Eq.~(\ref{atom_mass-energy_contributions}), the forces associated with the accelerations in (\ref{test-mass_acceleration-generic}) therefore violate the equivalence principle. 
Torsion pendula are extremely sensitive probes of equivalence-principle-violating forces. 
The E$\ddot{\textrm{o}}$t-Wash experiments are performed in the laboratory using Be-Ti and Be-Al test-mass pairs and constrain the differential radial accelerations of both test-mass pairs at the $\sim 10^{-13}$ level (and the differential horizontal accelerations at the $\sim 10^{-16}$ level) relative to the radial acceleration induced by Earth's gravitational field \cite{EotWash2008EP,EotWash2009EP}. 
The MICROSCOPE mission is performed in space at an altitude of $h \approx 700~\textrm{km}$ using the Pt-Ti test-mass pair and constrains the differential radial acceleration of the test-mass pair at the $\sim 10^{-14}$ level relative to the radial acceleration induced by Earth's gravitational field \cite{MICROSCOPE_2017EP}. 

Using Eqs.~(\ref{quad_VFCs-alpha}), (\ref{quad_VFCs-fermions}), (\ref{DW_amplitude-infinite_network}), (\ref{scalar_profile_weak_screening-A}), (\ref{scalar_profile_weak_screening-B}), (\ref{scalar_profile_strong_screening-A}), (\ref{scalar_profile_strong_screening-Bx}) and (\ref{test-mass_acceleration-generic}), as well as the fractional mass-energy differences of the electromagnetic, electron-mass and nucleon-mass contributions to the relevant test-mass pairs in Table~\ref{tab:frac_mass-energies}, we derive bounds on the quadratic interactions of the domain-wall-model scalar field $\phi$ with the electromagnetic field, electron and nucleons in Eq.~(\ref{quad_portal}). 
We present our bounds as coloured regions in Fig.~\ref{Fig:Infinite_DW_Network-quadratic} (blue = E$\ddot{\textrm{o}}$t-Wash experiments, green = MICROSCOPE mission). 
In deriving the E$\ddot{\textrm{o}}$t-Wash bounds, we have treated Earth as a structureless, featureless sphere, with a pendulum tilt angle of $\sim 10^{-3}$ radians (mainly due to Earth's rotation) away from the normal to Earth's surface. 
In the case when the scalar field is weakly screened inside Earth (and hence the principle of linear superposition applies), there may be a further increase in the sensitivity to $\Lambda'_X$ for $d < R_\oplus$ by up to a factor of $\approx 4$ from the consideration of the horizontal forces sourced by nearby laboratory and geographical features (for details of these features, see Ref.~\cite{Eot-Wash_2012_review} and references 17--21 therein). 
On the other hand, in the case when the scalar field is strongly screened in the bulk of Earth, the principle of linear superposition is no longer guaranteed to hold due to the non-linear nature of the governing differential equation [see Eq.~(\ref{Scalar_field_DE-general})], and a more detailed analysis is required to elucidate the role of the nearby geographical and laboratory features. 
We note that the MICROSCOPE bounds are the stronger of the two sets of torsion-pendulum-based bounds for the domain-wall transverse size parameters $d \gtrsim h \approx 700~\textrm{km}$ due to the better precision of the MICROSCOPE mission to radially-directed equivalence-principle-violating forces. 
However, for $d \lesssim h$, the MICROSCOPE bounds degrade rapidly, since the scalar-field gradient becomes exponentially suppressed at the altitude $h \approx 700~\textrm{km}$ in this case, while there is no such suppression for the ground-based E$\ddot{\textrm{o}}$t-Wash experiments, see Eqs.~(\ref{scalar_profile_weak_screening-A}--\ref{scalar_profile_strong_screening-Bx}). 
It is also worth noting that these same scalar-field-induced equivalence-principle-violating forces will also manifest themselves in atom interferometry experiments that compare the rates of fall of two different atomic species or isotopes; 
however, the precision of current atom interferometry experiments \cite{Rasel_2014_AI,Tino_2014_AI,Wuhan_2015_AI,Tino_2017_AI,Stanford_2020_AI} is not yet competitive with torsion-pendulum experiments.

\subsubsection{Comparison of clocks at different heights}
\label{Sec:Main_Clocks}
The spherically-symmetric scalar field profiles in Eqs.~(\ref{scalar_profile_weak_screening-A}), (\ref{scalar_profile_weak_screening-B}), (\ref{scalar_profile_strong_screening-A}) and (\ref{scalar_profile_strong_screening-Bx}), induced by the interactions (\ref{quad_portal},\ref{quad_portal-atoms}), imply that the apparent values of $\alpha$ and the fermion masses vary with altitude $h$, in accordance with Eqs.~(\ref{quad_VFCs-alpha},\ref{quad_VFCs-fermions}). 
Spectroscopy (clock-based) measurements are very sensitive probes of varying fundamental constants. 
The recent Tokyo Skytree experiment compared two Sr optical clocks separated by a height difference of $\Delta h \approx 450~\textrm{m}$, with one of the clocks located close to Earth's surface, and verified the fractional atomic frequency shift due to the gravitational potential difference between the two clocks at the $4 \times 10^{-18}$ level \cite{TokyoSkytree_2020}. 
The scalar field contribution to the fractional frequency shift between the two Sr optical clocks takes the form: 
\begin{equation}
\label{Sr_sensitivity_coefficients}
\frac{\Delta \nu_\textrm{Sr}}{\nu_\textrm{Sr}} \approx +2 \frac{\Delta \alpha}{\alpha} + \frac{\Delta m_e}{m_e} \approx \Delta(\phi^2) \left[ \frac{2}{\left( \Lambda'_\gamma \right)^2} + \frac{1}{\left( \Lambda'_e \right)^2} \right]  \, , 
\end{equation}
where we have neglected a small relativistic correction factor of $+0.06$ to the $\alpha$ sensitivity coefficient of the Sr clock transition \cite{Sr_sens_coef_2004}. 
In order to distinguish the effects of the scalar-field interactions (\ref{quad_portal}) from the usual gravitational redshift effect, we note that the clock comparison measurements are ``referenced'' against a combination of laser ranging and gravimeter measurements, which provide an independent prediction for the clock frequency shift. 
First of all, let us remark that the effect of the $\phi^2$ interactions on the height of the tower via changes in the atomic Bohr radius $a_\textrm{B} = 1/(m_e \alpha)$ is common to all of the clock comparison, laser ranging and gravimeter measurements, while the effects of the $\phi^2$ interactions on the size and apparent mass of Earth are common to the clock comparison and gravimeter measurements (and are practically irrelevant for the laser ranging measurements). 

The effects of the $\phi^2$ interactions (\ref{quad_portal}) on the time-of-flight laser ranging measurements via their effects on (i) the timekeeping element that measures the photon flight time $\Delta t = 2 \Delta h / c$, (ii) the laser frequency and wavelength, and (iii) the photon propagation speed, are generally negligible compared to the effects on the clock comparison measurements themselves. 
Regarding (i), the time-of-flight laser ranging measurements in the Tokyo Skytree experiment have a relative precision of $\sim 10^{-5}$, and so the effects of the $\phi^2$ interactions on the associated timekeeping element can be neglected (indeed the timekeeping element is practically insensitive to the usual gravitational redshift effect). 
Regarding (ii) and (iii), the $\phi^2$ interactions (\ref{quad_portal}) have no effect on the speed of light in vacuum and hence have no associated effect on the time of flight measured in the laser ranging measurements. 
The reason is that photons, which propagate freely through vacuum, have $|\v{E}| = |\v{B}|$ in terms of their electric- and magnetic-field components, and so the relevant coupling $\phi^2 F_{\mu \nu} F^{\mu \nu}$ vanishes, since $F_{\mu \nu} F^{\mu \nu} = 2 (|\v{B}|^2 - |\v{E}|^2) = 0$ in this case, while the $\phi^2$ couplings to the fermions are irrelevant. 
The $\phi^2$ interactions do affect the laser frequency and wavelength, but this is irrelevant for time-of-flight measurements in vacuum (though can be relevant in phase-shift measurements \cite{Stadnik_2019_DM-LIFO}). 
For photons travelling through air, the $\phi^2$ interactions have a small effect on the photon propagation speed via changes in the refractive index of air and changes in the laser frequency \cite{Stadnik_2019_DM-LIFO}. 
The $\phi^2$ interactions also have a small effect on the $\Delta h$ measurement via laser ranging in the Tokyo Skytree experiment, due to the fact that a small fraction of this measurement is conducted via spirit levelling, with the $\phi^2$ interactions affecting the lengths of the levelling rods via changes in $a_\textrm{B}$. 

The $\phi^2$ interactions (\ref{quad_portal}) also affect the acceleration measured by the two gravimeters in the Tokyo Skytree experiment, according to $\v{a} (\v{r}) = -\v{\nabla} U (\v{r}) - K'_X \v{\nabla} \beta_X (\v{r})  = -\v{\nabla} U'_\textrm{eff} (\v{r})$, where $U (\v{r})$ is the usual gravitational potential, $K'_X$ is the fractional mass-energy contribution of the underlying mass-energy component $X$ associated with the test mass ($K'_N \approx 1$, while $K'_{\gamma,e} \ll 1$), and $\beta_X (\v{r})$ are functions of the interaction type, strength and range, as well as details of screening of the scalar field by Earth, and are related to the functions in Eqs. (\ref{scalar_profile_weak_screening-A}) -- (\ref{scalar_profile_strong_screening-Bx}). 
In particular, the functions $\beta_X$ are the same functions that enter the transition frequency shift between clocks at different heights, according to $\Delta \nu / \nu = \Delta U + K_X \Delta \beta_X = \Delta U_\textrm{eff}$, where $K_X$ denotes the relative sensitivity coefficient of a given clock transition frequency to changes in the fundamental constant $X$, defined according to $\Delta \nu / \nu = K_X \Delta X / X$. 
When $K_X \approx K'_X$, the effects of the $\phi^2$ interactions on the clock comparison and gravimeter measurements are practically indistinguishable (since $U_\textrm{eff} \approx U'_\textrm{eff}$ in this case), and so the corresponding effective sensitivity coefficient $K_{X,\textrm{eff}} = K_X - K'_X$ will be suppressed in this case. 
Most optical and microwave clocks have $K_X \gg K'_X$ for the $\phi^2$ interactions with the electromagnetic field and the electron, giving $K_{X,\textrm{eff}} \approx K_X$ in this case. 
On the other hand, for the $\phi^2$ interaction with nucleons, $K'_X \approx +1$ and so $K_{X,\textrm{eff}} \approx K_X - 1$. 

Altogether, the effective contribution of the $\phi^2$ interactions (\ref{quad_portal}) to the fractional frequency shift between the two Sr optical clocks in the Tokyo Skytree experiment, referenced against a combination of laser ranging and gravimeter measurements, takes the form: 
\begin{equation}
\label{Sr_sensitivity_coefficients_EFFECTIVE}
\left( \frac{\Delta \nu_\textrm{Sr}}{\nu_\textrm{Sr}} \right)_\textrm{eff} \approx +2 \frac{\Delta \alpha}{\alpha} + \frac{\Delta m_e}{m_e} - \frac{\Delta m_N}{m_N} \approx \Delta(\phi^2) \left[ \frac{2}{\left( \Lambda'_\gamma \right)^2} + \frac{1}{\left( \Lambda'_e \right)^2} - \frac{1}{\left( \Lambda'_N \right)^2} \right]  \, . 
\end{equation}
Note the appreciable sensitivity to the nucleon interaction parameter in Eq.~(\ref{Sr_sensitivity_coefficients_EFFECTIVE}), which is absent in Eq.~(\ref{Sr_sensitivity_coefficients}), but arises in Eq.~(\ref{Sr_sensitivity_coefficients_EFFECTIVE}) due to referencing against the gravimeter measurements.

Besides ground-based clock comparison measurements, ground-to-space clock comparison measurements are also possible. 
The Gravity Probe A (GPA) mission in the 1970s compared two hydrogen microwave clocks, with one of the clocks located on the ground and the other clock located in space at an altitude of $h \approx 10^4~\textrm{km}$, and verified the fractional atomic frequency shift due to the gravitational potential difference between the two clocks at the $\sim 10^{-13}$ level \cite{GPA_H-maser_1980,GPA_H-maser_1979,GPA_H-maser_1989}. 
More recent spectroscopy data from hydrogen microwave clocks on board the Galileo satellites in eccentric orbits about Earth at altitudes of $h \sim (1 - 2) \times 10^4~\textrm{km}$ demonstrated improved precision at the $\sim 10^{-14}$ level \cite{Eccentric_Galileo_2018A,Eccentric_Galileo_2018B}. 
In both cases, the scalar field contribution to the fractional frequency shift between two H microwave clocks takes the form: 
\begin{equation}
\label{H_sensitivity_coefficients}
\frac{\Delta \nu_\textrm{H}}{\nu_\textrm{H}} \approx + 4 \frac{\Delta \alpha}{\alpha} + 2 \frac{\Delta m_e}{m_e} - \frac{\Delta m_N}{m_N} \approx \Delta(\phi^2) \left[ \frac{4}{\left( \Lambda'_\gamma \right)^2} + \frac{2}{\left( \Lambda'_e \right)^2} - \frac{1}{\left( \Lambda'_N \right)^2} \right]  \, . 
\end{equation}
In ground-to-space clock comparisons, the clock comparison measurements are referenced against a combination of laser ranging measurements 
and predominantly the orbital position data of Earth-bound satellites, such as LAGEOS and GRACE, which together provide an independent prediction for the clock frequency shifts. 
The discussion concerning the effective sensitivity coefficients $K_{X,\textrm{eff}}$ proceeds in a similar fashion as for the Tokyo Skytree experiment above (the lunar laser ranging measurements have a relative precision of $\sim 10^{-11}$ and so the effects of the $\phi^2$ interactions on the reference Cs microwave clock, which functions as the timekeeping element, can be neglected). 
The effective contribution of the $\phi^2$ interactions (\ref{quad_portal}) to the fractional frequency shift between two H microwave clocks, referenced against a combination of laser ranging measurements and orbital position data of Earth-bound satellites, takes the form: 
\begin{equation}
\label{H_sensitivity_coefficients_EFFECTIVE}
\left( \frac{\Delta \nu_\textrm{H}}{\nu_\textrm{H}} \right)_\textrm{eff} \approx + 4 \frac{\Delta \alpha}{\alpha} + 2 \frac{\Delta m_e}{m_e} - 2 \frac{\Delta m_N}{m_N} \approx \Delta(\phi^2) \left[ \frac{4}{\left( \Lambda'_\gamma \right)^2} + \frac{2}{\left( \Lambda'_e \right)^2} - \frac{2}{\left( \Lambda'_N \right)^2} \right]  \, . 
\end{equation}
Note the enhancement in the sensitivity to the nucleon interaction parameter in Eq.~(\ref{H_sensitivity_coefficients_EFFECTIVE}), compared with in Eq.~(\ref{H_sensitivity_coefficients}), due to the effect of the $\phi^2$ interaction with nucleons on the measured accelerations of the Earth-bound satellites. 

Before proceeding to our results, let us remark that it is instead possible to compare the ratios of a pair of different clock transition frequencies, where the two clocks have different sensitivities to the $\phi^2$ interactions (\ref{quad_portal}), at different heights. 
In this approach, the usual gravitational redshift effect, which is universal across all clock transition frequencies, cancels in the frequency ratio measurements, potentially greatly reducing the systematic uncertainties associated with independent knowledge of the local gravitational potential. 
This can be particularly advantageous when a clean independent prediction for an individual clock transition frequency shift might be hindered by complex local geographical features, such as near or on top of a large mountain or in underground experiments.


Using Eqs.~(\ref{DW_amplitude-infinite_network}), (\ref{scalar_profile_weak_screening-A}), (\ref{scalar_profile_weak_screening-B}), (\ref{scalar_profile_strong_screening-A}), (\ref{scalar_profile_strong_screening-Bx}), (\ref{Sr_sensitivity_coefficients_EFFECTIVE}) and (\ref{H_sensitivity_coefficients_EFFECTIVE}), we derive bounds on the quadratic interactions of the domain-wall-model scalar field $\phi$ with the electromagnetic field, electron and nucleons in Eq.~(\ref{quad_portal}). 
We present our bounds as coloured regions in Fig.~\ref{Fig:Infinite_DW_Network-quadratic} (red = Tokyo Skytree experiment, yellow = Galileo satellites and Gravity Probe A mission). 
In deriving the Tokyo Skytree bounds, we note that the back-action effects of the tower on the scalar field, including screening of the scalar field by the tower itself, can be neglected in the entire relevant parameter space. 
The main support structure of the tower is a cylindrical shell ``shin-bashira'', with a height of $\approx 375~\textrm{m}$ from the base of the tower, radius of $\approx 4~\textrm{m}$, shell thickness of $\approx 0.5~\textrm{m}$, and made of steel-reinforced concrete (density $\approx 2.5~\textrm{g/cm}^3$). 
The upper clock is located $\approx 12~\textrm{m}$ from the tower's symmetry axis, while the lower clock is located $\approx 40~\textrm{m}$ from the tower's symmetry axis. 
We note that the Tokyo Skytree bounds are the significantly stronger of the two types of clock-based bounds for the domain-wall transverse size parameters $d \lesssim \Delta h \approx 450~\textrm{m}$ due to the much better precision of modern state-of-the-art Sr optical clocks compared with H microwave clocks. 
However, for $d \gtrsim \Delta h$, the Tokyo Skytree bounds degrade substantially, since in this case the scalar field contribution to $\Delta \nu / \nu$ becomes suppressed by the factor $\sim \Delta h / R_\oplus \sim 10^{-4}$ in the weak screening regime and by the factor $\sim \Delta h / \textrm{min} \left( d , R_\oplus \right) \ll 1$ in the strong screening regime. 
On the other hand, there are no such suppression factors for the ground-to-space clock comparisons, in which one of the clocks is located at a distance $\gtrsim R_\oplus$ away from Earth's surface, and the two types of clock-based bounds become comparable in the limit $d \gg \Delta h $. 
We note that Cs microwave spectroscopy data from the near-future ACES-PHARAO mission \cite{ACES_2015}, which will place clocks on board the International Space Station ($h \approx 400~\textrm{km}$), may also be used in similar types of searches.

\subsubsection{Comparison of laboratory and astrophysical spectra}
\label{Sec:Main_Lab-astro-spectra}
In the presence of the interactions (\ref{quad_portal},\ref{quad_portal-atoms}), the value of the domain-wall-model scalar field $\phi$, and in turn the apparent values of $\alpha$ and the fermion masses, depend on the local environment. 
Near the surface of a sufficiently dense and small body (i.e., where the strong screening regime applies), $\phi_\textrm{lab}^2 \ll \phi_0^2$, see Eq.~(\ref{scalar_profile_strong_screening-A}). 
On the other hand, in a sufficiently dilute environment far away from the dense body (where the weak screening regime applies), $\phi_\textrm{astro}^2 \approx \phi_0^2$. 
In this case, $\Delta (\phi^2) = \phi_\textrm{astro}^2 - \phi_\textrm{lab}^2 \approx \phi_0^2$, which corresponds to the maximum allowable difference in $\phi^2$ and leads to the following variations of the fundamental constants via (\ref{quad_VFCs-alpha},\ref{quad_VFCs-fermions}): 
\begin{gather}
\frac{\Delta \alpha}{\alpha} \approx \frac{\phi_0^2}{\left( \Lambda'_\gamma \right)^2}  \label{VFCs_low-high_density_enviros-cf_alpha}  \, ,  \\ 
\frac{\Delta m_f}{m_f} \approx \frac{\phi_0^2}{\left( \Lambda'_f \right)^2}  \label{VFCs_low-high_density_enviros-cf_fermions}  \, . 
\end{gather}
The comparison of atomic and molecular spectra in ground-based laboratory and low-density astrophysical environments provides a powerful probe of such environmentally-dependent fundamental constants. 
Comparisons of Zn and Cr$^+$ transition frequencies in the laboratory and in extragalactic quasar absorption gas clouds, where typical particle densities are $\sim 10^{-3} - 10^{-1}~\textrm{cm}^{-3}$, constrain $\Delta \alpha / \alpha \lesssim 10^{-6}$ \cite{Astrophys_alpha_limit_2016}. 
Comparisons of NH$_3$ and other molecular transition frequencies in the laboratory and in intragalactic cold molecular gas clouds, where typical particle densities are $\sim 10^3 - 10^5~\textrm{cm}^{-3}$, constrain $\Delta (m_e/m_N) / (m_e/m_N) \lesssim 3 \times 10^{-8}$ \cite{Astrophys_mu_limit_2010}, while comparisons of H$_2$ transition frequencies in the laboratory and in extragalactic quasar absorption gas clouds constrain $\Delta (m_e/m_N) / (m_e/m_N) \lesssim 10^{-5}$ \cite{Astrophys_mu_limit_2011,Astrophys_mu_limit_2015}. 

Using Eqs.~(\ref{DW_amplitude-infinite_network}), (\ref{VFCs_low-high_density_enviros-cf_alpha}) and (\ref{VFCs_low-high_density_enviros-cf_fermions}), as well as the fractional mass-energy contributions to the interstellar medium's mass-energy content from the electromagnetic, electron-mass and nucleon-mass components in Table \ref{tab:frac_mass-energies}, we derive bounds on the quadratic interactions of the domain-wall-model scalar field $\phi$ with the electromagnetic field, electron and nucleons in Eq.~(\ref{quad_portal}). 
We present our bounds as the light grey regions in Fig.~\ref{Fig:Infinite_DW_Network-quadratic}. 
Note that these bounds apply to the regions of parameter spaces where $2 \rho_{\textrm{astro},X} / (\Lambda'_X)^2 \lesssim \lambda \phi_0^2$, otherwise $\phi_\textrm{astro}^2 = 0$ and $\Delta (\phi^2) = 0$ due to the inhibition of spontaneous symmetry breaking. 
We also mention that it is possible to look for changes in the fundamental constants that are correlated with differences in ambient density between different gas clouds. 

\subsubsection{Gravitational effects}
\label{Sec:Main_Grav-effects}
We have already discussed bounds on domain walls of cosmological origin via their gravitational effects, see Eq.~(\ref{Omega_walls_bound}). 
Let us now discuss the bounds from the gravitational effects of bubble-like defect structures surrounding dense bodies. 
The spherically-symmetric scalar field profiles in Eqs.~(\ref{scalar_profile_weak_screening-A}), (\ref{scalar_profile_weak_screening-B}), (\ref{scalar_profile_strong_screening-A}) and (\ref{scalar_profile_strong_screening-Bx}), induced by the interactions (\ref{quad_portal},\ref{quad_portal-atoms}), have a characteristic thickness of $\Delta r \sim \textrm{min}(d,R)$ extending out from the surface of the dense body and carry an energy density of $\rho_\textrm{core} \sim (\Delta \phi)^2 / (\Delta r)^2$ within this ``core region''. 
When the scalar field is strongly screened inside the dense body, the change in the scalar field amplitude across this core region is given by $(\Delta \phi)^2 \sim \phi_0^2$, while when the scalar field is weakly screened inside the dense body, $(\Delta \phi)^2 \ll \phi_0^2$. 
We focus on the former case, for which: 
\begin{equation}
\label{energy-density_core}
\rho_\textrm{core} \sim \frac{\phi_0^2}{\textrm{min} (d^2 , R^2)}  \, , 
\end{equation}
and the mass of the bubble within the core region is given by: 
\begin{equation}
\label{mass_bubble_core}
M_\textrm{core} \sim \frac{\phi_0^2 R^2}{\textrm{min} (d , R)}  \, . 
\end{equation}
Using Eq.~(\ref{DW_amplitude-infinite_network}) and assuming the same domain-wall model parameters as in Fig.~\ref{Fig:Infinite_DW_Network-quadratic}, we find that the core mass of a bubble defect surrounding Earth is: 
\begin{equation}
\label{mass_bubble_core_Earth}
M_\textrm{core} \sim 10^9~\textrm{kg} \times \textrm{max} (1 , d/R_\oplus)  \, , 
\end{equation}
while the core mass of a bubble defect surrounding the Sun is: 
\begin{equation}
\label{mass_bubble_core_Sun}
M_\textrm{core} \sim 10^{13}~\textrm{kg} \times \textrm{max} (1 , d/R_\odot)  \, . 
\end{equation}
In the limiting case when $d \gg \textrm{several} \times R$, the energy density associated with the bubble scalar-field configuration in the ``tail region'', where $R \lesssim r - R \lesssim d/2$, is: 
\begin{equation}
\label{energy-density_tail}
\rho_\textrm{tail} (r) \sim \frac{\phi_0^2 R^2}{r^4}  \, , 
\end{equation}
and the mass of the bubble within the tail region is: 
\begin{equation}
\label{mass_bubble_tail}
M_\textrm{tail} \sim M_\textrm{core}  \, , 
\end{equation}
with the dominant contribution to $M_\textrm{tail}$ coming from the innermost part of the tail region.

Comparison of the orbital position data of the Earth-bound LAGEOS satellite and those of lunar orbiters with independent lunar laser ranging data constrains the mass of an Earth-bound dark component between the radii $\approx 2 R_\oplus$ and $\approx 60 R_\oplus$ to be $M_\textrm{dark} < 4 \times 10^{-9} M_\oplus$ \cite{Adler_2008_Grav_LAGEOS-LLR}. 
Using Eqs.~(\ref{mass_bubble_core_Earth}) and (\ref{mass_bubble_tail}), we hence place the following bound on the domain-wall transverse size parameter: 
\begin{equation}
\label{bubble_grav_bound_LAGEOS-LLR}
d \lesssim 10^{14}~\textrm{m}  \, . 
\end{equation}
Planetary ephemeris orbital data constrain the mass of a Sun-bound dark component between the orbits of Earth and Saturn to be $M_\textrm{dark} \lesssim 10^{-10} M_\odot$ \cite{Pitjeva_2013_planetary_ephemeris_MNRAS,Pitjeva_2013_planetary_ephemeris_Astron-Lett}. 
Using Eqs.~(\ref{mass_bubble_core_Sun}) and (\ref{mass_bubble_tail}), we therefore place the following bound: 
\begin{equation}
\label{bubble_grav_bound_planet_ephemeris_orbital}
d \lesssim 10^{19}~\textrm{m}  \, . 
\end{equation}
Additionally, planetary ephemeris data pertaining to the perihelion shift of Mars' orbit constrain the energy density of a Sun-bound dark component at the orbital position of Mars to be $\rho_\textrm{dark} < 1.4 \times 10^{-20}~\textrm{g/cm}^3$ \cite{Pitjeva_2013_planetary_ephemeris_MNRAS,Pitjeva_2013_planetary_ephemeris_Astron-Lett}. 
Using Eqs.~(\ref{DW_amplitude-infinite_network}) and (\ref{energy-density_tail}), and assuming the same domain-wall model parameters as in Fig.~\ref{Fig:Infinite_DW_Network-quadratic}, we obtain the following bound: 
\begin{equation}
\label{bubble_grav_bound_planet_ephemeris_perihelion_shift}
d \lesssim 10^{16}~\textrm{m}  \, . 
\end{equation}
Note that, unlike the bounds (\ref{bubble_grav_bound_LAGEOS-LLR}) and (\ref{bubble_grav_bound_planet_ephemeris_orbital}) which apply to the regions of parameters $d' \lesssim R_\oplus$ and $d' \lesssim R_\odot$, respectively, the bound (\ref{bubble_grav_bound_planet_ephemeris_perihelion_shift}) only applies to the region of parameters $R_\textrm{Mars} \lesssim d' \lesssim R_\odot$, for which the scalar field is not strongly screened inside Mars. 
The bounds in Eqs.~(\ref{bubble_grav_bound_LAGEOS-LLR}), (\ref{bubble_grav_bound_planet_ephemeris_orbital}) and (\ref{bubble_grav_bound_planet_ephemeris_perihelion_shift}) do not apply, however, in the case when spontaneous symmetry breaking is inhibited by the interplanetary medium. 
We present our bounds as the cyan regions in Fig.~\ref{Fig:Infinite_DW_Network-quadratic}.

\vspace{200mm}

\section{``Dark matter'' networks of domain walls}
\label{Sec:TDM}
A number of recent works have reported on searches for ``dark matter'' networks of domain walls using clock- and cavity-based techniques \cite{Wcislo_2016_TDM-cavity,Roberts_2017_TDM-GPS,Wcislo_2018_TDM-cavity,Roberts_2019_TDM-clocks}, as well as magnetometry-based techniques \cite{Pustelny_GNOME_2013}, via transient signatures induced by passing domain walls. 
In these papers, the authors assume that such domain-wall networks saturate the local cold dark matter energy density $\rho_\textrm{DM}^\textrm{local} \approx 0.4~\textrm{GeV/cm}^3$. 
The authors further assume that these domain walls pass through Earth with a typical speed of $v_\textrm{TD} \sim 300~\textrm{km/s}$, which is characteristic of objects in our local galactic region, and with an average time between encounters of a wall with Earth, $\mathcal{T}$, ranging from hours to years. 
In this case, the relationship between the average energy density associated with such a network of domain walls, $\rho_\textrm{TDN}$, and the wall parameters is given by: 
\begin{equation}
\label{DW_amplitude-TDM_network}
\phi_0^2 \sim \rho_\textrm{TDN} v_\textrm{TD} \mathcal{T} d  \, . 
\end{equation}
Such ``dark matter'' networks of domain walls cannot be produced under the minimal cosmological assumptions outlined in Sec.~\ref{Sec:Main_Cosmology}. 
Specifically, let us mention that a number of assumptions made in the aforementioned searches \cite{Wcislo_2016_TDM-cavity,Roberts_2017_TDM-GPS,Wcislo_2018_TDM-cavity,Roberts_2019_TDM-clocks,Pustelny_GNOME_2013} for these ``dark matter'' networks of domain walls are dubious and unsubstantiated:

\textbf{(I)} The assumption that domain walls saturate the local cold dark matter abundance is dubious. 
Indeed, the present-day energy density associated with domain walls that stretch across the horizon size is constrained to be at least 10 orders of magnitude smaller than the local cold dark matter energy density from measurements of the CMB quadrupolar temperature anisotropy \cite{Spergel_1989_walls,Planck2018}, see Eq.~(\ref{Omega_walls_bound}). 
The general expectation is that networks of domain walls are ``stiff'', meaning that their energy densities inside and outside of galaxies are similar. 
Ref.~\cite{Pospelov_2013_TDM-magnetometers} has suggested that the energy density associated with a ``flexible'' network of domain walls may be enhanced inside galaxies by orders of magnitude compared to an average cosmological value. 
It might be possible to evade, at least to some extent, constraints from measurements of the CMB quadrupolar temperature anisotropy with such an unconventionally flexible network of walls. 

We point out that one possible mechanism for producing flexible domain walls arises in models of domain walls with quadratic-in-$\phi$ interactions of the type in (\ref{quad_portal}), but with the signs reversed. 
In this case, the adiabatic transverse wall profile coincides with the kink profile (\ref{wall_kink_solution},\ref{wall_kink_thickness}), but with the replacement $\phi_0 \to \phi'_0 = \sqrt{\phi_0^2 + 2 \rho_X / [\lambda ( \Lambda'_X )^2]}$, which increases the energy density inside the wall by the factor $(\phi'_0/\phi_0)^4$. 
The increase in energy associated with the domain walls in this case is facilitated by the transfer of energy between ordinary matter and the domain walls. 
Such a mechanism, however, by itself cannot lead to a network of walls that saturates the galactic dark matter abundance, since the total energy of ordinary cold matter in our galaxy is $\sim 5$ times less than that of galactic dark matter. 
It remains an open question as to how a flexible domain-wall network saturating the galactic dark matter abundance may arise.

\textbf{(II)} The assumption that the average time between encounters of a domain wall with Earth lies within the convenient human timescales from hours to years is dubious. 
Numerical simulations in Refs.~\cite{Spergel_1989_walls,Hindmarsh2003walls,Avelino2004walls,Avelino2005walls} indicate that only $\mathcal{O}(1 - 10)$ macroscopic domain walls with the potential (\ref{phi4_potential}) survive to the present day. 
The small number of surviving walls is due to the efficient processes of wall reconnection and annihilation when domain walls collide. 
Numerical simulations with more complicated potentials and more complicated types of hybrid wall networks \cite{Avelino2006walls,Avelino2010walls} also indicate a lack of stability of such networks against annihilation. 
Therefore, one would expect an average time between encounters of a wall with Earth of $\ggg 1~\textrm{year}$, well in excess of the assumptions made in earlier searches for domain walls.

\textbf{(III)} The assumption that domain walls pass through Earth with a non-relativistic speed of $v_\textrm{TD} \sim 300~\textrm{km/s}$ is dubious. 
Domain walls with the potential (\ref{phi4_potential}) have large spatial components in the associated energy-momentum tensor that give significant deviations from the non-relativistic equation of state and, furthermore, numerical simulations in Refs.~\cite{Spergel_1989_walls,Avelino2004walls,Avelino2005walls-semirelativistic} indicate that domain walls with the potential (\ref{phi4_potential}) move at semi-relativistic speeds. 
More complicated hybrid wall networks containing sufficiently massive junctions may become non-relativistic \cite{Avelino2006walls,Avelino2010walls}, but such wall networks cannot be realised within the single-scalar-field models assumed in the earlier searches \cite{Wcislo_2016_TDM-cavity,Roberts_2017_TDM-GPS,Wcislo_2018_TDM-cavity,Roberts_2019_TDM-clocks,Pustelny_GNOME_2013}. 

Furthermore, the assumption that the velocities of galactic domain walls follow a quasi-Maxwell-Boltzmann distribution in accordance with the standard halo model, as explicitly made by Refs.~\cite{Roberts_2017_TDM-GPS,Roberts_2019_TDM-clocks} within single-scalar-field models, is dubious. 
In models with a single scalar field, the domain walls and vacua tend to be collinearly aligned in a stacked formation. 
Slanting or oblique alignments of domain walls and vacua would lead to more frequent wall collisions and subsequent annihilation. 
In other words, within models of a single scalar field, we would expect domain walls to be travelling in roughly one and the same direction, rather than in random directions. 
The passage of domain walls from different directions can occur within more complicated models of domain walls involving more than one scalar field and more complicated potentials \cite{Bazeia2002_DWs-review}. 

We also note that Ref.~\cite{Roberts_2019_TDM-clocks}'s claim that their results apply to topological defects of all dimensionalities is inconsistent with their chosen model of a single real scalar field. 
As discussed in Sec.~\ref{Sec:Main_Theory}, models of topological strings and monopoles require at least one complex scalar field or several real scalar fields, respectively, as well as associated gauge fields.

\textbf{(IV)} In models of topological defects with the quadratic-in-$\phi$ interactions (\ref{quad_portal}), which were considered in the earlier searches \cite{Wcislo_2016_TDM-cavity,Roberts_2017_TDM-GPS,Wcislo_2018_TDM-cavity,Roberts_2019_TDM-clocks} (Ref.~\cite{Roberts_2019_TDM-clocks} also considered the analogous interaction with the electromagnetic field but with the sign reversed), the unwitting assumption made that such topological defects pass in an unperturbed manner through the strongly repulsive potential generated by Earth (and in some cases the apparatus itself) appears dubious and warrants further investigation. 
We discuss this issue further in Sec.~\ref{Sec:TDM_Transients} below.

\begin{figure*}
\centering
\includegraphics[width=8.5cm]{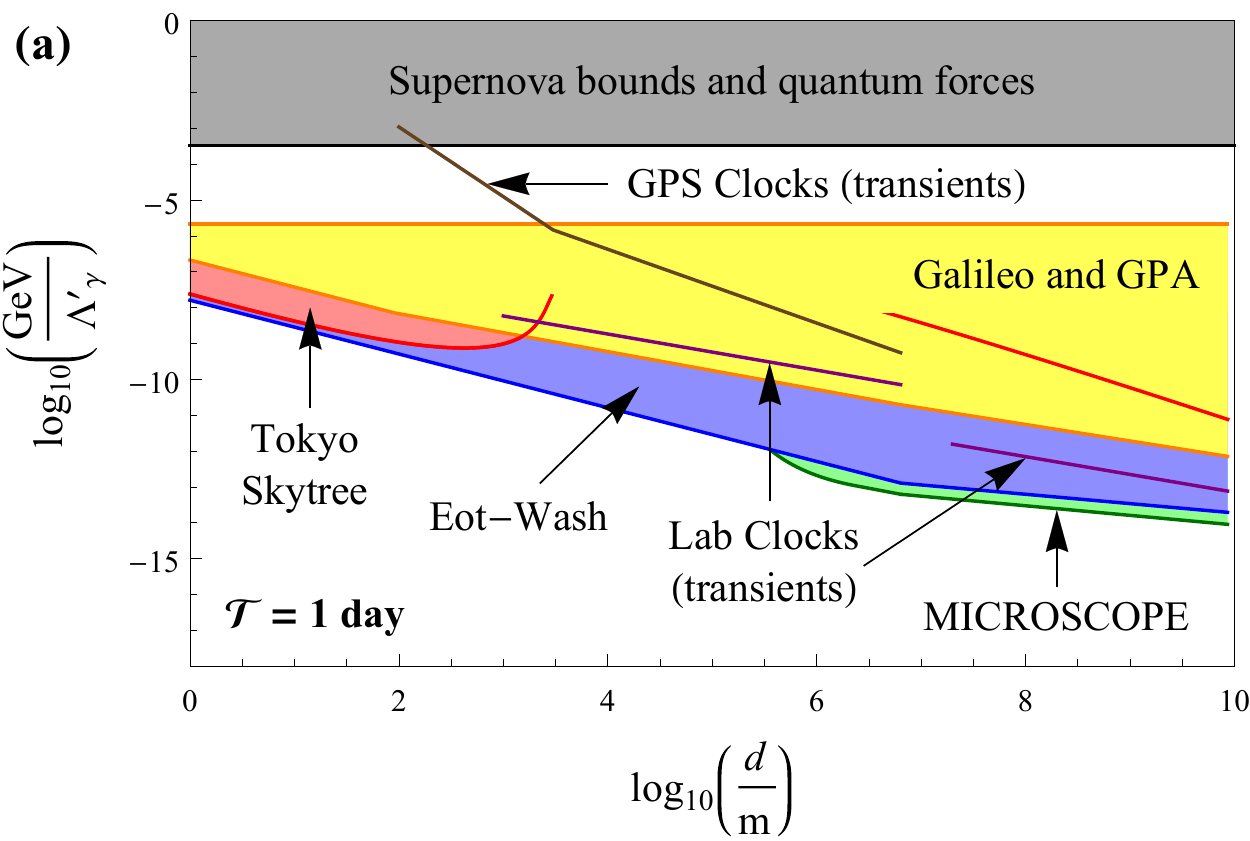}
\hspace{5mm}
\vspace{3mm}
\includegraphics[width=8.5cm]{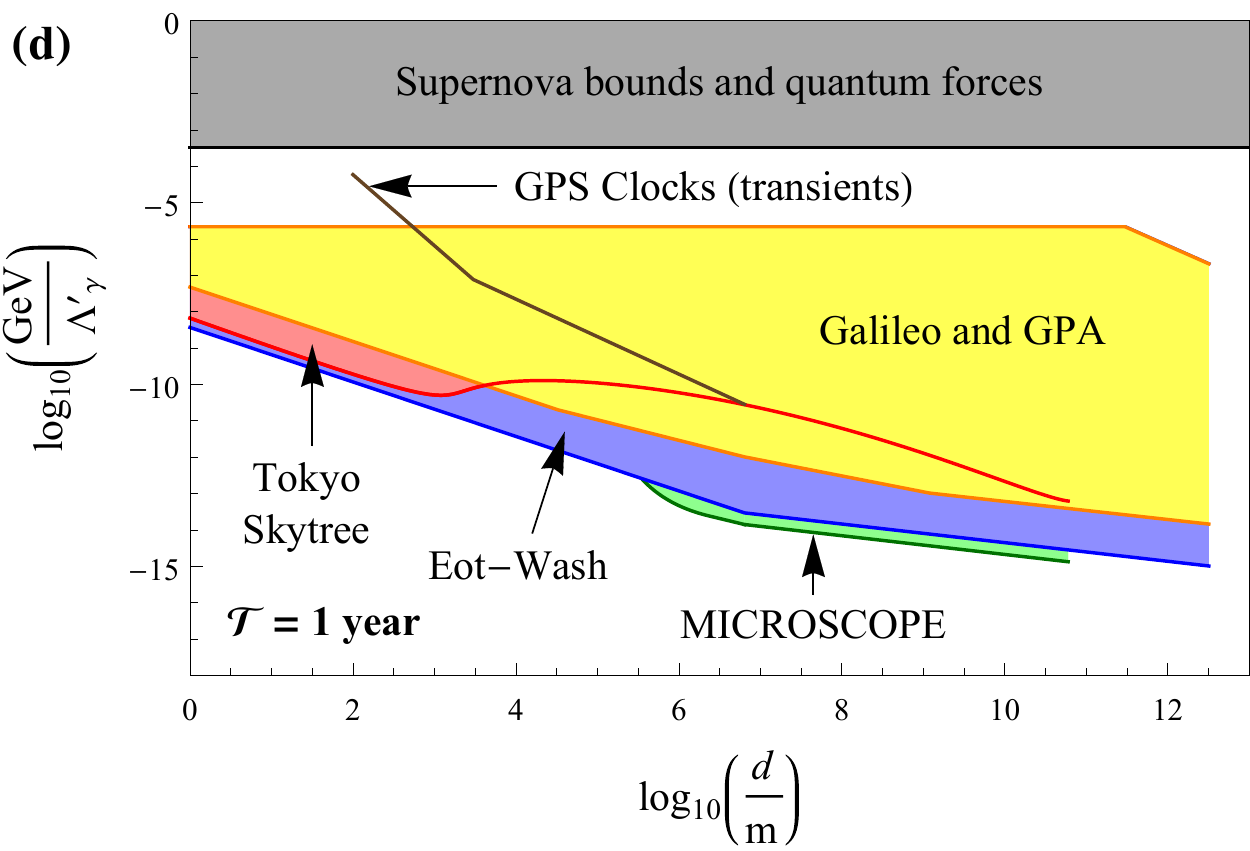}
\includegraphics[width=8.5cm]{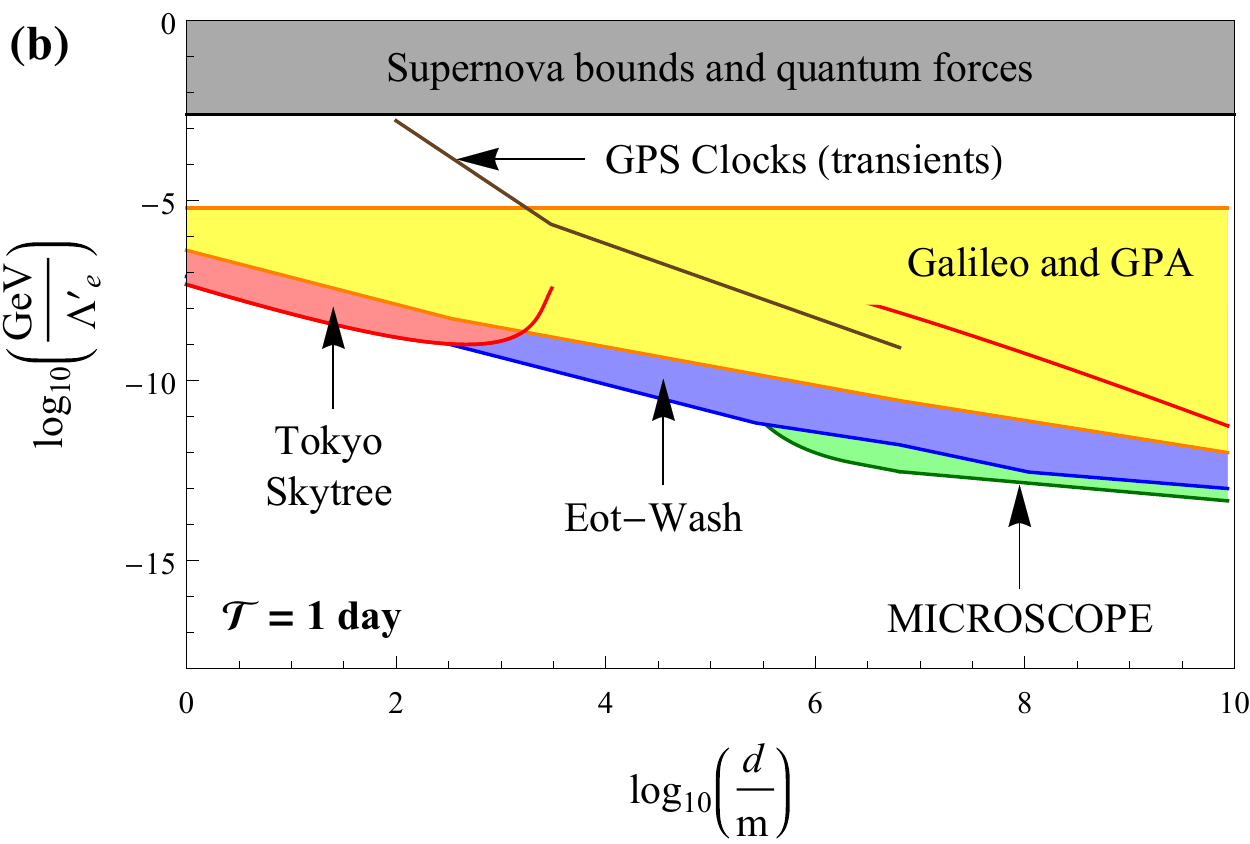}
\hspace{5mm}
\vspace{3mm}
\includegraphics[width=8.5cm]{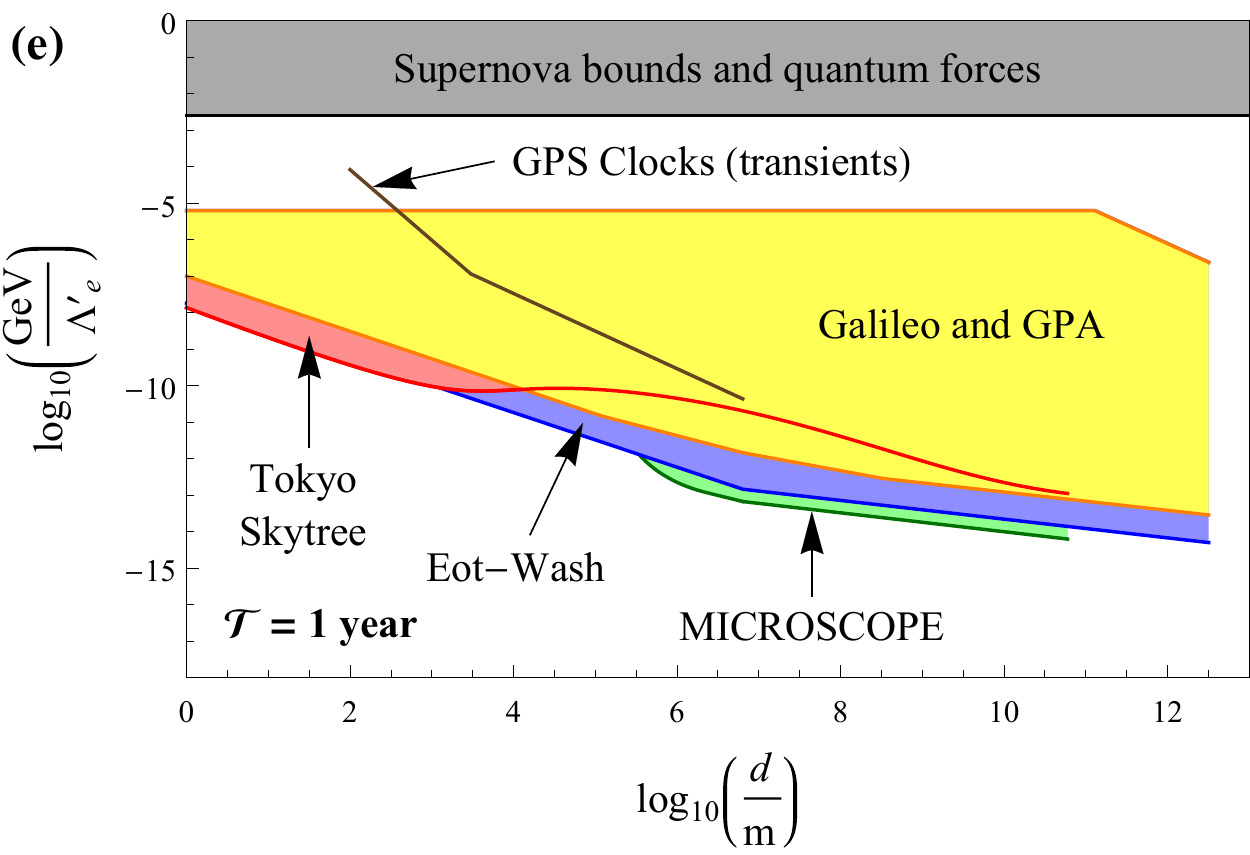}
\includegraphics[width=8.5cm]{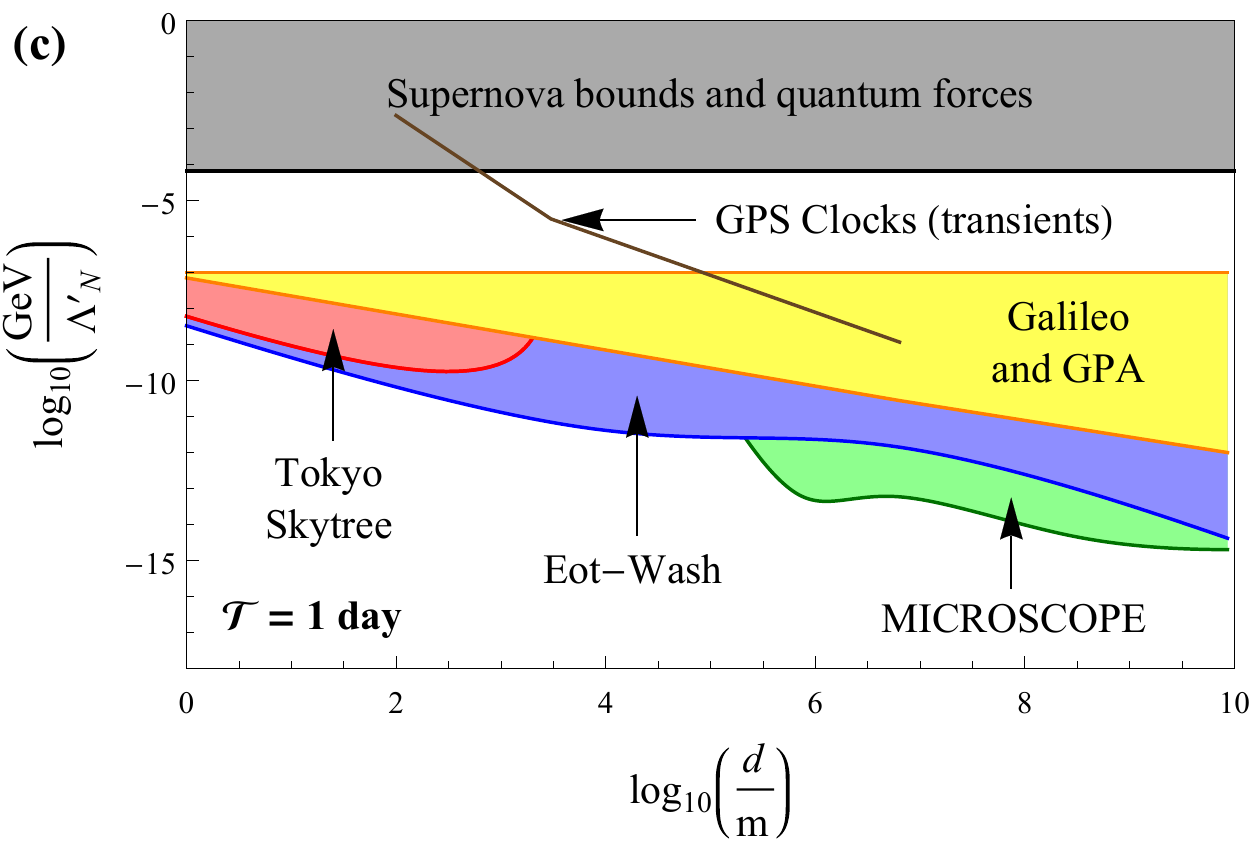}
\hspace{5mm}
\vspace{3mm}
\includegraphics[width=8.5cm]{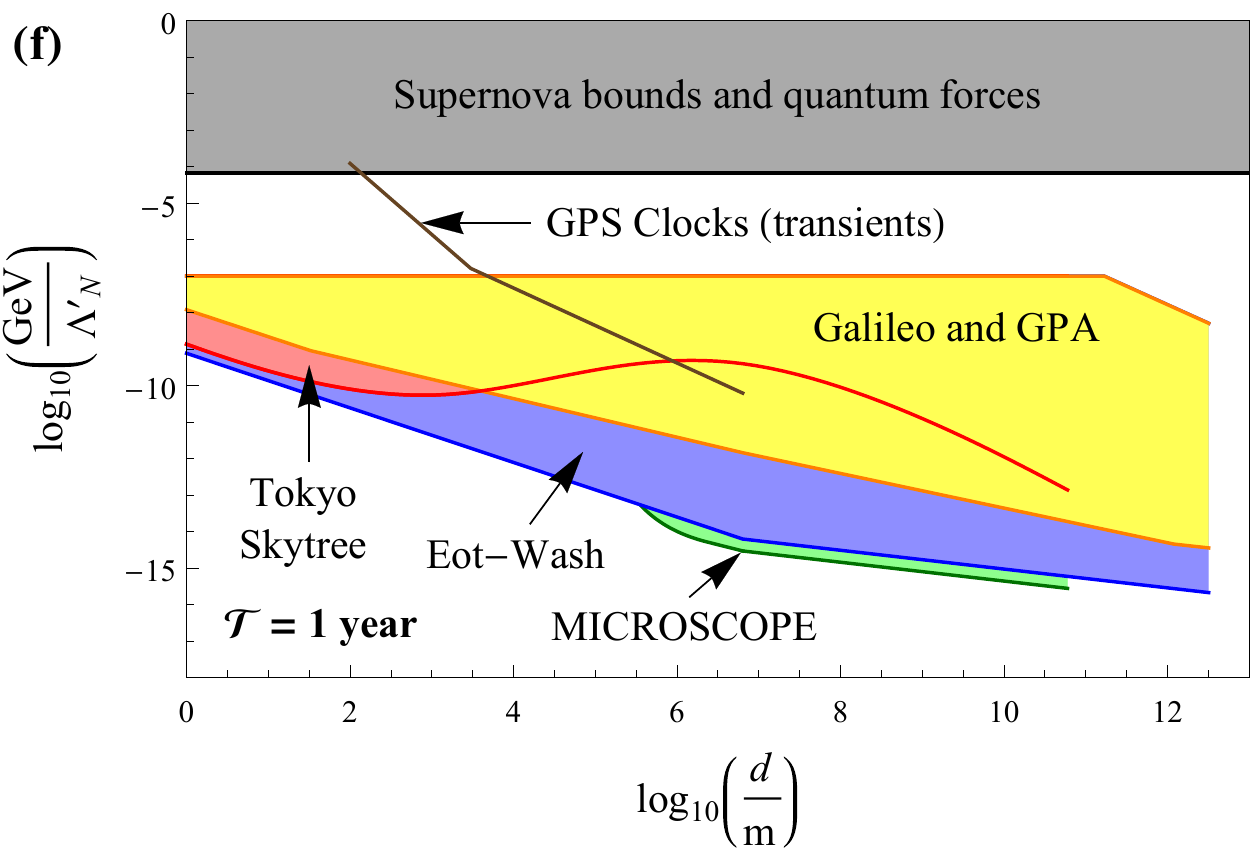}
\caption{ \normalsize (Colour online) 
Constraints on the quadratic interactions of a domain-wall-model scalar field $\phi$ with the \textbf{(a,d)} electromagnetic field (photon), \textbf{(b,e)} electron and \textbf{(c,f)} nucleons, as defined in (\ref{quad_portal}), as a function of the domain-wall transverse size parameter $d = \sqrt{2 / ( \lambda \phi_0^2 )}$, assuming that the domain-wall network has a local average energy density of $\rho_\textrm{TDN}^\textrm{local} \sim \rho_\textrm{DM}^\textrm{local} \approx 0.4~\textrm{GeV/cm}^3$ and that domain walls pass through Earth with a typical speed of $v_\textrm{TD} \sim 300~\textrm{km/s}$. 
Subfigures \textbf{(a,b,c)} correspond to an average time between encounters of a wall with Earth of $\mathcal{T} = 1~\textrm{day}$, while subfigures \textbf{(d,e,f)} correspond to an average time between encounters of a wall with Earth of $\mathcal{T} = 1~\textrm{year}$. 
The following regions of parameters are excluded by our present work from the consideration of:~comparison of Sr optical clocks at different heights in the Tokyo Skytree experiment (red region), comparison of ground- and space-based H microwave clocks in the Gravity Probe A (GPA) mission and on board the Galileo satellites (yellow region), Be-Ti and Be-Al torsion pendula in the ground-based E$\ddot{\textrm{o}}$t-Wash experiments (blue region), and Pt-Ti torsion pendulum in the space-based MICROSCOPE mission (green region). 
The region in dark grey corresponds to existing constraints from supernova energy-loss bounds \cite{Olive-Pospelov_2008-FCs+symmetron} and short-range tests of gravity \cite{Olive-Pospelov_2008-FCs+symmetron,Brax-Fichet_2019-symmetron}. 
The purple and brown lines correspond to previously reported bounds from transient searches using ground-based laboratory optical clocks \cite{Wcislo_2016_TDM-cavity,Roberts_2019_TDM-clocks} (the lower values of $d$ were probed in \cite{Wcislo_2016_TDM-cavity}, while the higher values of $d$ were probed in \cite{Roberts_2019_TDM-clocks}) and space-based microwave clocks within the GPS network \cite{Roberts_2017_TDM-GPS}, respectively, without account of screening effects. 
See the main text for more details. 
}
\label{Fig:TDN-quadratic_results}
\end{figure*}

For the sake of direct comparison to the earlier literature, in this section, we shall proceed under the same oversimplified assumptions made in the earlier papers \cite{Derevianko_2014_TDM-clocks,Wcislo_2016_TDM-cavity,Roberts_2017_TDM-GPS,Wcislo_2018_TDM-cavity,Roberts_2019_TDM-clocks}. 
Specifically, we assume the relation (\ref{DW_amplitude-TDM_network}) with $v_\textrm{TD} \sim 300~\textrm{km/s}$, $\mathcal{T} \sim 1~\textrm{day} - 1~\textrm{year}$, and the local average energy density $\rho_\textrm{TDN}^\textrm{local} \sim \rho_\textrm{DM}^\textrm{local} \approx 0.4~\textrm{GeV/cm}^3$. 
We also assume the same $\phi^4$ potential (\ref{phi4_potential}) that was considered in the theory paper \cite{Derevianko_2014_TDM-clocks}, on which the recent clock-based searches for transient signatures of passing domain walls were predicated, and focus on the same quadratic interactions of the domain-wall-model scalar field $\phi$ with the standard-model fields in (\ref{quad_portal}) that were considered in Refs.~\cite{Derevianko_2014_TDM-clocks,Wcislo_2016_TDM-cavity,Roberts_2017_TDM-GPS,Wcislo_2018_TDM-cavity,Roberts_2019_TDM-clocks}. 
We summarise our results based on the non-transient signatures of the domain-wall-model scalar field investigated in our present paper, and compare with the earlier results from clock- and cavity-based searches for transient signatures of passing domain walls, in Fig.~\ref{Fig:TDN-quadratic_results}. 
We find that our newly derived bounds via effects of non-transient variations of the fundamental constants are significantly more stringent than the previously reported bounds from transient signatures \textit{under the same set of assumptions}.

\subsection{Transient signatures}
\label{Sec:TDM_Transients}
Refs.~\cite{Wcislo_2016_TDM-cavity,Roberts_2017_TDM-GPS,Wcislo_2018_TDM-cavity,Roberts_2019_TDM-clocks} performed clock- and cavity-based searches for transient variations of the fundamental constants of the form (\ref{quad_VFCs-alpha},\ref{quad_VFCs-fermions}) induced by the passage of macroscopic domain walls through a network of detectors. 
These bounds are shown in Fig.~\ref{Fig:TDN-quadratic_results} as coloured lines (purple = ground-based laboratory optical clocks, brown = space-based microwave clocks within the GPS network). 
In these searches, it was assumed that the sought domain-wall-induced transient signals are well separated in time, which necessitates that the signal time $\Delta t \sim d / v_\textrm{TD}$ is sufficiently short, $\Delta t \ll \mathcal{T}$. 
The largest average time between encounters of a wall with Earth that can be probed in a statistically meaningful way with such types of searches is limited by the duration of the experiment $T_\textrm{exp}$, $\mathcal{T} \lesssim T_\textrm{exp}$.

\begin{figure}[t!]
\centering
\includegraphics[width=8.5cm]{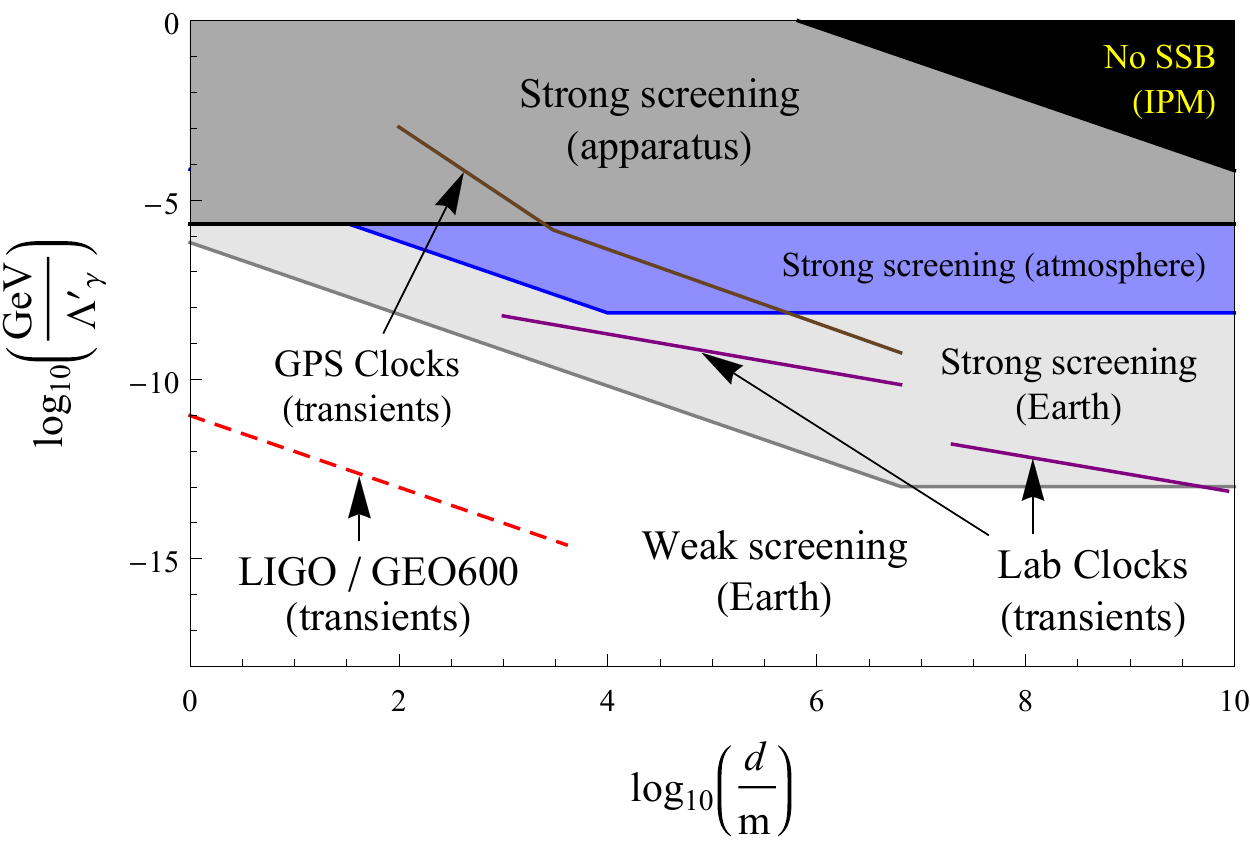}
\caption{ \normalsize (Colour online) 
Regions of domain-wall-model parameter space for the quadratic interaction of a scalar field $\phi$ with the electromagnetic field (photon), as defined in (\ref{quad_portal}), in which the scalar field is weakly screened near the surface of and inside Earth (white region), 
strongly screened near the surface of and inside Earth (light grey region), 
strongly screened by Earth's atmosphere (blue region), 
strongly screened inside an apparatus or satellite of size $\sim 60~\textrm{cm}$ and with a density comparable to Earth's average density (dark grey region), 
and where spontaneous symmetry breaking is inhibited altogether by the interplanetary and interstellar media (black region). 
The purple and brown lines correspond to previously reported bounds from transient searches using ground-based laboratory optical clocks \cite{Wcislo_2016_TDM-cavity,Roberts_2019_TDM-clocks} (the lower values of $d$ were probed in \cite{Wcislo_2016_TDM-cavity}, while the higher values of $d$ were probed in \cite{Roberts_2019_TDM-clocks}) and space-based microwave clocks within the GPS network \cite{Roberts_2017_TDM-GPS}, respectively, for a domain-wall network with a local average energy density of $\rho_\textrm{TDN}^\textrm{local} \sim \rho_\textrm{DM}^\textrm{local} \approx 0.4~\textrm{GeV/cm}^3$, a typical speed of passage of domain walls through Earth of $v_\textrm{TD} \sim 300~\textrm{km/s}$ and an average time between encounters of a wall with Earth of $\mathcal{T} = 1~\textrm{day}$, but without account of screening effects. 
The dashed red line denotes the estimated sensitivity of the current ground-based laser interferometers LIGO and GEO600 to transient signatures of passing domain walls, based on the rescaling of the estimates for $\mathcal{T} = 1~\textrm{year}$ presented in Ref.~\cite{Stadnik_2019_DM-LIFO} to $\mathcal{T} = 1~\textrm{day}$. 
}
\label{Fig:Screening_region+clock_transients-overlay}
\end{figure}

Comparing with Fig.~\ref{Fig:Screening_regions}, we see that the purple and brown curves in Fig.~\ref{Fig:TDN-quadratic_results} all lie in regions of domain-wall parameter spaces where the scalar field is strongly screened near the surface of and inside Earth, and in some cases also in regions of parameter spaces where the scalar field is strongly screened inside the apparatus itself (see Fig.~\ref{Fig:Screening_region+clock_transients-overlay} for an overlay in the case of the electromagnetic coupling for $\mathcal{T} = 1~\textrm{day}$). 
Thus the unwitting assumption made in the earlier searches \cite{Wcislo_2016_TDM-cavity,Roberts_2017_TDM-GPS,Wcislo_2018_TDM-cavity,Roberts_2019_TDM-clocks} that domain walls pass in an unperturbed manner through the strongly repulsive potential generated by Earth (and in some cases the apparatus itself) appears dubious. 
More specifically, there are at least two possible outcomes for a scalar-field domain wall incident on a strongly repulsive potential generated by Earth:~(i) part of the domain wall may tunnel through Earth with a diminished amplitude and/or (ii) part of the domain wall may envelop Earth and ``pinch off'', forming a bubble-like defect structure around Earth in the process --- this defect structure, which is distinct from the bubble-like defect structures discussed in Sec.~\ref{Sec:Main_Phenomenology}, may be either stable (if there is insufficient kinetic energy to overcome the potential barrier inside Earth) or metastable (in which case the bubble might undergo a number of oscillations before collapsing and radiating away its energy in the form of $\phi$ particles). 
The outcome may even depend on the properties of the incident domain wall, such as its initial speed and transverse size, the interaction strength, as well as the size and density of the target body.

The investigation of this problem, as well as the analogous problem concerning the strongly antiscreened regime for the same types of quadratic-in-$\phi$ interactions in (\ref{quad_portal}) but with the signs reversed, warrants numerical simulations, which go beyond the scope of the present work. 
These calculations are not only of interest for numerous terrestrial experiments \cite{Derevianko_2014_TDM-clocks,Stadnik_2015_DM-LIFO,Stadnik_2016_DM-cavities,Stadnik_2019_DM-LIFO,Wcislo_2016_TDM-cavity,Roberts_2017_TDM-GPS,Wcislo_2018_TDM-cavity,Roberts_2019_TDM-clocks}, but also for measurements involving networks of pulsars \cite{Stadnik_2014_TDM-pulsars}. 
(Note, however, that laser-interferometry-based searches for transient signatures of passing domain walls using existing gravitational-wave detectors such as LIGO and GEO600 \cite{Stadnik_2019_DM-LIFO} are already expected to be able to probe regions of parameter space that lie deep within the weak screening regime where back-action effects are negligible; see, e.g., the dashed red line in Fig.~\ref{Fig:Screening_region+clock_transients-overlay}.) 
Additionally, reanalysis of the data in Refs.~\cite{Wcislo_2016_TDM-cavity,Roberts_2017_TDM-GPS,Wcislo_2018_TDM-cavity,Roberts_2019_TDM-clocks} is warranted for the strongly screened and strongly antiscreened regimes. 
The interaction of scalar-field topological defects with strongly repulsive or strongly attractive potentials generated by dense bodies may also affect the evolution and propagation of topological defects at very late cosmological times. 
For example, if parts of domain walls are ``pinched off'' by dense bodies and subsequently radiate away, then the energy density associated with the domain-wall network would decrease in time at a faster rate. 
This possibility also warrants further investigation. 
Additionally, we note that in the case of sufficiently strong interactions, when spontaneous symmetry breaking is inhibited altogether within the Solar System and Galaxy by the interplanetary and interstellar media, there is no longer topological stability to support the free propagation of a domain wall through the Solar System or Galaxy.

Finally, let us briefly remark that if the velocities of domain walls are taken to be semi-relativistic, in accordance with the results of numerical simulations in Refs.~\cite{Spergel_1989_walls,Avelino2004walls,Avelino2005walls-semirelativistic}, instead of the assumption $v_\textrm{TD} \sim 300~\textrm{km/s}$ made in Refs.~\cite{Wcislo_2016_TDM-cavity,Roberts_2017_TDM-GPS,Wcislo_2018_TDM-cavity,Roberts_2019_TDM-clocks}, then bounds from experiments searching for transient signatures of passing domain walls would be shifted towards larger transverse wall size parameters, since the characteristic wall passage time is $\Delta t \sim d / v_\textrm{TD}$. 
On the other hand, bounds from experiments searching for non-transient signatures of domain walls would not be shifted in such a manner. 
The sensitivity of experiments searching for either transient or non-transient signatures of domain walls that depend on $\phi_0$ would be increased in the case of a larger value of $v_\textrm{TD}$, in accordance with Eq.~(\ref{DW_amplitude-TDM_network}). 
Searches for transient signatures would also benefit from being shifted up towards larger values of $d$ due to the associated increase in $\phi_0$ in this case, see Eq.~(\ref{DW_amplitude-TDM_network}).

\subsection{Non-transient signatures}
\label{Sec:TDM_Non-transients}

We now derive bounds on non-transient variations of the fundamental constants, which arise in the same scalar-field domain-wall model, via the torsion-pendulum-based and clock-based approaches discussed in Secs.~\ref{Sec:Main_TP} and \ref{Sec:Main_Clocks}, respectively. 
Like the transient searches described in Sec.~\ref{Sec:TDM_Transients} above, we assume that the domain walls are well separated, $\Delta t \ll \mathcal{T}$, and that $\Delta t \ll T_\textrm{exp}$. 
In this case, the effects of the domain-wall scalar field are quasi-non-transient on the experimental timescale, since terrestrial detectors are located very far from a domain wall for the vast majority of the time. 
Importantly, the sign of $\phi^2$ remains unchanged on either side of a domain wall, meaning that the sign of these quasi-non-transient effects remains unchanged on either side of a wall, and so there is no cancellation between contributions from different domains. 
Therefore, if we neglect the brief transients associated with the passage of domain walls through Earth (which is justified when $\Delta t \ll \mathcal{T}$ and $\Delta t \ll T_\textrm{exp}$), then it is possible to utilise our findings from Sec.~\ref{Sec:Main_Phenomenology} with minimal modification. 
In contrast to transient search methods, searches for quasi-non-transient signatures also retain their sensitivity in the regime $\mathcal{T} \gg T_\textrm{exp}$, when the average time between encounters of a wall with Earth greatly exceeds the duration of the experiment, since it is not necessary to be near a passing domain wall to experience non-transient signatures. 

Using Eqs.~(\ref{quad_VFCs-alpha}), (\ref{quad_VFCs-fermions}), (\ref{scalar_profile_weak_screening-A}), (\ref{scalar_profile_weak_screening-B}), (\ref{scalar_profile_strong_screening-A}), (\ref{scalar_profile_strong_screening-Bx}), (\ref{test-mass_acceleration-generic}), (\ref{Sr_sensitivity_coefficients_EFFECTIVE}), (\ref{H_sensitivity_coefficients_EFFECTIVE}) and (\ref{DW_amplitude-TDM_network}), as well as the fractional mass-energy differences of the electromagnetic, electron-mass and nucleon-mass contributions to the relevant test-mass pairs in Table~\ref{tab:frac_mass-energies}, we derive bounds on the quadratic interactions of the domain-wall-model scalar field $\phi$ with the electromagnetic field, electron and nucleons in Eq.~(\ref{quad_portal}). 
We present our bounds as coloured regions in Fig.~\ref{Fig:TDN-quadratic_results} [red = Tokyo Skytree experiment (clocks), yellow = Galileo satellites and Gravity Probe A mission (clocks), blue = E$\ddot{\textrm{o}}$t-Wash experiments (torsion pendula), green = MICROSCOPE mission (torsion pendulum)]. 
We see that our newly derived bounds via effects of non-transient variations of the fundamental constants are significantly more stringent than the previously reported bounds from different types of non-transient signatures and transient signatures under the same set of assumptions. 
In our analysis, we have restricted ourselves to the values of $d$ satisfying the condition $\Delta t \ll T_\textrm{exp}$, for which the signal is quasi-non-transient on the experimental timescale. 
In the case of the currently available Tokyo Skytree and MICROSCOPE mission datasets, for which $T_\textrm{exp} \approx 1~\textrm{week}$, it is possible to extend the respective analyses to the higher values of $d$ in Fig.~\ref{Fig:TDN-quadratic_results}(d-f) by additionally considering the transient effects associated with the passage of domain walls through Earth. 
We do not present bounds of the type discussed in Sec.~\ref{Sec:Main_Lab-astro-spectra}, since in the absence of a concrete mechanism leading to the enhancement of the domain-wall energy density inside galaxies by orders of magnitude compared to an average cosmological value, further assumptions about the domain-wall energy density in low-density astrophysical environments are required. 

Note the differences in the shapes of some of our bounds based on effects of non-transient variations of the fundamental constants in going from Fig.~\ref{Fig:Infinite_DW_Network-quadratic} to Fig.~\ref{Fig:TDN-quadratic_results}, due to the transition from the weak screening regime to the strong screening regime. 
In the weak screening regime, our bounds generally scale as $\Lambda'_X \propto d^{3/4}$ when $d \ll R_\oplus$ and as $\Lambda'_X \propto d^{1/4}$ when $d \gg R_\oplus$. 
On the other hand, in the strong screening regime, the limits from ground-based experiments degrade sharply due to the strong screening of the scalar field near Earth's surface, and the scaling of the limits with $d$ takes a more complicated form. 
Additionally, we note that unlike in Fig.~\ref{Fig:Infinite_DW_Network-quadratic}, there are no analogous bounds on the bubble-like defect structures surrounding Earth or the Sun in Fig.~\ref{Fig:TDN-quadratic_results} from the consideration of their gravitational effects, partly because the masses of these bubbles are now diminished by at least a factor of $\sim 10^{-3}$ compared to Eqs.~(\ref{mass_bubble_core_Earth},\ref{mass_bubble_core_Sun}) in Sec.~\ref{Sec:Main_Grav-effects}. 

Let us also remark that, in the more realistic case when the domain-wall network is stiff and hence can only make a sub-dominant contribution to the galactic dark matter abundance, the bounds from the $\phi_0$-dependent non-transient signatures considered in our present work would generally become even more stringent than the previously reported types of bounds from transient signatures. 
This is because the bounds on the parameters $\Lambda'_X$ from transient signatures scale as $\Lambda'_X \propto \rho_\textrm{TDN}^{1/2}$ when the effects of screening of the scalar field by Earth can be neglected, whereas our bounds on $\Lambda'_X$ from the effects of non-transient variations of the fundamental constants scale as $\Lambda'_X \propto \rho_\textrm{TDN}^{1/4}$ in the weak screening regime (and in the strong screening regime generally scale as $\Lambda'_X \propto \rho_\textrm{TDN}^{1/2}$). 
In Fig.~\ref{Fig:TDN-quadratic_results}, we have assumed that $\rho_\textrm{TDN}^\textrm{local} \sim \rho_\textrm{DM}^\textrm{local} \approx 0.4~\textrm{GeV/cm}^3$, whereas the maximum allowable energy density for a stiff domain-wall network is ten orders of magnitude smaller, see Eq.~(\ref{Omega_walls_bound}).

\vspace{200mm}

\section{Discussion}
\label{Sec:Discussion}
In this paper, we have investigated effects associated with the back-action of ambient matter on scalar field(s) in models of macroscopic topological defects composed of scalar field(s) that interact with standard-model fields via quadratic-in-$\phi$ couplings. 
We have shown that in these models there is an environmental dependence of the fundamental constants of nature, as well as spatial variations of the fundamental constants in the vicinity of a dense body such as Earth due to the formation of a bubble-like defect structure surrounding the dense body. 
Making only minimal cosmological assumptions, we have derived bounds on such non-transient variations of the fundamental constants arising in models of scalar-field domain walls with $\phi^2$ interactions from torsion-pendulum experiments that search for equivalence-principle-violating forces, experiments comparing the frequencies of ground- and space-based atomic clocks as well as ground-based clocks at different heights, and measurements comparing atomic and molecular transition frequencies in terrestrial and low-density astrophysical environments. 
We have also derived bounds on the bubble-like defect structures surrounding Earth and the Sun from the consideration of their gravitational effects on planets and satellites within the Solar System. 
These bounds are summarised in Fig.~\ref{Fig:Infinite_DW_Network-quadratic}. 
For the domain-wall transverse size parameters $d \gtrsim 700~\textrm{km}$, the most stringent upper bounds on the interaction parameters $\Lambda'_X$, defined in (\ref{quad_portal}), come from the space-based MICROSCOPE mission based on the torsion pendulum technique. 
On the other hand, for $d \lesssim 700~\textrm{km}$, the most stringent upper bounds on $\Lambda'_X$ come from the ground-based E$\ddot{\textrm{o}}$t-Wash experiments using torsion pendula, with comparable bounds on $\Lambda'_{X}$ for $d \lesssim 450~\textrm{m}$ coming from frequency-comparison measurements using optical atomic clocks at different heights in the Tokyo Skytree experiment. 
Noting that these upper bounds on $\Lambda'_X$ via effects of non-transient variations of the fundamental constants scale as $\Lambda'_X \propto \rho_\textrm{walls}^{1/4}$ in the weak screening regime (and generally as $\Lambda'_X \propto \rho_\textrm{walls}^{1/2}$ in the strong screening regime), where $\rho_\textrm{walls}$ is the average energy density associated with the domain-wall network, we see that our results in Fig.~\ref{Fig:Infinite_DW_Network-quadratic} constrain the present-day mass-energy fraction of the Universe due to a network of infinite domain walls produced shortly after the BBN or CMB epochs to be: 
\begin{equation}
\label{Infinite_DW_Network_Symmetron_Bound-Main}
\Omega_{\textrm{walls},0} \ll 10^{-10}  \, , 
\end{equation}
for the symmetron model with the $\phi^4$ potential (\ref{phi4_potential}) and $\phi^2$ interactions (\ref{quad_portal}), improving over the CMB quadrupolar temperature anisotropy bounds (\ref{Omega_walls_bound}) by at least 5 orders of magnitude. 
In arriving at the bound (\ref{Infinite_DW_Network_Symmetron_Bound-Main}), note the important role of comparisons of atomic and molecular spectra in ground-based laboratory and low-density astrophysical environments, which rule out parts of the parameter space where the scalar field becomes strongly screened inside conventional apparata (see Fig.~\ref{Fig:Screening_regions}) and so cannot otherwise be probed by the specific torsion-pendulum and clock experiments discussed in our present paper. 
We likewise expect very strong constraints on symmetron domain walls produced at most times proceeding the weak interaction freeze-out epoch. 
The interesting case when domain walls are produced at the very late times corresponding to redshifts $z \lesssim \mathcal{O} (10)$, when the Universe is no longer homogeneous, warrants particular attention in future investigations.

What are the ideal measurements to search for non-transient variations of the fundamental constants arising in models of scalar-field topological defects? 
For clock comparison experiments, it is preferable to have one of the clocks located near Earth's surface and the other clock located in space at an altitude $\Delta h \gtrsim R_\oplus$, in order to avoid suppression factors of the form $\sim \Delta h / R_\oplus \ll 1$ in the weak screening regime and $\sim \Delta h / \textrm{min} \left( d , R_\oplus \right) \ll 1$ in the strong screening regime, both of which arise for $d \gtrsim \Delta h$ when $\Delta h \ll R_\oplus$. 
It may be possible to further boost the sensitivity of clock-based measurements to scalar-field-induced non-transient variations of the fundamental constants by using ``non-standard'' clocks based on various transitions in highly-charged ions \cite{Berengut_2010_HCI-VFCs,Berengut_2011_HCI-VFCs}, nuclei \cite{Flambaum_2006_Th-VFCs} and molecules \cite{Flambaum_2007_mol-VFCs,Ye_2008_mol-VFCs,DeMille_2008_mol-VFCs}, in which the sensitivity to variations of one or more of the fundamental constants may be greatly enhanced relative to the ``standard'' types of clocks considered in our present work. 
In the case of torsion-pendulum experiments, there is an exponential loss in sensitivity for the defect transverse size parameters $d \lesssim h$, where $h$ is the altitude of the apparatus. 
Ground-based torsion-pendulum experiments performed close to Earth's surface, therefore, allow the evasion of this exponential loss in sensitivity for a broader range of defect transverse size parameters compared with space-based experiments. 
On the other hand, space-based torsion-pendulum experiments performed at an altitude $h \gtrsim R_\oplus$ avoid a possible suppression factor of the form $\sim h / \textrm{min} \left( d , R_\oplus \right) \ll 1$ in the strong screening regime, providing an advantage over ground-based experiments in this case. 
The minimum values of $\Lambda'_X$ that can be probed by clock- and torsion-pendulum-based experiments are ultimately limited by screening of the scalar field inside the apparatus itself (or by the interplanetary/interstellar medium). 
It may be possible to probe smaller values of $\Lambda'_X$ than those probeable by the specific torsion-pendulum and clock experiments considered in our present paper via experiments that test the equivalence principle on short distances, such as the Rot-Wash experiment \cite{Rot-Wash_exp_1999}, and experiments using smaller apparatus components; such experiments may also be used to probe the defect transverse size parameters $d \lesssim 1~\textrm{m}$ that are complementary to the macroscopic transverse size parameters considered in our present work. 

Regarding astrophysical measurements, if the parameter $\phi_0$ in the potential (\ref{phi4_potential}) or the analogous potentials (\ref{phi4_potential-strings},\ref{phi4_potential-monopoles}) remains constant in time (as assumed in our present work), then astrophysical spectral measurements of the type considered in our present paper offer a more powerful probe of the presently-considered effects of non-transient variations of the fundamental constants than astrophysical probes at earlier cosmological epochs, due to the better intrinsic precision of astrophysical spectral measurements. 
We note that the situation here is markedly different compared with models of scalar-field dark matter with the same $\phi^2$ interactions as in (\ref{quad_portal}), in which variations of the fundamental constants are linearly correlated with changes in the dark matter density and hence more sensitive astrophysical probes can involve intrinsically less precise measurements at much earlier cosmological epochs, such as BBN where the effects of varying fundamental constants are strongly enhanced due to the very large dark matter density during the BBN epoch \cite{Stadnik_2015_DM-VFCs}. 

If one departs from minimal cosmological assumptions and takes at face value the oversimplified model of a ``dark matter'' network of domain walls considered in Refs.~\cite{Derevianko_2014_TDM-clocks,Wcislo_2016_TDM-cavity,Roberts_2017_TDM-GPS,Wcislo_2018_TDM-cavity,Roberts_2019_TDM-clocks}, then our newly derived bounds from effects of non-transient variations of the fundamental constants in torsion-pendulum experiments give significantly more stringent bounds than previously reported clock- and cavity-based searches for transient signatures of passing domain walls in \cite{Wcislo_2016_TDM-cavity,Roberts_2017_TDM-GPS,Wcislo_2018_TDM-cavity,Roberts_2019_TDM-clocks} \textit{under the same set of assumptions}, with the recent clock comparison measurements at different heights within Tokyo Skytree also probing (via effects of non-transient variations of the fundamental constants) regions of parameter space that were inaccessible to previous searches for transient signatures, see Fig.~\ref{Fig:TDN-quadratic_results} for a summary. 
The possible degradation of the clock- and cavity-based bounds reported in \cite{Wcislo_2016_TDM-cavity,Roberts_2017_TDM-GPS,Wcislo_2018_TDM-cavity,Roberts_2019_TDM-clocks} due to the effects of the strongly repulsive potential generated by Earth on passing domain walls (which Refs.~\cite{Wcislo_2016_TDM-cavity,Roberts_2017_TDM-GPS,Wcislo_2018_TDM-cavity,Roberts_2019_TDM-clocks} neglected) would only further strengthen the importance of our presently-derived bounds from non-transient signatures. 
We also remark that the types of non-transient signatures of macroscopic domain walls and other topological defects considered in our present paper are quite general. 
In particular, these non-transient signatures arise even if the Universe contains only a single domain, i.e., in the absence of any domain walls of cosmological origin; 
in this case, our bounds in Fig.~\ref{Fig:Infinite_DW_Network-quadratic} may even be strengthened further, since the bound in (\ref{Omega_walls_bound}) no longer applies and so $\phi_0$ can now be significantly larger.

It is worth reminding here that in models of oscillating scalar-field dark matter with the same $\phi^2$ interactions as in (\ref{quad_portal}), existing bounds from clock-based searches for oscillating variations of the fundamental constants \cite{Stadnik_2015_DM-VFCs,Stadnik_2016_DM-clocks} are more stringent for a broad range of dark-matter particle masses $m_\phi$ than bounds from effects of non-oscillating variations of the fundamental constants that depend on the amplitude of the oscillating scalar field far away from dense bodies (also denoted by $\phi_0$) \cite{Hees_2018_DM-EP}, the latter of which share similarities with the effects considered in our present work. 
This difference in hierarchy compared with models of ``dark matter'' domain-wall networks can be traced back to the absence of velocity suppression factors in the amplitude of an oscillating scalar field, which takes the form $\phi_0^2 \sim \rho_\phi / m_\phi^2 \sim \rho_\phi T_\textrm{osc}^2$ [compare with Eq.~(\ref{DW_amplitude-TDM_network}), which takes the form $\phi_0^2 \sim \rho_\textrm{TDN} v_\textrm{TD} \mathcal{T} d \ll \rho_\textrm{TDN} v_\textrm{TD}^2 \mathcal{T}^2$ for the case of well-separated topological defects], and provides greater benefit to searches for oscillating signatures in the weak screening regime, which is accessible with current clock-based experiments that are sensitive to long periods of oscillation up to $T_\textrm{osc} \sim 10~\textrm{years}$. 
In particular, this absence of velocity suppression factors allows current clock-based experiments to already probe regions of parameter space that lie deep within the weak screening regime where oscillating signatures benefit from being practically unaffected by screening effects, whereas the $\phi_0$-dependent non-oscillating signatures (which rely on the deformation of the scalar field around Earth) suffer from the milder deformation of the scalar field in the weak screening regime. 
Let us also remark that there is a difference in terms of the screening mechanisms between models of scalar-field topological defects and those of oscillating scalar-field dark matter, when the scalar fields in both models have the same $\phi^2$ interactions (\ref{quad_portal}). 
In the case of, e.g., scalar-field domain walls with the $\phi^4$ potential in (\ref{phi4_potential}), where there is spontaneous symmetry breaking in a sufficiently dense body, the scalar-field amplitude is exponentially suppressed inside the body, provided that the dense body is sufficiently large (but otherwise of finite size), see Sec.~\ref{Sec:Appendix_strong}. 
On the other hand, in models of oscillating scalar-field dark matter with the usual $\phi^2$ potential, spontaneous symmetry breaking does not occur, and so there is only sub-exponential suppression of the amplitude of the scalar-field oscillations near the surface of and inside the dense body.

In our present paper, we have focused on models of domain walls with the quadratic-in-$\phi$ interactions (\ref{quad_portal}), which were also considered in the earlier papers \cite{Derevianko_2014_TDM-clocks,Wcislo_2016_TDM-cavity,Roberts_2017_TDM-GPS,Wcislo_2018_TDM-cavity,Roberts_2019_TDM-clocks}, including the choice of signs. 
In the case of domain walls with quadratic-in-$\phi$ interactions of the type in (\ref{quad_portal}) but with the signs reversed, the scalar field becomes antiscreened (rather than screened) in dense environments and in the vicinity of dense bodies. 
For the case when $2 \rho_X / (\Lambda'_X)^2 \ll \lambda \phi_0^2$, where the weak antiscreening regime applies, our weak-screening-regime results in Figs.~\ref{Fig:Infinite_DW_Network-quadratic} and \ref{Fig:TDN-quadratic_results} carry over unchanged, since the form of the scalar field deformation in Eqs.~(\ref{scalar_profile_weak_screening-A},\ref{scalar_profile_weak_screening-B}) remains unchanged except for the sign, see Eqs.~(\ref{Appendix:scalar_profile_weak_antiscreening-A},\ref{Appendix:scalar_profile_weak_antiscreening-B}). 
In the strong antiscreening regime, however, one expects to encounter qualitatively different behaviour compared with the strong screening regime. 
In particular, the strong antiscreening regime may open up interesting new opportunities for experiments comparing above-surface clocks with clocks located either underground (e.g., in a mine with up to a few km of rock coverage) or underwater (e.g., in a submarine with up to $\sim 10~\textrm{km}$ of water coverage), experiments with torsion pendula and atom interferometers located underground or underwater, and comparisons of atomic and molecular spectra in ground-based laboratory and high-density astrophysical environments (e.g., absorption lines originating from the surfaces of white dwarfs). 
Transportable clocks have recently been used to perform clock comparisons between above-surface and underground locations \cite{Grotti_2018_UG-clocks}, while investigations of the possible dependence of the fundamental constants on the local gravitational potential using white-dwarf spectra have been reported in Refs.~\cite{Berengut_2013_WD-constants,Ubachs_2014_WD-constants,Webb_2020_WD-constants}. 
We note that in the case of sufficiently strong couplings, the apparent changes in the electromagnetic fine-structure constant and fermion masses according to Eqs.~(\ref{quad_VFCs-alpha},\ref{quad_VFCs-fermions}), but with the signs of the variations reversed, can formally be $\Delta \alpha / \alpha < -1$ and $\Delta m_f / m_f < -1$. 
This apparently unphysical behaviour signals the breakdown of the applicability of perturbation theory and a transition to the non-perturbative regime. 
Furthermore, in the regime when $(\phi / \Lambda'_X)^2 > 1$, one would generally expect higher-dimensional operators to become important, regardless of the signs entering Eq.~(\ref{quad_portal}). 
We leave the investigation of these interesting questions to future work.

\begin{figure*}[t!]
\centering
\includegraphics[width=8.5cm]{Effective_potential_low_density_v2.pdf}
\hspace{5mm}
\includegraphics[width=8.5cm]{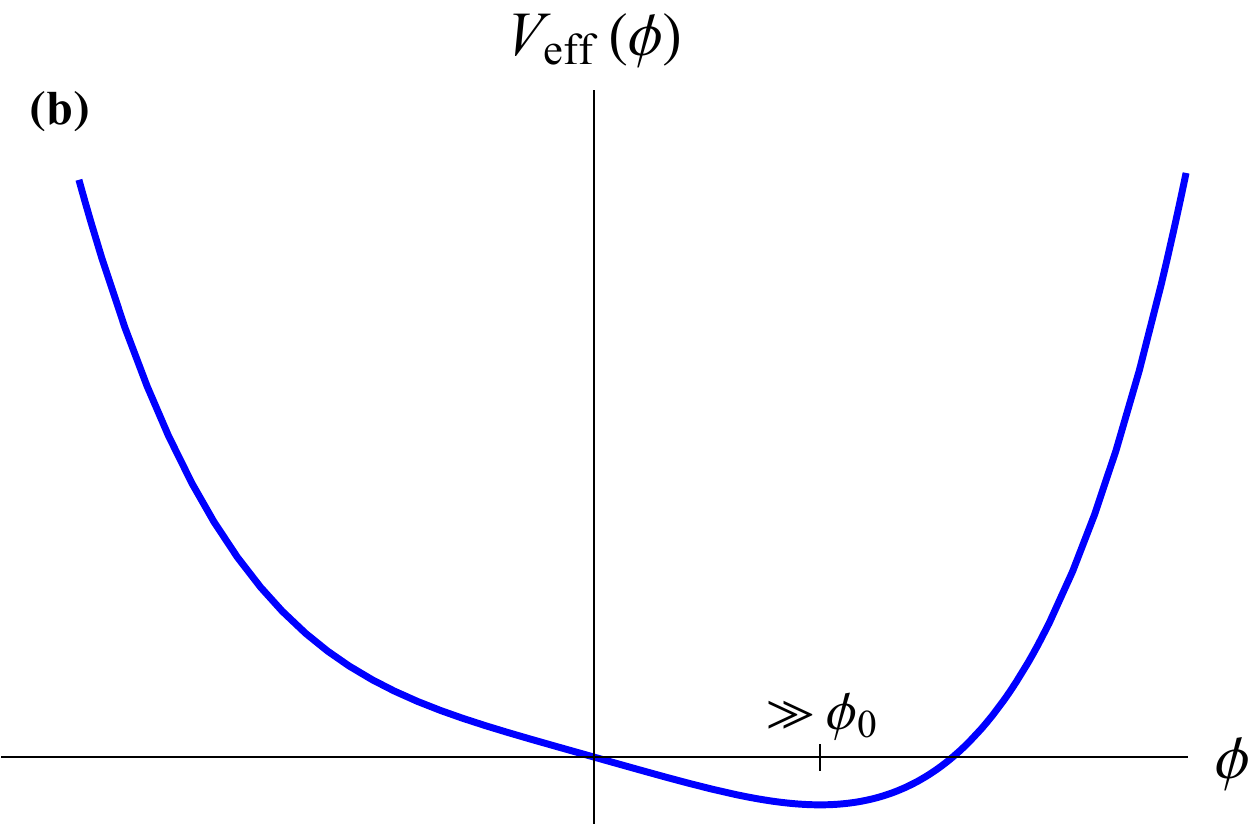}
\caption{ \normalsize (Colour online) 
Form of the effective potential (\ref{phi4_effective-potential-LINEAR}) with the lower sign choice in a \textbf{(a)} low-density environment and \textbf{(b)} high-density environment. 
}
\label{Fig:Effective_potentials-LINEAR}
\end{figure*}

Let us now discuss models of topological defects with other types of interactions. 
We begin with the linear interactions of a domain-wall scalar field $\phi$ with the standard-model fields: 
\begin{equation}
\label{linear_portal}
\mathcal{L}_\textrm{int}^\textrm{lin} = \pm \frac{\phi}{\Lambda_\gamma} \frac{F_{\mu\nu}F^{\mu\nu}}{4} \mp \sum_f \frac{\phi}{\Lambda_f} m_f \bar{f}f  \, , 
\end{equation}
which lead to apparent variations of the fundamental constants analogously to (\ref{quad_VFCs-alpha},\ref{quad_VFCs-fermions}). 
For the $\phi^4$ potential (\ref{phi4_potential}), the effective potential experienced by the scalar field in the presence of ambient matter reads: 
\begin{equation}
\label{phi4_effective-potential-LINEAR}
V_\textrm{eff} \left( \phi \right) = \frac{\lambda}{4} \left( \phi^2 - \phi_0^2 \right)^2 \pm \sum_{X=\gamma,e,N} \frac{\rho_X \phi}{ \Lambda_X }  \, .   
\end{equation}
In a low-density environment, the effective potential in (\ref{phi4_effective-potential-LINEAR}) has two minima at $\phi \approx \pm \phi_0$ as shown in Fig.~\ref{Fig:Effective_potentials-LINEAR}(a), while in a high-density environment, the effective potential has only a single minimum at $\phi \approx \mp [\rho_X / (\lambda \Lambda_X)]^{1/3}$ for the upper/lower sign choices in (\ref{linear_portal},\ref{phi4_effective-potential-LINEAR}) in the limiting case when $[\rho_X / (\lambda \Lambda_X)]^{1/3} \gg \phi_0$ as shown in Fig.~\ref{Fig:Effective_potentials-LINEAR}(b) for the lower sign choices in (\ref{linear_portal},\ref{phi4_effective-potential-LINEAR}). 
Since the scalar field is driven away from $\phi = \pm \phi_0$ to larger amplitudes in the presence of ambient matter, the scalar field therefore experiences antiscreening irrespective of the sign choices in (\ref{linear_portal},\ref{phi4_effective-potential-LINEAR}), in contrast to the screening effects that arise for the $\phi^2$ interactions (\ref{quad_portal}). 
In the case of the lower sign choices in (\ref{linear_portal},\ref{phi4_effective-potential-LINEAR}), if the couplings are sufficiently strong, then there will be a transition to the non-perturbative regime, similarly to the case of $\phi^2$ interactions of the type in (\ref{quad_portal}) but with the signs reversed. 
We note that the underlying mechanisms generating (anti)screening effects are different for linear-in-$\phi$ and quadratic-in-$\phi$ interactions, since linear-in-$\phi$ interactions give rise to a source term in the classical equation of motion for a scalar field, whereas quadratic-in-$\phi$ interactions give rise to a potential term in the classical equation of motion for a scalar field.

Interestingly, the cosmological production of topological defects may be inhibited altogether for sufficiently strong linear-in-$\phi$ interactions. 
For the $\phi^4$ potential (\ref{phi4_potential}), in the limit that tunneling through and thermally-induced excitations over the potential barrier separating the two minima $\phi \approx \pm \phi_0$ in a low-density environment can be neglected, the Universe settles in one and the same vacuum state $\phi \approx \mp \phi_0$ for the upper/lower sign choices in (\ref{linear_portal},\ref{phi4_effective-potential-LINEAR}). 
In other words, the Universe would contain only a single domain in this case, with no domain walls present. 
Nevertheless, even in the complete absence of topological defects of cosmological origin, there can still be a rich phenomenology from the types of non-transient effects of variations of the fundamental constants considered in our present paper, though topological defects with linear-in-$\phi$ interactions are generally strongly constrained from the consideration of equivalence-principle-violating forces that are independent of the parameter $\phi_0$ \cite{Varying_alpha_string_2005,Varying_alpha_monopole_2005,Varying_alpha_wall_2011}. 

Besides scalar-type linear-in-$\phi$ and quadratic-in-$\phi$ interactions which induce apparent variations of the fundamental constants, there are also derivative-type interactions of pseudoscalar fields that give rise to the precession of polarised spins \cite{Pospelov_2013_TDM-magnetometers,Stadnik_2014_axions}. 
These derivative-type interactions are expected to lead to (anti)screening of the pseudoscalar field in the presence of spin-polarised matter and in the vicinity of spin-polarised bodies, similarly to the (anti)screening of scalar fields in the presence of ordinary matter (regardless of it being spin polarised or unpolarised) for the case of scalar-type interactions. 
Unlike scalar-type interactions which produce isotropic scalar-field profiles around spherical bodies of ordinary matter, pseudoscalar-type interactions will lead to anisotropic pseudoscalar-field profiles around spherical bodies of spin-polarised matter, since the net spin vector of the spin-polarised body breaks the isotropy of space. 
Besides possible (anti)screening of pseudoscalar fields by spin-polarised components of apparata, pseudoscalar fields can also be (anti)screened by the spin-polarised geoelectrons inside Earth, which contains an estimated $\sim 10^{42}$ spin-polarised geoelectrons \cite{Hunter_2013_Geoelectrons-A,Hunter_2013_Geoelectrons-B}. 
Note that such (anti)screening effects are distinct from the screening mechanism considered in Ref.~\cite{Zolotorev_2016_Magnetic-screening}, in which pseudoscalar fields that interact specifically with electrons but are not necessarily related to topological defects can be screened by magnetic shielding made of a soft ferromagnetic or ferrimagnetic material. 
Regarding the possible effects of derivative-type interactions on the cosmological production of topological defects consisting of pseudoscalar fields, such interactions may be cosmologically relevant if some fraction of fermions in the early Universe was spin-polarised in the presence of primordial magnetic fields. 
In this case, the correlation features of the primordial magnetic fields may be imprinted into the resulting network of topological defects. 

We now briefly discuss models of topological defects other than domain walls. 
In the case of topological strings and monopoles with the $\phi^4$-type potentials in (\ref{phi4_potential-strings}) and (\ref{phi4_potential-monopoles}), respectively, and $\phi^2$-type interactions analogous to (\ref{quad_portal}), our results for topological domain walls with the $\phi^4$ potential (\ref{phi4_potential}) and the $\phi^2$ interactions (\ref{quad_portal}) carry over unchanged, modulo gauge-field effects. 
Further work investigating the role of the gauge fields in the case of topological strings and monopoles and their phenomenological implications is warranted. 
Regarding different possible forms of potentials, we expect the form of the bubble-like scalar field profile (\ref{scalar_profile_weak_screening-A},\ref{scalar_profile_weak_screening-B}) in the weak (anti)screening regime to be quite generic for a wide range of potentials admitting topological defects, since the linearised forms of the governing scalar-field differential equations (see Appendix~\ref{Sec:Appendix_weak}) generally reduce to a common form, thereby masking differences in the non-linear features of different effective potentials. 
On the other hand, in the strong (anti)screening regime, when the non-linear nature of the effective potential generally cannot be neglected, different forms of the scalar field potential may lead to different bubble-like scalar-field profile shapes. 
At the same time, the form of the scalar field potential plays an important role in determining the transverse profile (shape) of freely-propagating topological defects. 
Nevertheless, we generally expect that such changes in the shapes of bubble-like defects surrounding dense bodies and those of freely-propagating defects will be correlated and that the transverse size parameter $d$ will retain its role as one of the important parameters determining the sensitivity to $\Lambda'_X$. 
In our present paper, like in Refs.~\cite{Wcislo_2016_TDM-cavity,Roberts_2017_TDM-GPS,Wcislo_2018_TDM-cavity,Roberts_2019_TDM-clocks}, we have considered topological-defect networks containing only a single type of defect. 
In principle, a network of topological defects may contain more than one type of defect. 
In this case, non-transient signatures of the types considered in our present work may exhibit a temporal structure correlated with transitions between different vacua as topological defects pass through Earth. 
It is possible to search for such signatures arising in models that admit more than one type of defect by using dedicated search protocols, similar to those that would be required to search for transient signatures of a topological-defect network containing more than one type of defect.

Finally, let us address the possibility of non-topological solitons. 
Of particular interest are non-topological monopole-like solitons, such as Q-balls \cite{Q-ball_1976_2-field,Q-ball_1985_flat-directions}, which are a good candidate to explain the observed dark matter \cite{Q-ball_1997_DM,Q-ball_1998_DM}. 
Further work investigating the role of non-transient signatures, of the types considered in our present work, is warranted for various types of Q-balls, including Q-balls arising in two-field models \cite{Q-ball_1976_2-field} and in single-field models with suitable potentials \cite{Q-ball_1985_flat-directions,Q-ball_1997-specific_potential}. 
We expect the effects of a strongly repulsive or strongly attractive potential generated by Earth in the presence of $\phi^2$ interactions (and possibly other interactions) on the propagation of a passing non-topological soliton to be important; 
for example, a strongly repulsive potential generated by Earth would deflect an incident non-topological monopole. 
We also generally expect an increased importance of astrophysical measurements in models of non-topological monopoles compared with models of topological domain walls. 
Indeed, the average energy density associated with a network of monopole objects scales as $\rho_\textrm{monopoles} \propto a^{-3}$ with the scale factor, which decreases in time much faster than the average energy density associated with a network of domain walls (see Sec.~\ref{Sec:Main_Cosmology}). 
Therefore, the effects of varying fundamental constants induced by a network of non-topological monopoles with $\phi^2$ interactions will be strongly enhanced during the BBN and CMB epochs due to the very large energy density of the monopole network in the early Universe, similarly to models of scalar-field dark matter with $\phi^2$ interactions \cite{Stadnik_2015_DM-VFCs}. 
If the monopole scalar field(s) is already inhomogeneous during the BBN epoch, then the scalar-field inhomogeneities would be imprinted into the resulting spatial distributions of primordial light elements, in principle offering greater sensitivity to the underlying scalar-field interactions compared with the case when the monopole scalar field(s) is homogeneous during the BBN epoch. 
We remark that in the limiting case that the monopole scalar field(s) is extremely inhomogeneous during the BBN epoch, it may be possible to circumvent BBN constraints, since the primordial elements in the finite regions of the Universe at the rather small redshifts $z \lesssim \mathcal{O}(10)$, where primordial elemental abundances are measured, might not have arisen from regions of the Universe where monopole objects were located during the BBN epoch. 
However, since the number of different systems from which primordial elemental abundances are determined is quite large, in this case it would be extremely unlikely for even one such monopole to pass through Earth during our lifetime. 

In summary, we have pointed out that previous studies of macroscopic topological defects, which interact with standard-model fields via scalar-type couplings, overlooked a number of important effects associated with the back-action of ambient matter on the scalar field(s) comprising the topological defects. 
In particular, we have shown that such back-action effects produce an environmental dependence of the fundamental constants of nature, as well as spatial variations of the fundamental constants in the vicinity of dense bodies such as Earth. 
We have derived bounds on such non-transient variations of the fundamental constants from torsion-pendulum experiments that search for equivalence-principle-violating forces, experiments comparing the frequencies of ground- and space-based atomic clocks, as well as ground-based clocks at different heights, and measurements comparing atomic and molecular transition frequencies in terrestrial and low-density astrophysical environments. 
Our newly derived bounds on domain walls with the $\phi^2$ interactions in (\ref{quad_portal}) via their effects of non-transient variations of the fundamental constants are significantly more stringent than previously reported clock- and cavity-based limits on passing domain walls via transient signatures \cite{Wcislo_2016_TDM-cavity,Roberts_2017_TDM-GPS,Wcislo_2018_TDM-cavity,Roberts_2019_TDM-clocks} (regardless of their possible degradation due to previously neglected effects of the strongly repulsive potential generated by Earth on the passing domain walls) and previous bounds from different types of non-transient signatures (by about 10 orders of magnitude for wall thicknesses comparable to the size of Earth), under the same set of assumptions. 
We have also identified a number of interesting new problems for future investigation. 
In the case of experiments searching for transient signatures of domain walls that (may or may not) pass through Earth, numerical simulations studying the dynamics of domain walls incident on a strongly repulsive (or strongly attractive) potential generated by Earth are especially warranted. 
These simulations are directly relevant to the previous clock-based searches for transient signatures in Refs.~\cite{Wcislo_2016_TDM-cavity,Roberts_2017_TDM-GPS,Wcislo_2018_TDM-cavity,Roberts_2019_TDM-clocks}, which (unwittingly) probed regions of parameter space where the scalar field is strongly screened near the surface of and inside Earth. 
On the other hand, laser-interferometry-based searches for transient signatures of passing domain walls using existing gravitational-wave detectors such as LIGO and GEO600 \cite{Stadnik_2019_DM-LIFO} are already expected to be able to probe regions of parameter space that lie deep within the weak screening regime where back-action effects are negligible.

\vspace{200mm}

\appendix
\section{Appendix:~Derivation of scalar field profile around a spherical dense body}
\label{Sec:Appendix}

Here, we derive the scalar field profile around a homogeneous spherical dense body of radius $R$, at rest and surrounded by a vacuum containing no ambient matter, for the case of the scalar field potential (\ref{phi4_potential}) and interactions (\ref{quad_portal},\ref{quad_portal-atoms}). 
In this case, the time-independent differential equation for the scalar field reads: 
\begin{equation}
\label{Scalar_field_DE-general}
\frac{d^2 \phi}{d r^2} + \frac{2}{r} \frac{d \phi}{d r} - \lambda \left[ \phi^2 - \phi_0^2 + \frac{2 \rho_X (r)}{\lambda \left( \Lambda'_X \right)^2} \right] \phi(r) = 0  \, , 
\end{equation}
with the following density profile: 
\begin{equation}
\label{density_profile}
\rho_X (r) = 
\left\{\begin{array}{ll}
\rho_X  \, , &  0\leq r < R  \, ; \\
0  \, , &  r > R \, . 
\end{array}\right.
\end{equation}
We are not aware of an exact analytical solution to the non-linear differential equation (\ref{Scalar_field_DE-general}). 
Therefore, we consider in turn two limiting cases of physical importance, namely the ``weak screening'' regime and the ``strong screening'' regime, which give rise to qualitatively different scalar field profiles. 
We shall assume that the dense spherical body is located in a domain with vacuum value $+\phi_0$ (the overall signs of the scalar field profiles derived below are reversed if the dense body is located in a domain with vacuum value $-\phi_0$) and far away from any possible wall boundary joining two domains.

\subsection{``Weak screening'' regime}
\label{Sec:Appendix_weak}
The weak screening regime applies when the interactions of the scalar field with ambient matter in (\ref{quad_portal},\ref{quad_portal-atoms}) are sufficiently feeble to ensure that the condition $2 \rho_X / (\Lambda'_X)^2 \ll \lambda \phi_0^2$ is satisfied inside the dense body and/or the dense body is sufficiently small to prevent the scalar field from attaining its ``true'' minimum inside the dense body (since in this case it would be energetically unfavourable for the scalar field to appreciably change its amplitude across a region of space that is much smaller than the intrinsic wavelength of the field inside the dense body --- a consequence of the uncertainty principle). 
In the weak screening regime, the scalar field is only slightly perturbed by the dense body, allowing us to linearise the differential equation (\ref{Scalar_field_DE-general}) in all of space. 

Let us start with the simpler case when the condition $2 \rho_X / (\Lambda'_X)^2 \ll \lambda \phi_0^2$ is satisfied inside the dense body. 
Outside the dense body, we can linearise the differential equation (\ref{Scalar_field_DE-general}) about $\phi = \phi_0$ by approximating $\phi \approx \phi_0 + \delta \phi$ in this region to give the following linear differential equation: 
\begin{equation}
\label{Scalar_field_DE-Weak-A_outside}
\frac{d^2 (\delta \phi)}{d r^2} + \frac{2}{r} \frac{d (\delta \phi)}{d r} - 2 \lambda \phi_0^2 \, \delta \phi (r) \approx 0  \, , 
\end{equation}
which has the following solution that remains finite as $r \to \infty$: 
\begin{equation}
\label{Scalar_field_soln-Weak-A_outside}
\phi_\textrm{out} (r) \approx \phi_0 + \frac{A \exp \left( -2r / d \right)}{r}  \, , 
\end{equation}
where $d$ is the same size parameter as in Eq.~(\ref{wall_kink_thickness}): 
\begin{equation}
\label{wall_kink_thickness-App}
d = \sqrt{\frac{2}{\lambda}} \frac{1}{\phi_0}  \, . 
\end{equation}
Inside the dense body, we can linearise the differential equation (\ref{Scalar_field_DE-general}) about $\phi = \phi'_0 = \sqrt{\phi_0^2 - 2 \rho_X / [\lambda ( \Lambda'_X )^2]} \approx \phi_0 \{ 1 - \rho_X d^2 / [2 ( \Lambda'_X )^2] \}$ to give the following differential equation: 
\begin{equation}
\label{Scalar_field_DE-Weak-A_inside}
\frac{d^2 (\delta \phi)}{d r^2} + \frac{2}{r} \frac{d (\delta \phi)}{d r} - 2 \lambda (\phi'_0)^2 \, \delta \phi (r) \approx 0  \, , 
\end{equation}
which has the following solution that remains finite at $r = 0$: 
\begin{equation}
\label{Scalar_field_soln-Weak-A_inside}
\phi_\textrm{in} (r) \approx \phi'_0 + \frac{B \sinh \left( 2r / d' \right)}{r}  \, , 
\end{equation}
where the characteristic length scale, $d'$, over which the scalar field appreciably changes inside the dense body is defined analogously to Eq.~(\ref{wall_kink_thickness-App}): 
\begin{equation}
\label{char_length_scale-Weak-A_inside}
d' = \sqrt{\frac{2}{\lambda}} \frac{1}{\phi'_0} \approx d  \, . 
\end{equation}
The requirement of continuity of $\phi$ and $d \phi / dr$ at $r=R$ fixes the coefficients $A$ and $B$ in Eqs.~(\ref{Scalar_field_soln-Weak-A_outside}) and (\ref{Scalar_field_soln-Weak-A_inside}) to be: 
\begin{equation}
\label{coeffn-A_Weak-A_inside}
A = \frac{ \left( \phi'_0 - \phi_0 \right) d \exp \left( 2R/d \right) \left[ 2R \cosh \left( 2R/d' \right) - d' \sinh \left( 2R/d' \right) \right] }{ 2 \left[ d \cosh \left( 2R/d' \right) + d' \sinh \left( 2R/d' \right) \right] }  \, , 
\end{equation}
\begin{equation}
\label{coeffn-B_Weak-A_inside}
B = \frac{ \left( \phi_0 - \phi'_0 \right) d'  \left( d+2R \right)  }{ 2 \left[ d \cosh \left( 2R/d' \right) + d' \sinh \left( 2R/d' \right) \right] }  \, . 
\end{equation}
Noting that $d' \approx d$ and $\phi_0 - \phi'_0 \approx \phi_0 \rho_X d^2 / [2 ( \Lambda'_X )^2]$, we hence arrive at the scalar field profile outside of the spherical dense body as given in Eqs.~(\ref{scalar_profile_weak_screening-A},\ref{scalar_profile_weak_screening-B}): 
\begin{equation}
\label{Appendix:scalar_profile_weak_screening-A}
\phi_\textrm{out}(r) \approx \phi_0 - \frac{\phi_0 \rho_X d^2 R}{4 \left( \Lambda'_X \right)^2 r} \exp \left[ - \frac{2 \left( r - R \right)}{d} \right] ~~~\textrm{for} ~ d \ll R  \, , 
\end{equation}
\begin{equation}
\label{Appendix:scalar_profile_weak_screening-B}
\phi_\textrm{out}(r) \approx \phi_0 - \frac{2 \phi_0 \rho_X R^3}{3 \left( \Lambda'_X \right)^2 r} \exp \left[ - \frac{2 \left( r - R \right)}{d} \right] ~~~\textrm{for} ~ d \gg R  \, . 
\end{equation}
We thus see that the scalar field is weakly screened in the vicinity of and inside the dense body in this case.

In the case of the interactions (\ref{quad_portal},\ref{quad_portal-atoms}) but with the signs reversed, the sign of $\phi_0 - \phi'_0$ is reversed and the scalar field profile outside of the spherical dense body instead takes the following form: 
\begin{equation}
\label{Appendix:scalar_profile_weak_antiscreening-A}
\phi_\textrm{out}(r) \approx \phi_0 + \frac{\phi_0 \rho_X d^2 R}{4 \left( \Lambda'_X \right)^2 r} \exp \left[ - \frac{2 \left( r - R \right)}{d} \right] ~~~\textrm{for} ~ d \ll R  \, , 
\end{equation}
\begin{equation}
\label{Appendix:scalar_profile_weak_antiscreening-B}
\phi_\textrm{out}(r) \approx \phi_0 + \frac{2 \phi_0 \rho_X R^3}{3 \left( \Lambda'_X \right)^2 r} \exp \left[ - \frac{2 \left( r - R \right)}{d} \right] ~~~\textrm{for} ~ d \gg R  \, . 
\end{equation}
In this case, the scalar field is weakly antiscreened in the vicinity of and inside the dense body. 
We note that the forms of the scalar field profiles in (\ref{Appendix:scalar_profile_weak_screening-A}--\ref{Appendix:scalar_profile_weak_antiscreening-B}), which arise in the linearised regime, match those in a wide variety of scalar-field models with linear field equations.

Let us now consider the other case of interest when the condition $2 \rho_X / (\Lambda'_X)^2 \ll \lambda \phi_0^2$ is \textit{not} satisfied inside the dense body, but the dense body is sufficiently small to prevent the scalar field from attaining its ``true'' minimum inside the dense body. 
In the exterior region, the linearised scalar-field differential equation (\ref{Scalar_field_DE-Weak-A_outside}) and resulting solution (\ref{Scalar_field_soln-Weak-A_outside}) remain unchanged: 
\begin{equation}
\label{Scalar_field_soln-Weak-B_outside}
\phi_\textrm{out} (r) \approx \phi_0 + \frac{A \exp \left( -2r / d \right)}{r}  \, . 
\end{equation}
For the interior region, in the limiting case when $ \rho_X / (\Lambda'_X)^2 \gg \lambda \phi_0^2$, it turns out to be sufficient to formally linearise the differential equation (\ref{Scalar_field_DE-general}) about $\phi = 0$: 
\begin{equation}
\label{Scalar_field_DE-Weak-B_inside}
\frac{d^2 (\delta \phi)}{d r^2} + \frac{2}{r} \frac{d (\delta \phi)}{d r} - \lambda (\phi'_0)^2 \, \delta \phi (r) \approx 0  \, , 
\end{equation}
where $\phi'_0 = \sqrt{ 2 \rho_X / [\lambda ( \Lambda'_X )^2] - \phi_0^2} \approx \sqrt{ 2 \rho_X / [\lambda ( \Lambda'_X )^2]} $, and which has the following solution that remains finite at $r = 0$: 
\begin{equation}
\label{Scalar_field_soln-Weak-B_inside}
\phi_\textrm{in} (r) \approx \frac{B \sinh \left( r / d' \right)}{r}  \, , 
\end{equation}
where the characteristic length scale, $d'$, over which the scalar field appreciably changes inside the dense body is the same as in Eq.~(\ref{wall_thickness-dense_region}): 
\begin{equation}
\label{char_length_scale-Weak-B_inside}
d' = \frac{1}{\sqrt{\lambda} \phi'_0} \approx \frac{\Lambda'_X}{\sqrt{2 \rho_X}}  \, . 
\end{equation}
The requirement of continuity of $\phi$ and $d \phi / dr$ at $r=R$ fixes the coefficients $A$ and $B$ in Eqs.~(\ref{Scalar_field_soln-Weak-B_outside}) and (\ref{Scalar_field_soln-Weak-B_inside}) to be: 
\begin{equation}
\label{coeffn-A_Weak-B_inside}
A = - \frac{ \phi_0 d \exp \left( 2R/d \right) \left[ R \cosh \left( R/d' \right) - d' \sinh \left( R/d' \right) \right] }{  d \cosh \left( R/d' \right) + 2d' \sinh \left( R/d' \right)  }  \, , 
\end{equation}
\begin{equation}
\label{coeffn-B_Weak-B_inside}
B = \frac{ \phi_0 d' \left( d+2R \right)  }{  d \cosh \left( R/d' \right) + 2d' \sinh \left( R/d' \right)  }  \, . 
\end{equation}
Noting that $d' \ll d$, we find that the scalar field profile outside of the spherical dense body matches that in (\ref{Appendix:scalar_profile_weak_screening-B}) for the limiting case when $R \ll d' \ll d$. 
In other words, the scalar field is only weakly screened in the vicinity of and inside the dense body, since the dense body is too small to allow the scalar field to attain its ``true'' minimum near $\phi = 0$ inside the dense body [even though we have formally linearised the differential equation (\ref{Scalar_field_DE-general}) about $\phi = 0$ inside the dense body].

\subsection{``Strong screening'' regime}
\label{Sec:Appendix_strong}
The strong screening regime applies when the interactions of the scalar field with ambient matter in (\ref{quad_portal},\ref{quad_portal-atoms}) are sufficiently strong to ensure that the condition $2 \rho_X / (\Lambda'_X)^2 \gg \lambda \phi_0^2$ is satisfied inside the dense body \textit{and} the dense body is sufficiently large to ensure that the scalar field can attain its ``true'' minimum near $\phi = 0$ inside the dense body. 
In the strong screening regime, the scalar field is strongly affected by the dense body, necessitating a more delicate treatment than the treatment of the weak screening regime in Sec.~\ref{Sec:Appendix_weak} above. 
In particular, the two-region approximation used, e.g., in Refs.~\cite{Khoury_2010-symmetron,Vardanyan_2019-symmetron} is insufficient to capture the qualitative features of the screening of the scalar field near the surface of a dense body for $d \ll R$. 
Therefore, we employ a three-region approximation as outlined below. 
We focus on the simplest limiting case when the condition $\rho_X / (\Lambda'_X)^2 \gg \lambda \phi_0^2$ is satisfied inside the dense body. 
In this limiting case, $d' \ll R$ and $d' \ll d$, with $d'$ given by Eq.~(\ref{char_length_scale-Weak-B_inside}). 

Sufficiently far away from the dense body where $\phi \approx \phi_0$ --- denoted by ``region I'' --- we can linearise the differential equation (\ref{Scalar_field_DE-general}) about $\phi = \phi_0$, leading to the same solution as in (\ref{Scalar_field_soln-Weak-A_outside},\ref{Scalar_field_soln-Weak-B_outside}): 
\begin{equation}
\label{Scalar_field_soln-Strong_region-I}
\phi_\textrm{I} (r) \approx \phi_0 + \frac{A \exp \left( -2r / d \right)}{r}  \, , 
\end{equation}
where $d$ is given by (\ref{wall_kink_thickness-App}). 
Sufficiently deep inside the dense body where $\phi \ll \phi_0$ --- denoted by ``region III'' --- we can linearise the differential equation (\ref{Scalar_field_DE-general}) about $\phi = 0$, leading to the same solution as in (\ref{Scalar_field_soln-Weak-B_inside}): 
\begin{equation}
\label{Scalar_field_soln-Strong_region-III}
\phi_\textrm{III} (r) \approx \frac{D \sinh \left( r / d' \right)}{r}  \, , 
\end{equation}
where $d'$ is given by (\ref{char_length_scale-Weak-B_inside}). 
Additionally, there is also a transition region --- denoted by ``region II'' --- that joins regions I and III. 
Our aim is to exploit an approximately linear form of the governing differential equation (\ref{Scalar_field_DE-general}) in this transition region. 
\textit{A priori}, there are two possibilities for region II: 
(i) Region II lies inside the dense body in the spatial domain $r_* \le r \le R$, in which the scalar field amplitude lies in the range $\phi_0/\sqrt{2} \le \phi \lesssim \phi_0$; 
(ii) Region II lies outside the dense body in the spatial domain $R \le r \le r_*$, in which the scalar field amplitude lies in the range $0 \lesssim \phi \le \phi_0/\sqrt{2}$. 
It is straightforward to demonstrate that the requirement of continuity of $\phi$ and $d\phi/dr$ at $r = R$ prevents possibility (i). 
Therefore, only possibility (ii) is relevant. 
Linearising the differential equation (\ref{Scalar_field_DE-general}) in region II about $\phi = 0$ leads to the following differential equation: 
\begin{equation}
\label{Scalar_field_DE-Strong_region-II}
\frac{d^2 (\delta \phi)}{d r^2} + \frac{2}{r} \frac{d (\delta \phi)}{d r} + \lambda \phi_0^2 \, \delta \phi (r) \approx 0  \, , 
\end{equation}
which has the following solution: 
\begin{equation}
\label{Scalar_field_soln-Strong_region-II}
\phi_\textrm{II} (r) \approx \frac{B \sin \left( \sqrt{2} r / d \right)}{r} + \frac{C \cos \left( \sqrt{2} r / d \right)}{r}  \, . 
\end{equation}
Note that there are 5 unknown parameters (including $r_*$) in Eqs.~(\ref{Scalar_field_soln-Strong_region-I}), (\ref{Scalar_field_soln-Strong_region-III}) and (\ref{Scalar_field_soln-Strong_region-II}), supplemented by 5 conditions from the requirement of continuity of $\phi$ and $d \phi / d r$ at $r = R$ and $r = r_*$, as well as the requirement that $\phi_\textrm{I} (r_*) = \phi_\textrm{II} (r_*) = \phi_0 / \sqrt{2}$. 
The requirement of continuity of $\phi$ and $d \phi / d r$ at $r = R$ and $r = r_*$ fixes the coefficients $A$, $B$, $C$ and $D$ in Eqs.~(\ref{Scalar_field_soln-Strong_region-I}), (\ref{Scalar_field_soln-Strong_region-III}) and (\ref{Scalar_field_soln-Strong_region-II}) as functions of the parameter $r_*$: 
\begin{equation}
\normalsize
\label{coeffn-A_Strong}
A (r_*) = 
\frac{ \phi_0 \exp \left( \frac{2 r_*}{d} \right) \left\{ \sqrt{2} d \cos \left( \frac{R - r_*}{d / \sqrt{2}} \right) \left[ -r_* \cosh \left( \frac{R}{d'} \right) + d' \sinh \left( \frac{R}{d'} \right) \right] - \sin \left( \frac{R - r_*}{d / \sqrt{2}} \right) \left[ d^2 \cosh \left( \frac{R}{d'} \right) + 2 r_* d' \sinh \left( \frac{R}{d'} \right) \right] \right\} }
{ 2 \sin \left( \frac{R - r_*}{d / \sqrt{2}} \right) \left[ -d \cosh \left( \frac{R}{d'} \right) + d' \sinh \left( \frac{R}{d'} \right) \right] + \sqrt{2} \cos \left( \frac{R - r_*}{d / \sqrt{2}} \right) \left[ d \cosh \left( \frac{R}{d'} \right) + 2d' \sinh \left( \frac{R}{d'} \right) \right] }  \, , 
\end{equation}
\begin{equation}
\label{coeffn-B_Strong}
B (r_*) = 
\frac{ \phi_0 \left( d + 2 r_* \right) \left[ d \cos \left( \frac{\sqrt{2} R}{d} \right) \cosh \left( \frac{R}{d'} \right) + \sqrt{2} d' \sin \left( \frac{\sqrt{2} R}{d} \right) \sinh \left( \frac{R}{d'} \right) \right] }
{ 2 \sin \left( \frac{R - r_*}{d / \sqrt{2}} \right) \left[ -d \cosh \left( \frac{R}{d'} \right) + d' \sinh \left( \frac{R}{d'} \right) \right] + \sqrt{2} \cos \left( \frac{R - r_*}{d / \sqrt{2}} \right) \left[ d \cosh \left( \frac{R}{d'} \right) + 2d' \sinh \left( \frac{R}{d'} \right) \right] }  \, , 
\end{equation}
\begin{equation}
\label{coeffn-C_Strong}
C (r_*) = - 
\frac{ \phi_0 \left( d + 2 r_* \right) \left[ d \sin \left( \frac{\sqrt{2} R}{d} \right) \cosh \left( \frac{R}{d'} \right) - \sqrt{2} d' \cos \left( \frac{\sqrt{2} R}{d} \right) \sinh \left( \frac{R}{d'} \right) \right] }
{ 2 \sin \left( \frac{R - r_*}{d / \sqrt{2}} \right) \left[ -d \cosh \left( \frac{R}{d'} \right) + d' \sinh \left( \frac{R}{d'} \right) \right] + \sqrt{2} \cos \left( \frac{R - r_*}{d / \sqrt{2}} \right) \left[ d \cosh \left( \frac{R}{d'} \right) + 2d' \sinh \left( \frac{R}{d'} \right) \right] }  \, , 
\end{equation}
\begin{equation}
\label{coeffn-D_Strong}
D (r_*) = 
\frac{ \sqrt{2} \phi_0 \left(d + 2r_* \right) d' }
{ d \cosh \left( \frac{R}{d'} \right) \left[ \sqrt{2} \cos \left( \frac{R - r_*}{d / \sqrt{2}} \right) - 2 \sin \left( \frac{R - r_*}{d / \sqrt{2}} \right) \right] + 2d' \sinh \left( \frac{R}{d'} \right) \left[ \sqrt{2} \cos \left( \frac{R - r_*}{d / \sqrt{2}} \right) + \sin \left( \frac{R - r_*}{d / \sqrt{2}} \right) \right] }  \, . 
\end{equation}

We are particularly interested in the form of the scalar field profile outside of the spherical dense body. 
To proceed, we impose the fifth and final condition $\phi_\textrm{I} (r_*) = \phi_\textrm{II} (r_*) = \phi_0 / \sqrt{2}$. 
Let us first consider the simpler case when $d \gg r_* - R$. 
In this case, we find that: 
\begin{equation}
\label{scalar_profile_strong_screening_App-1}
\phi_\textrm{II}(r) \approx \frac{\phi_0 d'}{R} + \phi_0 \left( 1 - \frac{R}{r} \right)  ~~~\textrm{for} ~ r - R \ll r_* - R  \, , 
\end{equation}
\begin{equation}
\label{scalar_profile_strong_screening_App-1B}
\phi_\textrm{I}(r) \approx \phi_0 - \frac{\phi_0 R}{r} \exp \left[ - \frac{2 \left( r - r_* \right)}{d} \right]  ~~~\textrm{for} ~ r - R \gg r_* - R  \, , 
\end{equation}
where $r_* \approx R / (1 - 1 / \sqrt{2}) \approx 3.4 R$. 
Note that the condition $d \gg r_* - R$ necessitates $d \gg R$. 
Therefore, in the other limiting case when $d \ll R$, we must have $d \lesssim r_* - R$. 
We note that if $d \ll r_* - R$, then the scalar field would oscillate numerous times in region II, corresponding to a non-ground-state field configuration. 
Hence in the case when $d \ll R$, we must have $d \sim r_* - R$ for the ground-state field configuration, and we find that: 
\begin{equation}
\label{scalar_profile_strong_screening_App-2}
\phi_\textrm{II}(r) \sim \frac{\phi_0 d'}{d} + \frac{\phi_0 R}{d} \left( 1 - \frac{R}{r} \right)  ~~~\textrm{for} ~ r - R \ll d  \, , 
\end{equation}
\begin{equation}
\label{scalar_profile_strong_screening_App-2B}
\phi_\textrm{I}(r) - \phi_0 \sim - \frac{\phi_0 R}{r} \exp \left[ - \frac{2 \left( r - r_* \right)}{d} \right]  ~~~\textrm{for} ~ r - R \gg d  \, . 
\end{equation}
For the limiting cases when $d \ll R$ or $d \gg R$, we can combine Eqs.~(\ref{scalar_profile_strong_screening_App-1}), (\ref{scalar_profile_strong_screening_App-1B}), (\ref{scalar_profile_strong_screening_App-2}) and (\ref{scalar_profile_strong_screening_App-2B}) into the single unified form given in Eqs.~(\ref{scalar_profile_strong_screening-A},\ref{scalar_profile_strong_screening-Bx}): 
\begin{equation}
\label{scalar_profile_strong_screening-App-3}
\phi_\textrm{II}(r) \sim \frac{\phi_0 d' + \phi_0 R (1 - R/r)}{\textrm{min} \left( d , R \right)}  ~~~\textrm{for} ~ r - R \ll \textrm{min} \left( d , R \right)  \, , 
\end{equation}
\begin{equation}
\label{scalar_profile_strong_screening-App-3B}
\phi_\textrm{I}(r) - \phi_0 \sim - \frac{\phi_0 R}{r} \exp \left[ -\frac{2 \left( r - r_* \right)}{d} \right]  ~~~\textrm{for} ~ r - R \gg \textrm{min} \left( d , R \right)  \, , 
\end{equation}
where $r_* - R \sim \textrm{min} (d,R)$. 
We thus see that the scalar field is strongly screened near the surface of and inside the dense body in this case.

\vspace{200mm}

\section*{Acknowledgements}
I am grateful to Pedro Avelino, Dmitry Budker, Josh Eby, Sylvain Fichet, Victor Flambaum, Aurelien Hees, Hidetoshi Katori, Hitoshi Murayama, Alex Kusenko, Tom Melia, Joey Smiga, Tadahiro Takahashi, Tadayuki Takahashi and Volodymyr Takhistov for helpful discussions. 
I am grateful to Tadahiro Takahashi for helpful information about the geometry of the Tokyo Skytree experiment in Ref.~\cite{TokyoSkytree_2020}. 
I am grateful to Aurelien Hees for helpful information about the Galileo satellite measurements in Ref.~\cite{Eccentric_Galileo_2018A}. 
I am grateful to Michael Murphy and John Webb for helpful information about the gas densities in quasar absorption systems. 
This work was supported by the World Premier International Research Center Initiative (WPI), MEXT, Japan, and by the JSPS KAKENHI Grant Number JP20K14460.



\end{document}